\documentclass[aos,preprint]{imsart}

\RequirePackage{amsthm,amsmath}
\RequirePackage[numbers]{natbib}

\usepackage{amsfonts}
\usepackage{amssymb}
\usepackage{rotating}
\usepackage{bm}
\usepackage{color}

\newtheorem{thm}{Theorem}
\newtheorem{prop}{Proposition}
\newtheorem{lem}{Lemma}
\newtheorem{rmk}{Remark}
\newcommand{\RR}{\mathbb{R}}

\newcommand{\RN}{{\rm N}}
\newcommand{\RE}{{\rm E}}
\newcommand{\var}{{\rm Var}}
\newcommand{\bias}{{\rm Bias}}

\newcommand{\mx}{{m_{\rm max}}}
\newcommand{\bH}{\bm{H}}

\newcommand{\argmin}{\operatornamewithlimits{argmin}}

\begin{document}

\setcounter{equation}{0}
\setcounter{section}{0}
\setcounter{subsection}{0}
\setcounter{table}{0}
\setcounter{figure}{0}
\setcounter{page}{1}
\setcounter{thm}{1}
\setcounter{prop}{0}
\setcounter{lem}{0}
\renewcommand{\theequation}{S.\arabic{equation}}
\renewcommand{\thesection}{S.\arabic{section}}
\renewcommand{\thethm}{S.\arabic{thm}}
\renewcommand{\theprop}{S.\arabic{prop}}
\renewcommand{\thelem}{S.\arabic{lem}}
\renewcommand{\thetable}{S.\arabic{table}}

\begin{center}
{\bf \large Supplement to ``Efficient estimation in semivarying coefficient models for longitudinal/clustered data''}\\
by Ming-Yen Cheng, Toshio Honda, and Jialiang Li
\end{center}

\section{Additional simulation results}

\subsection{Nonparametric component estimates}

In Step 7 of our estimation procedure we give both local linear and spline approaches to estimation the nonparametric component after the efficient estimator $\widehat{\bm{\beta}}_{\widehat{\bm{\Sigma}}}$ is obtained. 
In this section we examine the finite sample performance via simulations. For comparison, we also computed the respective initial estimates, that is, the version using $\widehat{\bm{\beta}}_{\bm{I}}$ instead of 
$\widehat{\bm{\beta}}_{\widehat{\bm{\Sigma}}}$. We considered the same settings in Section \ref{sec:numerical}, and we used cross-validation to choose the bandwidth used in the local linear estimation. We computed the mean integrated square error (MISE) for all the function estimates and took their average. The results are given in Table \ref{psid}.

The figures in Table \ref{psid} indicate that it is clearly advantageous to update the nonparametric component after efficient estimation of the parametric component. In addition, we observe that the refine local linear and spline estimators perform roughly the same in terms of MISE. 

\begin{table}[htbp]
\begin{center}
\caption{ \it MISE for simulation studies. }
\label{psid}
\begin{tabular}{lcccc}
\hline\hline
& \multicolumn{2}{c}{Local linear estimate} & \multicolumn{2}{c}{Spline estimate}\\
\cline{2-5}
& Initial & Refined & Initial & Refined\\
n=100&\\
$\rho=.4$&.0449&.0354&.0492&.0376\\
$\rho=.8$&.0691&.0597&.0639&.0593\\
n=200&\\
$\rho=.4$&.0390&.0315&.0415&.0355\\
$\rho=.8$&.0595&.0589&.0584&.0576\\
\end{tabular}
\end{center}
\end{table}

\subsection{Parametric component estimates}

We note that we adjusted the covariance function $\widehat\sigma(s,t)$ by setting all negative eigenvalues to be zero. We also considered a strictly positive threshold $\lambda_L=0.05$ and set all eigenvalues lower than $\lambda_L$ to be zero. The estimator using this covariance estimate is denoted by ``Positive'' in Table \ref{table-1s}.
The ``positive'' estimator includes an adjustment when estimating the covariance function by setting eigenvalues lower than a positive cut-off to be zero while the efficient estimator only adjusts the negative eigenvalues. Therefore, it is slightly more biased than the efficient estimator. In all the considered cases, the crude and positive estimators are still more efficient than the working  independence estimator. 

Recall that in all the numerical analysis reported in the paper, $h_1$ and $h_2$ were selected via the commonly used leave-one-subject-out cross-validation, and the bandwidth $h_3$ used in the estimation of the covariance structure were selected as $h_3=2 h_1$. 
To examine effects of the bandwidth choice, we considered various choices of $h_3$ in the numerical studies and obtained quite similar results. Under the column ``Different $h_3$'', we report the results for another case when $h_3=1.5 h_1$, which are similar to those obtained when $h_3=2 h_1$.

Our procedure does not require any iteration. In practice it may be interesting to refine the estimation of coefficients and covariances using iterations and obtain a final estimation upon convergence. We report the numerical results under the ``Iterative'' column. The bias and SE are very close to those obtained without iteration.

\begin{table}[htbp]
\begin{center}
\caption{Estimation results of 200 simulations. ``Positive'' means we set a positive threshold for the covariance eigenvalues; ``Different $h_3$'' means using a different choice of $h_3$ in our efficient estimation; ``Iterative'' indicates an iterative estimation approach. } \label{table-1s}
{\scriptsize
\begin{tabular}{rrrllllll}
\hline
&&&\multicolumn{2}{c}{Positive}&\multicolumn{2}{c}{Different $h_3$}&\multicolumn{2}{c}{Iterative}\\
$n$& $\rho$ & & bias & SE  & bias & SE  & bias & SE\\
\hline
100& 0.4& $\beta_{1}$ &.0173& .0411  & -.0152& .0375 & -.0146 & .0361\\
     &&$\beta_{2}$ &.0176& .0423  & -.0098& .0375 & -.0095 & .0352\\
     &&$\beta_{3}$ &.0205& .0425 & -.0122 & .0369 & -.0099 & .0360\\
     &&$\beta_{4}$ &-.0096& .0425 & .0098& .0373 & -.0086 & .0362\\
200& 0.4& $\beta_{1}$ &-.0113&.0329 & .0056& .0274  & .0045 & .0228\\
     &&$\beta_{2}$ &-.0164&.0334 & -.0099& .0274 & -.0066 & .0219\\
&&$\beta_{3}$ &.0120&.0323 & .0072& .0273 & .0034 & .0259\\
&&$\beta_{4}$ &-.0095 & .0329  & -.0043& .0276 & -.0035 & .0274\\
100& 0.8& $\beta_{1}$ &.0202&.0366  & .0082& .0336 & .0065& .0325\\
     &&$\beta_{2}$ &.0163&.0378 & -.0075 & .0335 & -.0034& .0323\\
     &&$\beta_{3}$ &.0197& .0372  & .0166& .0337 & .0121& .0328\\
     &&$\beta_{4}$ &-.0168&.0354 & -.0182& .0338 & .0157 & .0325\\
200& 0.8& $\beta_{1}$ &-.0044& .0214  & -.0124& .0202& .0056& .0199\\
     &&$\beta_{2}$ &.0036& .0215& .0138 & .0200 & -.0049& .0199\\
&&$\beta_{3}$ & .0042&.0215 & .0165 & .0204 & .0052 & .0178\\
&&$\beta_{4}$ & -.0038& .0214 & -.0148 & .0200 & -.0050 & .0179\\
\hline
\hline
\end{tabular}
}
\end{center}
\end{table}

\section{Proofs of Propositions
\ref{prop:prop1}-\ref{prop:prop3} and Lemma \ref{lem:lem0}}

In this section, we outline the proofs of
Propositions \ref{prop:prop1}-\ref{prop:prop3}
and present the proof of Lemma \ref{lem:lem0}.
When $m_i$ is uniformly bounded, we have the same results
for general link functions by just following closely the arguments of 
\cite{CZH2014S}. We outline the results at the end of this supplement.
Note that the sub-Gaussian error assumption is necessary in that case.
We outline the proofs of Propositions \ref{prop:prop1}-\ref{prop:prop3}
since we allow some of the $m_i$'s to diverge as in Assumptions A1 and A2.
\newpage

\medskip
\noindent
{\it Proof of Proposition \ref{prop:prop1}. }\,\,
First we consider the properties of $\Gamma_{\bm{V}}$. The $(k,l)$th element
of $n^{-1}\bm{H}_{11\cdot 2}$ is given by
\[
\langle X_k- \bm{Z}^T\widehat{\bm{\varphi}}_{\bm{V}k},
X_l - \bm{Z}^T \widehat{\bm{\varphi}}_{\bm{V}l}
\rangle_n^V.
\]
From Lemma \ref{lem:lem0} (v)-(vii), we have
\begin{align*}
\langle X_k- \bm{Z}^T\widehat{\bm{\varphi}}_{\bm{V}k},
X_l - \bm{Z}^T \widehat{\bm{\varphi}}_{\bm{V}l} \rangle_n^V 
& = \langle X_k- \bm{Z}^T \bm{\varphi}_{\bm{V}k}^*,
X_l - \bm{Z}^T \bm{\varphi}_{\bm{V}l}^* \rangle_n^V + o_p(1)\\
& = \langle X_k- \bm{Z}^T \bm{\varphi}_{\bm{V}k}^*,
X_l - \bm{Z}^T \bm{\varphi}_{\bm{V}l}^* \rangle^V + o_p(1).
\end{align*}
This and (\ref{eqn:e236}) imply that for some positive constants
$C_1$ and $C_2$, we have
\begin{align}
C_1 & \le \lambda_{\rm min}( n^{-1} \bm{H}_{11\cdot 2} ) \le
\lambda_{\rm max}( n^{-1} \bm{H}_{11\cdot 2} ) \le C_2 \nonumber\\ 
\intertext{and hence}
\frac{1}{nC_2} & \le \lambda_{\rm min}(\bH^{11})
\le \lambda_{\rm max}( \bH^{11} ) \le \frac{1}{nC_1}
\label{eqn:r703}
\end{align}
with probability tending to 1. 
Note that
\[
\var ( \widehat{\bm{\beta}}_{\bm{V}} \, | \, \{ \bm{X}_{ij} \},
\{ \bm{Z}_{ij} \}, \{ T_{ij} \}) = \Gamma_{\bm{V}} 
\]
and Theorem 1 of \cite{HZZ2007} implies that
$\Gamma_{\bm{V}} - \bH^{11}$ 
is nonnegative definite when $\bH^{11}$ is defined with $\bm{V}_i=\bm{\Sigma}_i$.
Hence for some positive constant $C_3$,  we have
\[
\lambda_{\rm min}( \Gamma_{\bm{V}}) \ge \frac{C_3}{n}
\]
with probability tending to 1.

Now we prove the asymptotic normality of
\begin{eqnarray*}
\lefteqn{ \widehat{\bm{\beta}}_{\bm{V}} - \RE \{
\widehat{\bm{\beta}}_{\bm{V}} \, | \, \{ \bm{X}_{ij} \},
\{ \bm{Z}_{ij} \}, \{ T_{ij} \} \} }\\
& = & \bH^{11} \Big( \sum_{i=1}^n\bm{X}_i^T \bm{V}_i^{-1}
\underline{\epsilon}_i - \bH_{12}\bH_{22}^{-1}
\sum_{i=1}^n \bm{W}_i^T \bm{V}_i^{-1} \underline{\epsilon}_i
\Big).
\end{eqnarray*}
As in the proof of Theorem 2 of \cite{HZZ2007}, we take
$c\in \RR^p$ such that $|c|=1$ and write
\[
c^T ( \widehat{\bm{\beta}}_{\bm{V}} - \RE \{
\widehat{\bm{\beta}}_{\bm{V}} \, | \, \{ \bm{X}_{ij} \},
\{ \bm{Z}_{ij} \}, \{ T_{ij} \} \}  )
= \sum_{i=1}^n a_i\eta_i\qquad {\rm (say)},
\]
where
\[
a_i^2 = c^T \bH^{11}(\underline{\bm{X}}_i - \underline{\bm{W}}_i
\bH_{22}^{-1}\bH_{21})^T \bm{V}_i^{-1}\bm{\Sigma}_i \bm{V}_i^{-1}
(\underline{\bm{X}}_i - \underline{\bm{W}}_i \bH_{22}^{-1}\bH_{21})
\bH^{11}c
\]
and $\{ \eta_i \}$ is a sequence of conditionally independent
random variables with
\[
\RE \{ \eta _i \, | \, \{ \bm{X}_{ij} \},
\{ \bm{Z}_{ij} \}, \{ T_{ij} \} \}  = 0
\quad {\rm and}\quad \var (\eta _i \, | \, \{ \bm{X}_{ij} \},
\{ \bm{Z}_{ij} \}, \{ T_{ij} \})=1.
\]
We have from (\ref{eqn:r703}) and Lemma \ref{lem:lem0} (vii)
that
\[
\max_{1\le i \le n} a_i^2
= O_p \Big( \frac{ m_{\rm max}^2 }{n^2}
\sum_{k=1}^p \| X_k - \bm{Z}^T
\widehat{\bm{\varphi}}_{\bm{V}k}\|_\infty^2 \big)
= O_p \Big(  \frac{ m_{\rm max}^2 }{n^2} \Big).
\]
On the other hand, we have for some positive constant $C_4$,
\[
\sum_{i=1}^na_i^2 = c^T \Gamma_{\bm{V}} c \ge \frac{C_4}{n}
\]
with probability tending to 1. Hence we have established
\[
\frac{\max_{1\le i \le n}a_i^2}{\sum_{i=1}^na_i^2}
=O_p( n^{-1}m_{\rm max}^2)= o_p(1)
\]
and it follows from the standard argument that
\begin{equation}
\Big( \sum_{i=1}^n a_i^2 \Big)^{-1/2} \sum_{i=1}^n a_i\eta_i 
\stackrel{d}{\to} \RN ( 0, 1).
\label{eqn:r706}
\end{equation}

Finally we evaluate the conditional bias:
\[
\bias_\beta = \RE \{ \widehat{\bm{\beta}}_{\bm{V}} \, | \,  \{ \bm{X}_{ij} \},
\{ \bm{Z}_{ij} \}, \{ T_{ij} \} \} -\bm{\beta}_0
\]
Take $\widetilde{\bm{g}} \in \bm{G}_B$ such that
$ \| \bm{g}_0 - \widetilde{\bm{g}} \|_{G,\infty}
=O(K_n^{-2})$ and set
\[
\bm{\delta}_0 = \bm{g}_0 - \widetilde{\bm{g}}
\quad{\rm and}\quad \delta_0 = \bm{Z}^T\bm{\delta}_0.
\]
Note that
\[
\| \delta_0 \|_\infty = O(K_n^{-2})\quad
{\rm and}\quad \| \delta_0 \|^V = O(K_n^{-2}).
\]
We also take $\widetilde{\bm{\varphi}}_{\bm{V}k}\in \bm{G}_B$ such that
$\| \bm{\varphi}_{\bm{V}k}^* - \widetilde{\bm{\varphi}}_{\bm{V}k}\|_{G,\infty}
=O(K_n^{-2})$.
Then we have the following  expression for the conditional bias:
\[
\bias_\beta = n \bH^{11}(S_1, \ldots, S_p )^T,
\]
where
\begin{align*}
S_k & = \langle X_k, \delta_0 - \bm{Z}^T\widehat{\Pi}_{\bm{V}n}
\delta_0 \rangle_n^V
=  \langle X_k -\bm{Z}^T\widetilde{\bm{\varphi}}_{\bm{V}k},
\delta_0 - \bm{Z}^T\widehat{\Pi}_{\bm{V}n}
\delta_0 \rangle_n^V
\\
& = \langle X_k - \bm{Z}^T \bm{\varphi}_{\bm{V}k}^*, \delta_0 - \bm{Z}^T\Pi_{\bm{V}n}
\delta_0 \rangle_n^V\\
& \qquad + \langle X_k - \bm{Z}^T \bm{\varphi}_{\bm{V}k}^*,
\bm{Z}^T\Pi_{\bm{V}n} \delta_0 - \bm{Z}^T\widehat{\Pi}_{\bm{V}n}
\delta_0 \rangle_n^V\\
& \qquad + \langle \bm{Z}^T \bm{\varphi}_{\bm{V}k}^*
- \bm{Z}^T\widetilde{\bm{\varphi}}_{\bm{V}k},
\delta_0 - \bm{Z}^T\widehat{\Pi}_{\bm{V}n}
\delta_0 \rangle_n^V\\
& = S_{1k}+ S_{2k}+ S_{3k} \qquad {\rm (say)}.
\end{align*}
Note that
\[
\RE \{ S_{1k} \} = 0 \quad {\rm and} \quad
\RE \{ S_{1k}^2 \} = O\Big( \frac{(\| X_k - \bm{Z}^T 
\bm{\varphi}_{\bm{V}k}^* \|^V)^2}{K_n^3n} \Big)
\]
since $S_{1k}$ is a sum of independent random variables,
$ \bm{\varphi}_{\bm{V}k}^* = \Pi_{\bm{V}} X_k $, $\delta_0 = \bm{Z}^T
\bm{\delta}_0$, and
\begin{align*}
\| \delta_0 - \bm{Z}^T\Pi_{\bm{V}n} \delta_0 \|_\infty
& \le \| \delta_0 \|_\infty + C K_n^{1/2}\| 
\bm{Z}^T\Pi_{\bm{V}n} \delta_0 \|^V\\
& \le \| \delta_0 \|_\infty + CK_n^{1/2}\| 
\delta_0 \|^V = O(K_n^{-3/2}).
\end{align*}
Hence we have
\[
S_{1k} = O_p(1/(nK_n^3)^{1/2}) = o_p (n^{-1/2}).
\]
Now we deal with $S_{2k}$. From Lemma \ref{lem:lem0} (vi) and the fact that
$\| \delta_0 - \bm{Z}^T \Pi_{\bm{V}n}\delta_0 \|_\infty = O(K_n^{-3/2})$, we have
\begin{eqnarray*}
\lefteqn{
\| \bm{Z}^T\Pi_{\bm{V}n} \delta_0 - \bm{Z}^T\widehat{\Pi}_{\bm{V}n}
\delta_0 \|_n^V
}\\
& = & 
\sup_{ \bm{g}\in \bm{G}_B }
\frac{| \langle \delta_0 - \bm{Z}^T \Pi_{\bm{V}n} \delta_0,
\bm{Z}^T \bm{g}\rangle_n^V -
\langle \delta_0 - \bm{Z}^T \Pi_{\bm{V}n} \delta_0,
\bm{Z}^T \bm{g}\rangle^V |
}{\| \bm{Z}^T\bm{g}\|_n^V }\\
& = & O_p \Big( K_n^{-3/2}\sqrt{\frac{K_n}{n}} \Big)
\, = \,  O_p( K_n^{-1}n^{-1/2} ).
\end{eqnarray*}
Thus we have
\[|S_{2k}| = o_p(n^{-1/2}).\]
We also have
\begin{align*} |S_{3k}|
&\le \| \delta_0 \|_n^V 
\| \bm{Z}^T (\bm{\varphi}_{\bm{V}k}^*
- \widetilde{\bm{\varphi}}_{\bm{V}k}) \|_n^V
= O_p(K_n^{-4}) = o_p (n^{-1/2}) 
\end{align*}
since
$\| \delta_0 - \bm{Z}^T\widehat{\Pi}_{\bm{V}n}
\delta_0 \|_n^V \le \| \delta_0 \|_n^V$.
Hence we have 
\begin{equation*}
\bias_\beta = o_p( n^{-1/2} )\,.
\end{equation*}
The desired result follows from (\ref{eqn:r706}) and the above equality. 

\bigskip
\noindent
As for Proposition \ref{prop:prop2}, there is almost no change in
calculation of the score functions in \cite{HZZ2007} and
\cite{CZH2014} and we omit the outline. 
This is because $m_i$ is bounded for any fixed $n$.

\bigskip
\noindent
{\it Proof of Proposition \ref{prop:prop3}. }\,\,
When $\bm{V}_i = \bm{\Sigma}_i$, we have
\[
\Gamma_{\bm{V}} = \bH^{11} = (\bH_{11\cdot 2})^{-1}
\quad {\rm and}\quad
\bm{\varphi}_{\bm{\Sigma} k}^* = \bm{\varphi}_{eff, k}^*  .
\]
Lemma 1 (vii) implies that
\begin{equation*}
\frac{1}{n}\Gamma_{\bm{V}}^{-1} = \frac{1}{n} \bH_{11\cdot 2}
= \frac{1}{n}\RE \{ \bm{l}_{\bm{\beta}}^* (\bm{l}_{\bm{\beta}}^*)^T \}
+ o_p(1) = \Omega_{\bm{\Sigma}} + o_p(1).
\end{equation*}
The desired result follows from the above result 
and Proposition \ref{prop:prop1}.
\\


\vspace{0.1in}
\noindent 
{\it Proof of Lemma \ref{lem:lem0}.} \,\,
The proof consists of seven parts. \\
(i) Recall that
\[
( \| \bm{Z}^T\bm{g} \|^V )^2
=\frac{1}{n}\RE \Big\{
\sum_{i=1}^n \underline{(\bm{Z}^T\bm{g})}_i^T
\bm{V}_i^{-1}
\underline{(\bm{Z}^T\bm{g})}_i\Big\}.
\]
We have from Assumptions A4 and A5 that
\begin{align}
\lefteqn{
\frac{C_1}{n}
\RE \Big\{ \sum_{i=1}^n\frac{1}{m_i}\sum_{j=1}^{m_i}\bm{g}^T(T_{ij})
\bm{Z}_{ij}\bm{Z}_{ij}^T\bm{g}(T_{ij})
\Big\} }\label{eqn:e770}\\
& \le ( \| \bm{Z}^T\bm{g} \|^V )^2
\le \frac{C_2}{n}
\RE \Big\{ \sum_{i=1}^n\sum_{j=1}^{m_i}\bm{g}^T(T_{ij})
\bm{Z}_{ij}\bm{Z}_{ij}^T\bm{g}(T_{ij})
\Big\}
\nonumber
\end{align}
for some positive constants $C_1$ and $C_2$.
Assumptions A2 and A3 imply that for some positive constants
$C_3$ and $C_4$,
\begin{align}
C_3 \sum_{l=1}^q\int g_l^2(t)dt
& \le
\frac{1}{n} \RE \Big\{ \sum_{i=1}^n \frac{1}{m_i}
\sum_{j=1}^{m_i}
\bm{g}^T(T_{ij}) \bm{Z}_{ij}\bm{Z}_{ij}^T \bm{g}(T_{ij})
\Big\}\label{eqn:e772}\\
& \le
\frac{1}{n} \RE \Big\{ \sum_{i=1}^n 
\sum_{j=1}^{m_i}
\bm{g}^T(T_{ij}) \bm{Z}_{ij}\bm{Z}_{ij}^T \bm{g}(T_{ij})
\Big\}\le C_4 \sum_{l=1}^q\int g_l^2(t)dt.
\nonumber
\end{align}
The desired result follows from (\ref{eqn:e770}) and (\ref{eqn:e772}).

\noindent
(ii) This is a well-known result in the literature
of spline regression. See for example A.2 of \cite{HWZ2004}.

\noindent
(iii)
The result in (ii) implies
\[
\| \bm{X}^T\bm{\beta} + \bm{Z}^T\bm{g} \|_\infty^2
\le C K_n\Big( |\bm{\beta}|^2 + \| \bm{g} \|_{G,2}^2
\Big)
\]
for some positive constant $C$. Recall that $p$ and $q$
are fixed in this paper.
On the other hand, we have from Assumptions A1-3 and A5 that
for some positive constants $C_1$, $C_2$, and $C_3$,
\begin{align*}
\lefteqn{ ( \| \bm{X}^T\bm{\beta}+
\bm{Z}^T\bm{g} \|^V)^2}\\
& \qquad  \ge \frac{C_1}{n}
\RE \Big\{ \sum_{i=1}^n \frac{1}{m_i}\sum_{j=1}^{m_i}
( \bm{\beta}^T\, \bm{g}^T(T_{ij}) )
\begin{pmatrix}
\bm{X}_{ij}\bm{X}_{ij}^T & \bm{X}_{ij}\bm{Z}_{ij}^T\\
\bm{Z}_{ij}\bm{X}_{ij}^T & \bm{Z}_{ij}\bm{Z}_{ij}^T
\end{pmatrix}
\begin{pmatrix}
\bm{\beta} \\
\bm{g}(T_{ij})
\end{pmatrix} \Big\} \\
& \qquad \ge \frac{C_2}{n}
\RE \Big\{ \sum_{i=1}^n \frac{1}{m_i} \sum_{j=1}^{m_i}
( \bm{\beta}^T\, \bm{g}^T(T_{ij}) )
\begin{pmatrix}
\bm{\beta} \\
\bm{g}(T_{ij})
\end{pmatrix} \Big\}
\ge C_3 |\bm{\beta}|^2 + \| \bm{g} \|_{G,2}^2.
\end{align*}
Besides, we have for some positive constants $C_1$
and $C_2$,
\[
(\| v \|^V )^2 \le \frac{C_1}{n}\sum_{i=1}^n 
\sum_{j=1}^{m_i} |v_{ij}|^2 \le C_2 \| v \|_\infty.
\]
Hence the desired results are established.

\noindent
(iv) For $\bm{g}_1\in \bm{G}_B$ and $\bm{g}_2\in \bm{G}_B$,
we have
\begin{equation*}
\langle \bm{Z}^T\bm{g}_1,\bm{Z}^T\bm{g}_2 \rangle_n^V 
=\bm{\gamma}_1^T\Big\{ 
\frac{1}{n}\sum_{i=1}^n \underline{\bm{W}}_i^T
\bm{V}_i^{-1} \underline{\bm{W}}_i\Big\}
\bm{\gamma}_2 = \bm{\gamma}_1^T \overline{\Delta}_{\bm{V}n}
\bm{\gamma}_2 \quad {\rm (say)},
\end{equation*}
where $\overline{\Delta}_{\bm{V}n}$ is a $qK_n \times qK_n$ matrix
and $\bm{\gamma}_1$ and $\bm{\gamma}_2$ correspond to
$\bm{g}_1$ and $\bm{g}_2$, respectively. Elements of
$ \frac{1}{n}\sum_{i=1}^n \underline{\bm{W}}_i^T
\bm{V}_i^{-1} \underline{\bm{W}}_i $
are written as
\begin{equation}
\frac{1}{n}\sum_{i=1}^n\sum_{j_1,j_2}
v_i^{j_1j_2} B_{k_1}(T_{ij_1})B_{k_2}(T_{ij_2})
Z_{ij_1l_1}Z_{ij_2l_2} = \overline{\Delta}_{\bm{V}n}^{(k_1,l_1,k_2,l_2)}
\quad {\rm (say)},
\label{eqn:e776}
\end{equation}
where $v_i^{j_1j_2}$ is defined in (\ref{eqn:e508}),
$1\le k_1, k_2 \le K_n$, and $1\le l_1,l_2 \le q$.
By evaluating the variance of (\ref{eqn:e776})
and using the Bernstein inequality for independent
bounded random variables, and Assumptions A1 and A2, we have
uniformly in $k_1$, $k_2$, $l_1$, and $l_2$,
\begin{align}
\overline{\Delta}_{\bm{V}n}^{(k_1,l_1,k_2,l_2)}
- \RE ( \overline{\Delta}_{\bm{V}n}^{(k_1,l_1,k_2,l_2)} )
= O_p\Big( \sqrt{\frac{\log n}{nK_n^2}} \Big)
\quad {\rm if } \ B_{k_1}(t) B_{k_2}(t) \equiv 0 
\label{eqn:e778}\\
\intertext{and}
\overline{\Delta}_{\bm{V}n}^{(k_1,l_1,k_2,l_2)}
- \RE ( \overline{\Delta}_{\bm{V}n}^{(k_1,l_1,k_2,l_2)} )
= O_p\Big( \sqrt{\frac{\log n}{nK_n}} \Big)
\quad {\rm if } \ B_{k_1}(t) B_{k_2}(t) \not \equiv 0 .
\label{eqn:e780}
\end{align}
By exploiting (\ref{eqn:e778}), (\ref{eqn:e780}), and
the local property of the B-spline basis, we obtain
\begin{equation}
\max \{ | \lambda_{\rm min}( \overline{\Delta}_{\bm{V}n}
- \RE ( \overline{\Delta}_{\bm{V}n} ) ) |, | \lambda_{\rm max}
( \overline{\Delta}_{\bm{V}n} - \RE ( \overline{\Delta}_{\bm{V}n} ) ) | \}
= O_p\Big( \sqrt{\frac{\log n}{n}}\Big).
\label{eqn:e782}
\end{equation}
We also have
\begin{equation}
\frac{C_1}{K_n}\le \lambda_{\rm min}( \RE ( \overline{\Delta}_{\bm{V}n} ) )
\le \lambda_{\rm max}( \RE ( \overline{\Delta}_{\bm{V}n} ) )
\le \frac{C_2}{K_n}
\label{eqn:e784}
\end{equation}
since Assumptions A2 and A3 yields
\begin{align*}
\lefteqn{ \frac{C_3}{n}\sum_{i=1}^n \frac{1}{m_i}\sum_{j=1}^{m_i}
( \bm{Z}_{ij} \otimes \bm{B}(T_{ij}) )^T
( \bm{Z}_{ij} \otimes \bm{B}(T_{ij}) ) }\\
& \quad \qquad \le \overline{\Delta}_{\bm{V}n} \le 
\frac{C_4}{n}\sum_{i=1}^n \sum_{j=1}^{m_i}
( \bm{Z}_{ij} \otimes \bm{B}(T_{ij}) )^T
( \bm{Z}_{ij} \otimes \bm{B}(T_{ij}) )
\end{align*}
for some positive constants $C_3$ and $C_4$. See
the proof of Lemma A.3 of \cite{HWZ2004}. Hence the desired
result follows from (\ref{eqn:e782}) and (\ref{eqn:e784}).

\noindent
(v) This follows from (iv) and (vi).

\noindent
(vi) Using Assumptions A1 and A2 we have
\begin{align*}
\langle \delta_n,Z_lB_k \rangle_n^V
& = \frac{1}{n}\sum_{i=1}^n\sum_{j_1,j_2}
\delta_{n,ij_1}v_i^{j_1j_2}Z_{ij_2l}B_k(T_{ij_2})\\
\intertext{and}
\var ( \langle \delta_n,Z_lB_k \rangle_n^V  ) 
& \le \frac{C_1 \| \delta_n \|_\infty^2 }{n^2}
\sum_{i=1}^n m_i^2\sum_{j_1, j_2}\RE
\{  B_k^2(T_{ij_1})B_k^2(T_{ij_2}) \}
\le \frac{C_2 \| \delta_n \|_\infty^2 }{nK_n} 
\end{align*}
for some positive constants $C_1$ and $C_2$.
Hence we have 
\[
\sum_{l=1}^q\sum_{k=1}^{K_n}
\var ( \langle \delta_n,Z_lB_k \rangle_n^V  ) 
 \le \frac{C}{n}\| \delta_n \|_\infty^2
\]
for some positive constant $C$ and the desired result follows
from (\ref{eqn:e784}).

\noindent
(vii) Take $\widetilde{\bm{\varphi}}_{\bm{V}k}
\in \bm{G}_B$ such that
$\| \widetilde{\bm{\varphi}}_{\bm{V}k} - \bm{\varphi}_{\bm{V}k}^* \|_{G,\infty}
=O(K_n^{-2})$. Then we have for some positive $C$,
\begin{eqnarray}
\lefteqn{ \| \bm{Z}^T ( \overline{\bm{\varphi}}_{\bm{V}k}
- \bm{\varphi}_{\bm{V}k}^*) \|_\infty}\label{eqn:r701}\\
& \le & \| \bm{Z}^T ( \overline{\bm{\varphi}}_{\bm{V}k}
- \widetilde{\bm{\varphi}}_{\bm{V}k} ) \|_\infty + \|
\bm{Z}^T ( \widetilde{\bm{\varphi}}_{\bm{V}k} - \bm{\varphi}_{\bm{V}k}^*) \|_\infty
\nonumber\\
& \le & C \sqrt{K_n} \| \bm{Z}^T ( \overline{\bm{\varphi}}_{\bm{V}k}
- \widetilde{\bm{\varphi}}_{\bm{V}k} ) \|^V + \|
\bm{Z}^T ( \widetilde{\bm{\varphi}}_{\bm{V}k} - \bm{\varphi}_{\bm{V}k}^*) \|_\infty
\nonumber\\
& \le & C  \sqrt{K_n} \| \bm{Z}^T ( \bm{\varphi}_{\bm{V}k}^*
- \widetilde{\bm{\varphi}}_{\bm{V}k} ) \|^V + \|
\bm{Z}^T ( \widetilde{\bm{\varphi}}_{\bm{V}k} - \bm{\varphi}_{\bm{V}k}^*) \|_\infty\nonumber\\
& = & O(K_n^{-3/2}).\nonumber
\end{eqnarray}
Here we used the fact that $\overline{\bm{\varphi}}_{\bm{V}k}
= \Pi_{\bm{V}n}X_k \in \bm{G}_B$ and $\bm{\varphi}_{\bm{V}k}^*
= \Pi_{\bm{V}}X_k$. 
Inequality (\ref{eqn:r701}) implies $\| \bm{Z}^T
\overline{\bm{\varphi}}_{\bm{V}k} \|_\infty= O(1)$ and we have only to evaluate
$ \bm{Z}^T ( \overline{\bm{\varphi}}_{\bm{V}k}
- \widehat{\bm{\varphi}}_{\bm{V}k}) $. We should just follow the arguments
on p.16 of \cite{CZH2014S} by replacing $\varphi_{k,n}^* $
and $\widehat{\varphi}_{k,n}$ with $\bm{Z}^T
\overline{\bm{\varphi}}_{\bm{V}k} $ and
$\bm{Z}^T \widehat{\bm{\varphi}}_{\bm{V}k} $ since the arguments
employ (iv) and (vi) and don't depend on $m_i$.
Then we have \[
 \| \bm{Z}^T( \overline{\bm{\varphi}}_{\bm{V}k}  -
\widehat{\bm{\varphi}}_{\bm{V}k}) \|_\infty = o_p(1),
\quad 
\| \bm{Z}^T( \overline{\bm{\varphi}}_{\bm{V}k}  -
\widehat{\bm{\varphi}}_{\bm{V}k})  \|_n^V = O_p(\sqrt{K_n/n}),
\]
\[
{\rm and}\quad 
\| \bm{Z}^T( \overline{\bm{\varphi}}_{\bm{V}k}  -
\widehat{\bm{\varphi}}_{\bm{V}k}  ) \|^V = O_p(\sqrt{K_n/n}).
\]
The desired results follow from the above equations
and (\ref{eqn:r701}).
\\

\section{Proof of Proposition \ref{prop:prop4}}
In the proof, we repeatedly use arguments
based on exponential inequalities, truncation, and
division of regions into small rectangles to prove uniform convergence
results as in \cite{Masry1996x}.  
We do not give the details of these arguments
since they are standard ones in nonparametric kernel methods.
Since we impose Assumption A2 and we do not use $\bm{\Sigma}_i$ or
$\bm{V}_i$ in the construction of $\widehat{\bm{g}}(t)$, $\widehat{\sigma^2}(t)$,
and $\widehat{\sigma}(s,t)$, we see the effects of diverging $m_i$ 
explicitly only when applying the exponential inequality for generalized U-statistics.
Recall that we assume three times continuous differentiability
of the relevant functions in this proposition.

The proof consists of four parts: (i) representation of
$\widehat{\bm{g}}(t)$, (ii) representation of $\widehat{\epsilon}_{ij}$, (iii)
representation of $\widehat{\sigma^2}(t)$, and
(iv) representation of $\widehat{\sigma}(s,t)$.

\noindent
(i) Representation of $\widehat{\bm{g}}(t)$.
Applying the third order Taylor series expansion to $\bm{g}_0(t)$,
we have
\begin{equation}
\bm{Z}_{ij}^T \bm{g}_0(T_{ij})
= \bm{Z}_{ij}^T \Big\{ \bm{g}_0(t) + h_1\frac{T_{ij}-t}{h_1}
\bm{g}_0'(t) + \frac{h_1^2}{2}
\Big(\frac{T_{ij}-t}{h_1}\Big)^2\bm{g}_0''(t)
\Big\}+ O(h_1^3),
\label{eqn:e700}
\end{equation}
where $\bm{g}_0'(t) = (g_{01}'(t), \ldots, g_{0q}'(t))^T$
and $\bm{g}_0''(t) = (g_{01}''(t), \ldots, g_{0q}''(t))^T$.
By plugging (\ref{eqn:e700}) into (\ref{eqn:e303}), we have
uniformly in $t$,
\begin{align}
\widehat{\bm{g}}(t) & =  
\bm{g}_0(t) + D_q (\widehat{L}_1(t))^{-1}\widehat{L}_2(t)
( \bm{\beta}_0 - \widehat{\bm{\beta}}_I )\label{eqn:e703}\\
& \qquad + \frac{h_1^2}{2}D_q (\widehat{L}_1(t))^{-1}
\widehat{L}_3(t)\bm{g}_0''(t)+ 
D_q (\widehat{L}_1(t))^{-1}E_0(t)
+ O_p(h_1^3),\nonumber
\end{align}
where $\widehat{L}_1(t) =A_{1n}(t)$ defined after (\ref{eqn:e303}),
\begin{align*}
\widehat{L}_2(t)& = \frac{1}{N_1h_1}
\sum_{i=1}^n \sum_{j=1}^{m_i}
\bm{Z}_{ij}\otimes
\begin{pmatrix}
1\\
\frac{T_{ij}-t}{h_1}
\end{pmatrix}
K\Big( \frac{T_{ij}-t}{h_1} \Big)
\bm{X}_{ij}^T, \\
\widehat{L}_3(t)& = \frac{1}{N_1h_1}
\sum_{i=1}^n \sum_{j=1}^{m_i}
( \bm{Z}_{ij} \bm{Z}_{ij}^T )
\otimes
\begin{pmatrix}
(\frac{T_{ij}-t}{h_1})^2 \\
(\frac{T_{ij}-t}{h_1})^3
\end{pmatrix}
K\Big( \frac{T_{ij}-t}{h_1} \Big),\\
E_0(t) & = \frac{1}{N_1h_1}
\sum_{i=1}^n \sum_{j=1}^{m_i}
\bm{Z}_{ij}\otimes
\begin{pmatrix}
1\\
\frac{T_{ij}-t}{h_1}
\end{pmatrix}
K\Big( \frac{T_{ij}-t}{h_1} \Big) \epsilon_{ij}.
\end{align*}
By following standard arguments such as those in \cite{Masry1996x},
we obtain for $j=1,2,3$, 
\begin{align}
\widehat{L}_j(t) & = L_j(t) + O_p\Big(
\sqrt{\frac{\log n }{nh_1}}\Big)
\quad {\rm uniformly\ in}\ t,
\label{eqn:e705}\\
\intertext{where $L_j= \RE \{ \widehat{L}_j(t) \}$, and}
E_0 (t) & = O_p\Big(\sqrt{\frac{\log n }{nh_1}}\Big)
\quad {\rm uniformly\ in}\ t.
\label{eqn:e707}
\end{align}
Assumption A2 implies that
\begin{equation}
C_1 \bm{I}_{2q} \le L_1(t) \le C_2 \bm{I}_{2q}
\label{eqn:e709}
\end{equation}
for some positive constants $C_1$ and $C_2$. From
(\ref{eqn:e703})-(\ref{eqn:e709}), we have
uniformly in $t$,
\begin{align}
\widehat{\bm{g}}(t) & =  
\bm{g}_0(t) + D_q (L_1(t))^{-1}L_2(t)
( \bm{\beta}_0 - \widehat{\bm{\beta}}_I )
+ \frac{h_1^2}{2} D_q (L_1(t))^{-1}L_3(t)\bm{g}_0''(t)
\label{eqn:e711}\\
& \qquad \qquad + D_q (L_1(t))^{-1}E_0(t)
+ O_p(h_1^3) + O_p \Big( \frac{\log n}{nh_1}\Big) + 
O_p \Big( h_1^2 \sqrt{\frac{\log n}{nh_1}} \Big)\nonumber\\
& = \bm{g}_0(t) + L_4(t)( \bm{\beta}_0 - \widehat{\bm{\beta}}_I )
+ h_1^2 L_5(t)\bm{g}_0''(t) + L_6(t) E_0(t)\nonumber\\
& \qquad \qquad +  O_p(h_1^3) + O_p \Big( \frac{\log n}{nh_1}\Big) + 
O_p \Big( h_1^2 \sqrt{\frac{\log n}{nh_1}} \Big)\quad  {\rm (say)}.
\nonumber
\end{align}
Note that all the elements of $L_j(t)$, $j=4,5,6$, are
bounded functions of $t$.

\noindent
(ii) Representation of $\widehat{\epsilon}_{ij}$.
We have
\begin{equation*}
\widehat{\epsilon}_{ij}= \epsilon_{ij}
+ \bm{X}_{ij}^T ( \bm{\beta}_0 - \widehat{\bm{\beta}}_I)
+ \bm{Z}_{ij}^T ( \bm{g}_0 (T_{ij}) - \widehat{\bm{g}}
(T_{ij}) ). 
\end{equation*}
By plugging (\ref{eqn:e711}) into the above equality, 
we obtain uniformly in $i$ and $j$,
\begin{align}
\widehat{\epsilon}_{ij} & =
\epsilon_{ij}+ ( \bm{X}_{ij}^T - \bm{Z}_{ij}^TL_4(T_{ij}))
( \bm{\beta}_0 - \widehat{\bm{\beta}}_I) - h_1^2
\bm{Z}_{ij}^TL_5(T_{ij})\bm{g}''(T_{ij})\label{eqn:e715}\\
& \qquad \qquad - \bm{Z}_{ij}^TL_6(T_{ij})E_0(T_{ij})
+ O_p(h_1^3) + O_p \Big( \frac{\log n}{nh_1}\Big) + 
O_p \Big( h_1^2 \sqrt{\frac{\log n}{nh_1}} \Big)\nonumber\\
& = \epsilon_{ij}+ M_{ij}^{(1)}( \bm{\beta}_0 - \widehat{\bm{\beta}}_I) 
+ h_1^2 M_{ij}^{(2)}\bm{g}''(T_{ij})
+ M_{ij}^{(3)}E_0 (T_{ij}) \nonumber\\
& \qquad \qquad + O_p(h_1^3) + O_p \Big( \frac{\log n}{nh_1}\Big) + 
O_p \Big( h_1^2 \sqrt{\frac{\log n}{nh_1}} \Big)\quad {\rm (say)}.
\nonumber
\end{align}
Note that all the elements of $M_{ij}^{(1)}$, $M_{ij}^{(2)}$, and
$M_{ij}^{(3)}$ are uniformly bounded functions of 
$\bm{X}_{ij}$, $\bm{Z}_{ij}$, and $T_{ij}$.

\noindent
(iii) Representation of $\widehat{\sigma^2}(t)$.
We have uniformly in $i$ and $j$,
\begin{align}
(\widehat{\epsilon}_{ij} )^2 & =
\epsilon_{ij}^2 - \sigma^2(T_{ij}) + \sigma^2(T_{ij})
+ 2 \epsilon_{ij} M_{ij}^{(3)}E_0(T_{ij}) \label{eqn:e717}\\
& \qquad \qquad 
+ 2 \epsilon_{ij} M_{ij}^{(1)} ( \bm{\beta}_0 - \widehat{\bm{\beta}}_I )
+ 2 \epsilon_{ij} h_1^2 M_{ij}^{(2)} \bm{g}_0''(T_{ij})\nonumber\\
& \qquad \qquad + O_p(h_1^3) + O_p \Big( \frac{\log n}{nh_1}\Big) + 
O_p \Big( h_1^2 \sqrt{\frac{\log n}{nh_1}} \Big).\nonumber
\end{align}
Recall that $M_{ij}^{(l)}$, $l=1,2,3$, are defined in (\ref{eqn:e715}).
It is easy to see that the contributions of
$ 2 \epsilon_{ij} M_{ij}^{(1)} ( \bm{\beta}_0 - \widehat{\bm{\beta}}_I ) $
and
$ 2 \epsilon_{ij} h_1^2 M_{ij}^{(2)} \bm{g}''(T_{ij}) $
to $\widehat{\sigma^2}(t)$ are
\[
O_p\Big(  \frac{1}{\sqrt{n}}\sqrt{\frac{\log n }{n h_2} }\Big)
\quad
{\rm and}\quad
O_p\Big( h_1^2 \sqrt{ \frac{\log n }{n h_2} } \Big)
\]
uniformly in $t$, respectively. Thus we have only to
consider $ \epsilon_{ij}^2 - \sigma^2(T_{ij}) $,
$ \sigma^2(T_{ij}) $, and $  2 \epsilon_{ij} M_{ij}^{(3)}
E_0(T_{ij})$ in (\ref{eqn:e717}). 

Setting $\widehat{L}_7(t)=A_{2n}(t)$, which is defined after (\ref{eqn:e306}), we have 
for some positive constants $C_1$ and $C_2$,
\begin{equation}
\widehat{L}_7(t) = L_7(t) + O_p \Big(
\sqrt{ \frac{\log n }{n h_2} } \Big)
\quad {\rm and} \quad
C_1 \bm{I}_2 \le L_7(t) \le C_2 \bm{I}_2
\label{eqn:e718}
\end{equation}
uniformly in $t$, where $L_7 (t) = \RE \{ \widehat{L}_7(t) \}$.
Now we have uniformly in $t$,
\begin{align}
\widehat{\sigma^2}(t)  & =
(1\, 0)( \widehat{L}_7(t) )^{-1}
(E_1(t) + \bias_1(t) + R_1 (t)) \label{eqn:e719}\\
& \qquad \qquad + O_p(h_1^3) 
+ O_p \Big( \frac{\log n}{nh_1}\Big) + 
O_p \Big( h_1^2 \sqrt{\frac{\log n}{nh_1}} \Big),
\nonumber
\end{align}
where $E_1(t)$ is defined in Proposition \ref{prop:prop4},
$\bias_1(t)$ is the term of $\sigma^2 (T_{ij})$, and $R_1(t)$ is the
term of $2\epsilon_{ij}M_{ij}^{(3)}E_0(T_{ij})$.
It is easy to see that uniformly in $t$,
\begin{equation}
E_1(t) = O_p \Big(  \sqrt{\frac{\log n}{nh_2}} \Big).
\label{eqn:e720}
\end{equation}
By applying the Taylor series expansion, we have
\[
\sigma^2 (T_{ij}) = \sigma^2 (t) + h_2(\sigma^2)'(t)
\frac{T_{ij}-t}{h_2}+ \frac{h_2^2}{2}(\sigma^2)''(t)
\Big( \frac{T_{ij}-t}{h_2} \Big)^2 + O(h_2^3).
\]
Therefore $\bias_1(t)$ can be represented as
\begin{align*}
\bias_1(t)
& = \widehat{L}_7(t)
\begin{pmatrix}
\sigma^2(t) \\
h_2(\sigma^2)'(t)
\end{pmatrix} +
\frac{h_2^2(\sigma^2)''(t)}{2N_1h_2}
\sum_{i=1}^n \sum_{j=1}^{m_i}
\begin{pmatrix}
(\frac{T_{ij}-t}{h_2})^2 \\
(\frac{T_{ij}-t}{h_2})^3
\end{pmatrix}
K\Big( \frac{T_{ij}-t}{h_2} \Big)\\
& \qquad \qquad + O_p (h_2^3).
\end{align*}
uniformly in $t$.
Setting
\[
\widehat{L}_8(t)
= \frac{1}{N_1h_2}
\sum_{i=1}^n \sum_{j=1}^{m_i}
\begin{pmatrix}
(\frac{T_{ij}-t}{h_2})^2 \\
(\frac{T_{ij}-t}{h_2})^3
\end{pmatrix}K\Big( \frac{T_{ij}-t}{h_2} \Big),
\]
we have uniformly in $t$,
\[
\widehat{L}_8(t) = L_8(t) + O_p \Big(
\sqrt{ \frac{\log n }{n h_2} } \Big),
\]
where $L_8 (t) = \RE \{ \widehat{L}_8(t) \}$
and $L_8(t)$ is a bounded vector function of $t$.
Hence we have uniformly in $t$,
\begin{equation}
\bias_1(t)
= \widehat{L}_7(t)
\begin{pmatrix}
\sigma^2(t) \\
h_2(\sigma^2)'(t)
\end{pmatrix} +
\frac{h_2^2(\sigma^2)''(t)}{2}
L_8(t)+ O_p (h_2^3)+ 
O_p \Big(h_2^2 \sqrt{ \frac{\log n }{n h_2} } \Big).
\label{eqn:e721}
\end{equation}

Next we deal with $R_1(t)$, which can be written as
\begin{equation}
\frac{1}{N_1^2h_1h_2}
\sum_{a,b}\sum_{i=1}^n\sum_{j=1}^{m_i}\sum_{i'=1}^n\sum_{j'=1}^{m_{i'}}
\epsilon_{ij}\epsilon_{i'j'}
A_{ab,ij}B_{ab,i'j'}K_a\Big( \frac{T_{ij}-t}{h_2}\Big)
K_b\Big( \frac{T_{i'j'}-T_{ij}}{h_1}\Big),
\label{eqn:e723}
\end{equation}
where $K_l(t) = t^lK(t)$, $a=0,1$, and $b=0,1$. Note that
$A_{ab,ij}$ and $B_{ab,ij}$ are uniformly bounded functions
of $\bm{X}_{ij}$, $\bm{Z}_{ij}$, and $T_{ij}$.
We evaluate 
\begin{align}
\lefteqn{\frac{1}{N_1^2h_1h_2}
\sum_{i=1}^n\sum_{j=1}^{m_i}\sum_{i'=1}^n\sum_{j'=1}^{m_{i'}}
\epsilon_{ij}\epsilon_{i'j'}
A_{ab,ij}B_{ab,i'j'}K_a\Big( \frac{T_{ij}-t}{h_2}\Big)
K_b\Big( \frac{T_{i'j'}-T_{ij}}{h_1}\Big)}\label{eqn:e725} \\
& \qquad = \frac{1}{N_1^2h_1h_2}\sum_{i=1}^n\sum_{j=1}^{m_i}
\epsilon_{ij}^2A_{ab,ij}B_{ab,ij}K_a\Big( \frac{T_{ij}-t}{h_2}\Big)
K_b( 0 ) \nonumber \\
& \qquad \qquad + \frac{1}{N_1^2h_1h_2}
\sum_{i=1}^n\sum_{j\ne j'}
\epsilon_{ij}\epsilon_{ij'}
A_{ab,ij}B_{ab,ij'}K_a\Big( \frac{T_{ij}-t}{h_2}\Big)
K_b\Big( \frac{T_{ij'}-T_{ij}}{h_1}\Big)
\nonumber \\
& \qquad \qquad + \frac{1}{N_1^2h_1h_2}
\sum_{i\ne i'}\sum_{j, j'}
\epsilon_{ij}\epsilon_{i'j'}
A_{ab,ij}B_{ab,i'j'}K_a\Big( \frac{T_{ij}-t}{h_2}\Big)
K_b\Big( \frac{T_{i'j'}-T_{ij}}{h_1}\Big)
\nonumber \\
& \qquad = R_{1ab}^{(1)}(t) +  R_{1ab}^{(2)}(t) 
+ R_{1ab}^{(3)}(t)\quad {\rm (say)}.  \nonumber 
\end{align}
Note that we cannot apply classical exponential inequalities
for U-statistics since kernel functions depend on $i$ and $i'$
and observations are not identical.
It is easy to see that uniformly in $t$,
\begin{equation}
R_{1ab}^{(1)}(t) = O_p((nh_1)^{-1})
\quad {\rm and}\quad
R_{1ab}^{(2)}(t) = O_p(n^{-1}).
\label{eqn:e727}
\end{equation}
We evaluate $ R_{1ab}^{(3)}(t)$ by using an exponential inequality as the one given in (3.5) of \cite{GLZ2000x}
with 
$A=C_1(\log n)^km_{\rm max}^2/(n^2h_1h_2)$,
\[
B^2 = C_2\frac{(\log n)^{2k}m_{\rm max}^2}{n^{3}h_1h_2}
(h_1^{-1}+ h_2^{-1}),
\]
$C=C_3/(n h_1^{1/2}h_2^{1/2})$, and
$x= M\log n/(n h_1^{1/2}h_2^{1/2})$ in the inequality and standard arguments
in nonparametric regression as in \cite{Masry1996x}. Note that we used a kind 
of truncation technique to handle $\epsilon_{ij}$ and that we have to take
sufficiently large $k$ and $M$ here. Hence we have
\[
R_{1ab}^{(3)}(t) = O_p \Big( \frac{\log n}{ n (h_1h_2)^{1/2} } \Big).
\]
The above equation and (\ref{eqn:e723})-(\ref{eqn:e727}) imply that
\begin{equation}
R_1 (t) = O_p \Big( \frac{\log n}{n (h_1h_2)^{1/2} } \Big) +
O_p\Big( \frac{1}{nh_1} \Big)
\label{eqn:e729}
\end{equation}
uniformly in $t$.
It follows from (\ref{eqn:e718})-(\ref{eqn:e721}) and (\ref{eqn:e729})
that
\begin{align*}
\widehat{\sigma^2}(t) & = \sigma^2 (t)
+ (1\, 0)( L_7(t) )^{-1}E_1(t) 
+ \frac{h_2^2}{2}(1\, 0)( L_7(t) )^{-1}L_8(t)(\sigma^2)''(t)\\
& \qquad \qquad + O_p(h_1^3) + O_p(h_2^3) 
+ O_p \Big( \frac{\log n}{nh_1}\Big)
+ O_p \Big( \frac{\log n}{nh_2}\Big).
\end{align*}
The expression of $\widehat{\sigma^2}(t)$ in Proposition \ref{prop:prop4}
also follows from the above expression.

\noindent
(iv) Representation of $\widehat{\sigma}(s,t)$.
We can proceed almost in the same way as when we deal with $\widehat{\sigma^2}(t)$.
First 
we have uniformly in $i$, $j$, and $j'$,
\begin{align}
\widehat{\epsilon}_{ij}\widehat{\epsilon}_{ij'} 
& = \epsilon_{ij}\epsilon_{ij'} - \sigma (T_{ij},T_{ij'})
+ \sigma (T_{ij},T_{ij'}) 
\nonumber\\
& \qquad \quad +
\epsilon_{ij} M_{ij'}^{(3)}E_0(T_{ij'})+\epsilon_{ij'} M_{ij}^{(3)}E_0(T_{ij})
\nonumber\\
& \qquad \quad +
\epsilon_{ij}M_{ij'}^{(1)} ( \bm{\beta}_0 - \widehat{\bm{\beta}}_I )
+ \epsilon_{ij'}M_{ij}^{(1)} ( \bm{\beta}_0 - \widehat{\bm{\beta}}_I )
\label{eqn:e731}\\
& \qquad \quad +
\epsilon_{ij} h_1^2 M_{ij'}^{(2)} \bm{g}_0''(T_{ij'})
+ \epsilon_{ij'} h_1^2 M_{ij}^{(2)} \bm{g}_0''(T_{ij})
\label{eqn:e732}\\
& \qquad \quad + O_p(h_1^3) + O_p \Big( \frac{\log n}{nh_1}\Big) + 
O_p \Big( h_1^2 \sqrt{\frac{\log n}{nh_1}} \Big).\nonumber
\end{align}
It is easy to see that the contributions of (\ref{eqn:e731})
and (\ref{eqn:e732}) to $\widehat{\sigma}(s,t)$ are
\[
O_p\Big(  \frac{1}{\sqrt{n}}\sqrt{\frac{\log n }{n h_3^2} }\Big)
\quad
{\rm and}\quad
O_p\Big( h_1^2 \sqrt{ \frac{\log n }{n h_3^2} } \Big)
\]
uniformly in $s$ and $t$, respectively. Therefore we have only to
consider $ \epsilon_{ij}\epsilon_{ij'} - \sigma(T_{ij},T_{ij'}) $,
$ \sigma(T_{ij},T_{ij'} ) $, and $ \epsilon_{ij} M_{ij'}^{(3)}
E_0(T_{ij'})+ \epsilon_{ij'} M_{ij}^{(3)}E_0(T_{ij})$ in
$\widehat{\epsilon}_{ij}\widehat{\epsilon}_{ij'}$. 

Setting $\widehat{L}_9(s,t)=A_{3n}(s,t)$, which is defined after (\ref{eqn:e309}), we have 
for some positive constants $C_1$ and $C_2$,
\begin{equation}
\widehat{L}_9(s,t) = L_9(s,t) + O_p \Big(
\sqrt{ \frac{\log n }{n h_3^2} } \Big)
\quad {\rm and} \quad
C_1 \bm{I}_3 \le L_9(s,t) \le C_2 \bm{I}_3
\label{eqn:e796}
\end{equation}
uniformly in $s$ and $t$, where $L_9 (s,t) 
= \RE \{ \widehat{L}_9(s,t) \}$.
Now we have uniformly in $s$ and $t$,
\begin{align}
\widehat{\sigma}(s, t)  & =
(1\,0\, 0)( \widehat{L}_9(s,t) )^{-1}
(E_2(s,t) + \bias_2(s,t) + R_2 (s,t)) \label{eqn:e733}\\
& \qquad \qquad + O_p(h_1^3) 
+ O_p \Big( \frac{\log n}{nh_1}\Big) + 
O_p \Big( h_1^2 \sqrt{\frac{\log n}{nh_1}} \Big),
\nonumber
\end{align}
where $E_2(s,t)$ is defined in Proposition \ref{prop:prop4},
$\bias_2(s,t)$ is the term of $\sigma (T_{ij},T_{ij'})$, and
$R_2(s,t)$ is the term of $\epsilon_{ij}
M_{ij'}^{(3)}E_0(T_{ij'}) + \epsilon_{ij'}
M_{ij}^{(3)}E_0(T_{ij})$.
It is easy to see that uniformly in $s$ and $t$,
\begin{equation}
E_2(s,t) = O_p \Big(  \sqrt{\frac{\log n}{nh_3^2}} \Big).
\label{eqn:e734}
\end{equation}

Setting
\begin{align*}
\lefteqn{
\widehat{L}_{10}(s,t)}\\
& = \frac{1}{N_2h_3^2}
\sum_{i=1}^n \sum_{j\ne j'}
\begin{pmatrix}
1\\
\frac{T_{ij}-s}{h_3} \\
\frac{T_{ij'}-t}{h_3}
\end{pmatrix}
\Big( \Big( \frac{T_{ij}-s}{h_3} \Big)^2\ 
\frac{2(T_{ij}-s)(T_{ij'}-t)}{h_3^2}\
\Big( \frac{T_{ij'}-t}{h_3} \Big)^2 \Big)\\
& \qquad \qquad \qquad \qquad \times
K\Big( \frac{T_{ij}-s}{h_3} \Big)
K\Big( \frac{T_{ij'}-t}{h_3} \Big),
\end{align*}
we have uniformly in $s$ and $t$,
\[
\widehat{L}_{10}(s,t) = L_{10}(s,t) + O_p \Big(
\sqrt{ \frac{\log n }{n h_3^2} } \Big),
\]
where $L_{10} (s,t) = \RE \{ \widehat{L}_{10}(s,t) \}$ 
which is a bounded matrix function of $(s,t)$.
Then we have, as in the proof of the representation of $\widehat{\sigma^2}(t)$, uniformly in $s$ and $t$
\begin{align}
\bias_2(s,t)
& = \widehat{L}_9(s,t)
\begin{pmatrix}
\sigma (s, t) \\
h_3\frac{\partial \sigma}{\partial s}(s,t)\\
h_3\frac{\partial \sigma}{\partial t}(s,t)
\end{pmatrix} +
\frac{h_3^2}{2}
L_{10}(s,t)
\begin{pmatrix}
\frac{\partial^2 \sigma}{\partial s^2}(s,t) \\
\frac{\partial^2 \sigma}{\partial s \partial t}(s,t)\\
\frac{\partial^2 \sigma}{\partial t^2}(s,t)
\end{pmatrix}\label{eqn:e735}\\
& \qquad \qquad + O_p (h_3^3)+ 
O_p \Big(h_3^2 \sqrt{ \frac{\log n }{n h_3^2} } \Big).
\nonumber
\end{align}

Finally we deal with $R_2(s,t)$ in the same way as in the proof of the representation 
of $\widehat{\sigma^2}(t)$. We use the same exponential inequality
for U-statistics. We should consider
\begin{align}
\lefteqn{ \frac{1}{N_1N_2h_1h_3^2}
\sum_{i_1=1}^n\sum_{i_2=1}^n\sum_{j_1\ne j_2}\sum_{j_3}
\epsilon_{i_1j_1}\epsilon_{i_2j_3} 
A_{abc,i_1j_2}B_{abc,i_2j_3} } \label{eqn:e736}\\
& \qquad \qquad \times K_a\Big( \frac{T_{i_2j_3}-T_{i_1j_2}}{h_1} \Big)
K_b\Big( \frac{T_{i_1j_1}-t}{h_3} \Big)
K_c\Big( \frac{T_{i_1j_2}-s}{h_3} \Big),
\nonumber
\end{align}
where $K_l(t) = t^lK(t)$, $a=0,1$, $b=0,1$, and $c=0,1$.
Note that $A_{abc,ij}$ and $B_{abc,ij}$ are uniformly bounded
functions of $\bm{X}_{ij}$, $\bm{Z}_{ij}$, and $T_{ij}$. This is
a generalized U-statistics when we remove the summands of
$i_1=i_2$ and we recall (\ref{eqn:e100}) when we evaluate
(\ref{eqn:e736}).
It is easy to see that uniformly in $s$ and $t$,
\begin{align}
\lefteqn{ \frac{1}{N_1N_2h_1h_3^2}
\sum_{i_1=1}^n\sum_{j_1\ne j_2}\sum_{j_3}
\epsilon_{i_1j_1}\epsilon_{i_1j_3} 
A_{abc,i_1j_2}B_{abc,i_1j_3} } \label{eqn:e737}\\
& \qquad \qquad \times K_a\Big( \frac{T_{i_1j_3}-T_{i_1j_2}}{h_1} \Big)
K_b\Big( \frac{T_{i_1j_1}-t}{h_3} \Big)
K_c\Big( \frac{T_{i_1j_2}-s}{h_3} \Big) = O_p \Big( \frac{1}{nh_1}\Big).
\nonumber
\end{align}
In the same way as when dealing with $R_{1ab}^{(3)}(t)$, we obtain
\begin{align}
\lefteqn{ \frac{1}{N_1N_2h_1h_3^2}
\sum_{i_1\ne i_2}\sum_{j_1\ne j_2}\sum_{j_3}
\epsilon_{i_1j_1}\epsilon_{i_2j_3} 
A_{abc,i_1j_2}B_{abc,i_2j_3} } \label{eqn:e738}\\
& \qquad \qquad \times K_a\Big( \frac{T_{i_2j_3}-T_{i_1j_2}}{h_1} \Big)
K_b\Big( \frac{T_{i_1j_1}-t}{h_3} \Big)
K_c\Big( \frac{T_{i_1j_2}-s}{h_3} \Big)=O_p 
\Big( \frac{\log n}{nh_1^{1/2}h_3}\Big)
\nonumber
\end{align}
with 
$A=C_1(\log n)^k m_{\rm max}^3/ ( n^2 h_1 h_3^2)$,
$B=C_2(\log n)^k m_{\rm max}^2/ ( n^{3/2} h_1^{1/2} h_3^2)$,
$C=C_3 / (n h_1^{1/2} h_3 )$, and
$x= M\log n/(n h_1^{1/2} h_3 )$ 
in the exponential inequality. Note that we should choose
sufficiently large $k$ and $M$. 
It follows from (\ref{eqn:e737}) and (\ref{eqn:e738})
that uniformly in $s$ and $t$,
\begin{equation}
R_2(s,t) = O_p \Big( \frac{\log n}{nh_1^{1/2}h_3}\Big).
\label{eqn:e739}
\end{equation}
Note that we cannot relax the assumption of $\mx = O(n^{1/8})$ 
in Assumption A1 when we derive (\ref{eqn:e739}).
It follows from  (\ref{eqn:e796})- (\ref{eqn:e735}) and (\ref{eqn:e739}) 
that uniformly in $s$ and $t$,
\begin{align*}
\lefteqn{\widehat{\sigma}(s,t)- \sigma (s,t)}\\
 & = 
(1\,0\, 0)( L_9(s,t) )^{-1}E_2(s,t) 
+ \frac{h_3^2}{2}(1\,0\, 0)( L_9(s,t) )^{-1}L_{10}(s,t)
\begin{pmatrix}
\frac{\partial^2 \sigma}{\partial s^2}(s,t) \\
\frac{\partial^2 \sigma}{\partial s \partial t}(s,t)\\
\frac{\partial^2 \sigma}{\partial t^2}(s,t)
\end{pmatrix}\\
& \qquad \qquad + O_p(h_1^3) + O_p(h_3^3) 
+ O_p \Big( \frac{\log n}{nh_1}\Big)
+ O_p \Big( \frac{\log n}{nh_3^2}\Big).
\end{align*}
The expression given in Proposition \ref{prop:prop4} follows from
the above expression.

\section{Proofs of Lemmas \ref{lem:lem1}-\ref{lem:lem7}}

\vspace{0.1in}


First we state some results on $\widehat{\bm{\Sigma}}_i$. Set
\begin{equation}
\overline{\delta_n} = h_2^2 + h_3^2
+ \sqrt{\frac{\log n}{nh_2}}+\sqrt{\frac{\log n}{nh_3^2}}.
\label{eqn:r742}
\end{equation}
Then we have from Proposition \ref{prop:prop4} that uniformly
in $i$, 
\[
\max\{ | \lambda_{\rm min}
(\bm{\Sigma}_i- \widehat{\bm{\Sigma}}_i) |,
|\lambda_{\rm max} (\bm{\Sigma}_i- \widehat{\bm{\Sigma}}_i)|
\}  = O_p (m_i\overline{\delta_n}).
\]
Recall that
\begin{align*}
\widehat{\bm{\Sigma}}_i^{-1}- \bm{\Sigma}_i^{-1}
& = \widehat{\bm{\Sigma}}_i^{-1}
(\bm{\Sigma}_i- \widehat{\bm{\Sigma}}_i)\bm{\Sigma}_i^{-1}\\
& =  \bm{\Sigma}_i^{-1}(\bm{\Sigma}_i- \widehat{\bm{\Sigma}}_i)\bm{\Sigma}_i^{-1}
+ \widehat{\bm{\Sigma}}_i^{-1}(\bm{\Sigma}_i- \widehat{\bm{\Sigma}}_i)
\bm{\Sigma}_i^{-1}(\bm{\Sigma}_i- \widehat{\bm{\Sigma}}_i)\bm{\Sigma}_i^{-1}.
\end{align*}
We have from Assumption A4 and Proposition \ref{prop:prop4}
that uniformly in $i$,
\begin{align}| 
\bm{\Sigma}_i^{-1}( \bm{\Sigma}_i - \widehat{\bm{\Sigma}}_i ) \bm{\Sigma}_i^{-1} |_{\rm max}
& = O_p(m_i \overline{\delta_n}) , \label{eqn:r715}\\  |
  \widehat{\bm{\Sigma}}_i^{-1}(\bm{\Sigma}_i- \widehat{\bm{\Sigma}}_i)
\bm{\Sigma}_i^{-1}(\bm{\Sigma}_i- \widehat{\bm{\Sigma}}_i)\bm{\Sigma}_i^{-1} |_{\rm max} & = 
O_p(m_i^2 \overline{\delta_n}^2), \label{eqn:r721}
\end{align}
where $|A|_{\rm max}=\max_{i,j}|a_{ij}|$ for any matrix $A=(a_{ij})$. Besides, it follows from Assumption A4 that we have
uniformly in $i$, 
\begin{equation}
\max\{ | \lambda_{\rm min}
(\bm{\Sigma}_i^{-1}(\bm{\Sigma}_i- \widehat{\bm{\Sigma}}_i)\bm{\Sigma}_i^{-1}) |, | \lambda_{\rm max}
(\bm{\Sigma}_i^{-1}(\bm{\Sigma}_i- \widehat{\bm{\Sigma}}_i)\bm{\Sigma}_i^{-1}) |
\} = O_p(m_i \overline{\delta_n}). \label{eqn:r718}
\end{equation}
We also have the same result for
$ \widehat{\bm{\Sigma}}_i^{-1}(\bm{\Sigma}_i- \widehat{\bm{\Sigma}}_i)
\bm{\Sigma}_i^{-1}(\bm{\Sigma}_i- \widehat{\bm{\Sigma}}_i)\bm{\Sigma}_i^{-1}$ as in (\ref{eqn:r718})
with $m_i \overline{\delta_n}$ replaced by $(m_i \overline{\delta_n})^2$.
Proposition \ref{prop:prop4} also implies 
each element of $\bm{\Sigma}_i^{-1}
(\bm{\Sigma}_i- \widehat{\bm{\Sigma}}_i)\bm{\Sigma}_i^{-1}$ has the form of
\begin{equation}
D_{i}^{(1)}(\underline{T}_i)h_2^2 + D_{i}^{(2)}(\underline{T}_i)h_3^2
+\sum_{j=1}^{m_i}D_{ij}^{(3)}(\underline{T}_i)E_1(T_{ij})
+\sum_{j \ne j'}D_{ijj'}^{(4)}(\underline{T}_i)
E_2(T_{ij},T_{ij'}) + D_{i}^{(5)},
\label{eqn:e612}
\end{equation}
where
\[
D_{i}^{(5)} =m_i O_p\Big(
h_1^3+ h_2^3+ h_3^3 + \frac{\log n}{nh_1}
+ \frac{\log n}{nh_2} + \frac{\log n}{nh_3^2}
\Big)
\]
uniformly in $i$. 

We state the following two useful facts before we start proving Lemmas
\ref{lem:lem1}-\ref{lem:lem7}, both hold uniformly in
$l$:
\begin{equation}
\frac{1}{n}\sum_{i=1}^nm_i^3 \sum_{j=1}^{m_i}|W_{ijl}|
=O_p( K_n^{-1} )\,,
\label{eqn:r733}
\end{equation}
\begin{equation}
\mbox{and }\quad \frac{1}{n}\sum_{i=1}^nm_i^2 \sum_{j_1=1}^{m_i}
\sum_{j_2=1}^{m_i} |W_{ij_1l}||\epsilon_{ij_2}|
=O_p( K_n^{-1} )\,,
\label{eqn:r736}
\end{equation}
where $W_{ijl}$ denotes the $l$th element of $\bm{W}_{ij}$.
We can prove them in the same way, except that we need a kind of truncation argument when showing (\ref{eqn:r736}), and we outline the proof
of (\ref{eqn:r733}) in the following. 
To prove (\ref{eqn:r733}), 
we evaluate the expectation and variance and apply the Bernstein inequality.
First note that we have uniformly in $l$,
\[
\RE \Big\{ 
n^{-1}\sum_{i=1}^nm_i^3 \sum_{j=1}^{m_i}|W_{ijl}|
\Big\} = O( K_n^{-1} ).
\]
This follows from the local property of the B-spline basis and Assumption A2.
In addition, since we have from Assumption A2 that
\begin{eqnarray*}
&& \RE \Big\{ \frac{m_{\rm max}^2}{n^2}
\sum_{i=1}^n m_i^4 \sum_{j=1}^{m_i}|W_{ijl}|^2 + \frac{m_{\rm max}^3}{n^2}
\sum_{i=1}^n m_i^3 \sum_{j_1\ne j_2} |W_{ij_1l}| |W_{ij_2l}|
\Big\} \\
 & & = O\Big( \frac{m_{\rm max}^2}{nK_n} + \frac{m_{\rm max}^3}{nK_n^2} \Big),
\end{eqnarray*}
the variance is bounded from above by $C_1n^{-19/20}$ uniformly in $l$.
Each summand is bounded from above by $C_2m_{\rm max}^4/n=O(n^{-1/2})$.
Hence (\ref{eqn:r733}) and the uniformity in $l$
follow from the Bernstein inequality.
\\

\medskip
\noindent {\it Proof of Lemma \ref{lem:lem1}. }\,\, 
We can verify the result on $n^{-1}h_{12,kl}$
by using the local property
of the B-spline basis and the Bernstein inequality
for independent bounded random variables.
Since
\begin{align*}
\frac{1}{n}(\widehat{\bH}_{12}- \bH_{12})
&  =  \frac{1}{n}\sum_{i=1}^n
\underline{\bm{X}}_i^T
\{ \bm{\Sigma}_i^{-1}(\bm{\Sigma}_i- \widehat{\bm{\Sigma}}_i)\bm{\Sigma}_i^{-1}
\} \underline{\bm{W}}_i\\
& \qquad \quad + \frac{1}{n}\sum_{i=1}^n
\underline{\bm{X}}_i^T
\{ \widehat{\bm{\Sigma}}_i^{-1}(\bm{\Sigma}_i- \widehat{\bm{\Sigma}}_i)
\bm{\Sigma}_i^{-1}(\bm{\Sigma}_i- \widehat{\bm{\Sigma}}_i)\bm{\Sigma}_i^{-1}
\} \underline{\bm{W}}_i,
\end{align*}
the desired result on $n^{-1}( \widehat{h}_{12,kl}
- h_{12,kl}) $ follows from (\ref{eqn:r715}),
(\ref{eqn:r721}),  and (\ref{eqn:r733}).
The results on the Euclidean norm follow from
those on the elements. Hence the proof is complete.\\

\noindent {\it Proof of Lemma \ref{lem:lem2}}. \,\,
We have from Assumption A4 that
\begin{align}
\frac{C_1}{n}\sum_{i=1}^n \frac{1}{m_i}
\underline{\bm{W}}_i^T
\underline{\bm{W}}_i & \le \frac{1}{n}\bH_{22} \le
\frac{C_2}{n}\sum_{i=1}^n \underline{\bm{W}}_i^T
\underline{\bm{W}}_i \label{eqn:e740}\\
\intertext{for some positive constants $C_1$ and $C_2$ and for $k=0,1$,}
\frac{1}{n}\sum_{i=1}^n 
\frac{1}{m_i^k}\underline{\bm{W}}_i^T
\underline{\bm{W}}_i &
=\frac{1}{n}\sum_{i=1}^n\frac{1}{m_i^k}
\sum_{j=1}^{m_i}
(\bm{Z}_{ij}\bm{Z}_{ij}^T)\otimes
(\bm{B}(T_{ij})\bm{B}^T(T_{ij})).\nonumber
\end{align}
Thus the first result follows from Assumptions A2 and A3
and the standard arguments on B-spline bases as in the
proofs of Lemmas A.1 and A.2 of \cite{HWZ2004}.

Since we have
\begin{align*}
\frac{1}{n}(\widehat{\bH}_{22}- \bH_{22})
 & =  \frac{1}{n}\sum_{i=1}^n
\underline{\bm{W}}_i^T
\{ \bm{\Sigma}_i^{-1}(\bm{\Sigma}_i- \widehat{\bm{\Sigma}}_i)\bm{\Sigma}_i^{-1}
\} \underline{\bm{W}}_i\\
& \quad \qquad + \frac{1}{n}\sum_{i=1}^n
\underline{\bm{W}}_i^T
\{ \widehat{\bm{\Sigma}}_i^{-1}(\bm{\Sigma}_i- \widehat{\bm{\Sigma}}_i)
\bm{\Sigma}_i^{-1}(\bm{\Sigma}_i- \widehat{\bm{\Sigma}}_i)\bm{\Sigma}_i^{-1}
\} \underline{\bm{W}}_i,
\end{align*}
the second result follows from (\ref{eqn:r718}), 
the inequalities similar to (\ref{eqn:e740}), and Assumptions A2 and
A3.
The third result follows from the first and second results. Finally we deal with the fourth result. Note that
\begin{eqnarray}
\lefteqn{(n^{-1}\widehat{\bH}_{22})^{-1}
- (n^{-1}\bH_{22})^{-1}}\label{eqn:e743}\\
 & = &(n^{-1}\bH_{22})^{-1}( n^{-1}\bH_{22} - n^{-1}\widehat{\bH}_{22})
(n^{-1}\bH_{22})^{-1}\nonumber \\
& & + (n^{-1}\widehat{\bH}_{22})^{-1}
( n^{-1}\bH_{22} - n^{-1}\widehat{\bH}_{22})\nonumber\\
& & \qquad \times (n^{-1}\bH_{22})^{-1}
( n^{-1}\bH_{22} - n^{-1}\widehat{\bH}_{22})(n^{-1}\bH_{22})^{-1}.
\nonumber
\end{eqnarray}
By using the first, second, and third results and
(\ref{eqn:e743}), we obtain the fourth one.
Hence the proof is complete.\\

\noindent {\it Proof of Lemma \ref{lem:lem3}}. \,\,
The first result follows from (\ref{eqn:r718}).
The second one follows from Lemmas \ref{lem:lem1} and \ref{lem:lem2}.
The last one follows from the first two.\\

\noindent {\it Proof of Lemma \ref{lem:lem4}}. \,\,
The first result follows from the fact
\[
\frac{C_1}{n}\sum_{i=1}^n \frac{1}{m_i}\underline{\bm{W}}_i^T
\underline{\bm{W}}_i  
\le \frac{1}{n}\sum_{i=1}^n \underline{\bm{W}}_i^T
\bm{\Sigma}_i^{-1}\underline{\bm{W}}_i
\le
\frac{C_2}{n}\sum_{i=1}^n \underline{\bm{W}}_i^T
\underline{\bm{W}}_i
\]
for some positive constants $C_1$ and $C_2$.
Next note that
\begin{eqnarray}
\lefteqn{
\frac{1}{\sqrt{n}}\sum_{i=1}^n
\underline{\bm{W}}_i^T
\{ \bm{\Sigma}_i^{-1}(\bm{\Sigma}_i- \widehat{\bm{\Sigma}}_i)\bm{\Sigma}_i^{-1}
\} \, \underline{\epsilon}_i
}\label{eqn:e746} \\
& = & \frac{1}{\sqrt{n}}\sum_{i=1}^n
\underline{\bm{W}}_i^T
\{ \bm{\Sigma}_i^{-1}(\bm{\Sigma}_i- \widehat{\bm{\Sigma}}_i)\bm{\Sigma}_i^{-1}
\} \, \underline{\epsilon}_i\nonumber\\
& & + \frac{1}{\sqrt{n}}\sum_{i=1}^n
\underline{\bm{W}}_i^T
\{ \widehat{\bm{\Sigma}}_i^{-1}(\bm{\Sigma}_i- \widehat{\bm{\Sigma}}_i)
\bm{\Sigma}_i^{-1}(\bm{\Sigma}_i- \widehat{\bm{\Sigma}}_i)\bm{\Sigma}_i^{-1}
\} \, \underline{\epsilon}_i.\nonumber
\end{eqnarray}
By employing (\ref{eqn:r721}) and (\ref{eqn:r736}),
we can prove
the stochastic order of
the elements of the second term of the right-hand side
is uniformly 
$
O_p(\sqrt{n}K_n^{-1}(h_2^4 + h_3^4 + \log n /(nh_2) 
+ \log n /(nh_3^2))).
$
Thus the norm of this $qK_n$-dimensional
vector has the stochastic
order of
\begin{equation} 
\sqrt{\frac{n}{K_n}}
O_p\Big( h_2^4 + h_3^4 + \frac{\log n}{nh_2}
+ \frac{\log n}{nh_3^2} \Big).
\label{eqn:e754}
\end{equation}

According to Proposition \ref{prop:prop4}, the first term of the right-hand side of (\ref{eqn:e746}) can be decomposed into
\begin{equation}
\frac{1}{\sqrt{n}}\sum_{i=1}^n \underline{\bm{W}}_i^T
Q_{1i} \underline{\epsilon}_i +
\frac{1}{\sqrt{n}}\sum_{i=1}^n \underline{\bm{W}}_i^T
Q_{2i} \underline{\epsilon}_i + 
\frac{1}{\sqrt{n}}\sum_{i=1}^n \underline{\bm{W}}_i^T
Q_{3i} \underline{\epsilon}_i\,,
\label{eqn:e752}
\end{equation}
where $ Q_{1i} $ corresponds to the first and second terms in (\ref{eqn:e612}), 
$ Q_{2i} $ corresponds to the third and fourth terms in (\ref{eqn:e612}), and $ Q_{3i} $ corresponds to the fifth term
in (\ref{eqn:e612}).
Proposition \ref{prop:prop4} implies
\[
Q_{1i}=Q_{1i}^{(2)}h_2^2+ Q_{1i}^{(3)}h_3^2,
\]
where we have for $s=2,3$,
\[
\max \{ |\lambda_{\rm min}( Q_{1i}^{(s)} )|, |
\lambda_{\rm max}( Q_{1i}^{(s)} )|
\} = O(m_i)
\]
uniformly in $i$. Besides $Q_{1i}^{(s)}$ depends
only on $\underline{T_i}$ for $s=2,3$.
The $(k,l)$ element of $Q_{2i}$ has the form of
\[
\sum_{j=1}^{m_i} \sigma_i^{kj}\sigma_i^{lj}
E_1(T_{ij}) + \sum_{j\ne j'}\sigma_i^{kj}\sigma_i^{lj'}
E_2(T_{ij}, T_{ij'}),
\]
where $\bm{\Sigma}_i^{-1} = (\sigma_i^{kl})$. Note that
uniformly in $l$ and $i$, 
\[
\sum_{k=1}^{m_i}(\sigma_i^{kl})^2 = O(1).
\]
Uniformly in $i$, the elements of $Q_{3i}$, $D_i^{(5)}$ in (\ref{eqn:e612}),
have the order
of
\[
m_iO_p\Big(
h_1^3+ h_2^3 + h_3^3 + \frac{\log n}{nh_1}
+ \frac{\log n}{nh_2} + \frac{\log n}{nh_3^2} 
\Big)\,.
\]

We can prove as in the proof of Lemma \ref{lem:lem2}
that for $s=2,3$,
\[
\frac{C_1}{K_n}\bm{I}_{qK_n}
\le  {\rm Cov}\Big( n^{-1/2}\sum_{i=1}^n \underline{\bm{W}}_i^T
Q_{1i}^{(s)} \underline{\epsilon}_i \Big)
\le \frac{C_2}{K_n}\bm{I}_{qK_n}
\]
for some positive constants $C_1$ and $C_2$. Hence we have
\begin{equation}\Big|
n^{-1/2}\sum_{i=1}^n \underline{\bm{W}}_i^T
Q_{1i}\underline{\epsilon}_i \Big| = O_p( h_2^2 + h_3^2 ).
\label{eqn:e753}
\end{equation}
Similarly to the second term in the right-hand side of (\ref{eqn:e746}), we can
demonstrate by using (\ref{eqn:r736}) that
\begin{equation}\Big|
n^{-1/2}\sum_{i=1}^n \underline{\bm{W}}_i^T
Q_{3i}\underline{\epsilon}_i \Big| = 
\sqrt{\frac{n}{K_n}}
O_p \Big( h_1^3 + h_2^3 + h_3^3 +\frac{\log n}{nh_1} + 
\frac{\log n}{nh_2} + \frac{\log n}{nh_3^2} \Big).
\label{eqn:e755}
\end{equation}
Finally we evaluate the second term of (\ref{eqn:e752}) and it
has a structure of V-statistics. By exploiting the
structure, we evaluate the expectations and the variances of
the elements by using Assumption A2. Then we have
\begin{equation*}\Big|
n^{-1/2}\sum_{i=1}^n \underline{\bm{W}}_i^T
Q_{2i}\underline{\epsilon}_i \Big| =
O_p\Big(\frac{1}{\sqrt{nh_2}}+ \frac{1}{\sqrt{nh_3^2}} + 
\frac{1}{\sqrt{nK_n}h_2} + \frac{1}{\sqrt{nK_n} h_3^2} \Big).
\end{equation*}
The second result follows from (\ref{eqn:e754}), 
(\ref{eqn:e753}), (\ref{eqn:e755}), and the above equality. \\ 

\noindent {\it Proof of Lemma \ref{lem:lem5}}. \,\,
This lemma can be proved in the same way as
Lemma \ref{lem:lem4} and the details are omitted.\\

\noindent {\it Proof of Lemma \ref{lem:lem6}}. \,\,
From the definition of $\bm{\gamma}^*$ given after (\ref{eqn:e603}), we have
\begin{equation*} 
\max_{1\leq j\leq m_i} | \bm{W}_{ij}^T\bm{\gamma}^*- \bm{Z}_{ij}^T\bm{g}_0(T_{ij}) |
= O_p(K_n^{-2})
\end{equation*}
uniformly in $i$.
The above equality and (\ref{eqn:r733}) imply 
that the elements of
\[
\frac{1}{n}\sum_{i=1}^n\underline{\bm{W}}_i^T\bm{\bm{\Sigma}}_i^{-1}(
\underline{\bm{W}}_i\bm{\gamma}^* - \underline{(\bm{Z}^T\bm{g}_0)}_i)
\]
is uniformly $O_p(K_n^{-3})$ and the first result follows from
this.
As for the second result, first we note that
\[|
\widehat{\bm{\bm{\Sigma}}}_i^{-1} - \bm{\Sigma}_i^{-1}|_{\rm max}
= O_p( m_i \overline{\delta_n} )
\]
uniformly in $i$ from (\ref{eqn:r715}) and (\ref{eqn:r721}).
Recall that $  \overline{\delta_n}$ is defined in (\ref{eqn:r742}).
Thus the elements of $
\underline{\bm{W}}_i^T (\widehat{\bm{\Sigma}}_i^{-1} - \bm{\Sigma}_i^{-1})
( \underline{\bm{W}}_i\bm{\gamma}^* - \underline{(\bm{Z}^T\bm{g}_0)}_i)
$
are bounded uniformly in $l$ by
\[
CK_n^{-2} \overline{\delta_n} m_i^2 \sum_{j=1}^{m_i} |W_{ijl}|
\]
with probability tending to 1 for some positive constant
$C$.
Hence the second result follows from (\ref{eqn:r733}).
\\

\noindent {\it Proof of Lemma \ref{lem:lem7}}. \,\, This lemma can be proved in the same way as
Lemma \ref{lem:lem6} and the details are omitted.
\\

\section{Theoretical results for general link functions}

We state the results of Section \ref{sec:efficiency}
for general link functions when $m_i$ is uniformly
bounded and $\underline{\epsilon}_i$ satisfies the
sub-Gaussian assumption, Assumption A6$^\prime$ here. 
Note that we have no counterpart of Theorem \ref{thm:thm1}
for general link functions even when $m_i$ is uniformly bounded.

Let  
$v_1$ and $v_2$ be two processes each taking a scalar stochastic value at $T_{ij}$, $i=1,\ldots,n$, $j=1,\ldots,m_i$. Then we define two inner products of $v_1$ and $v_2$ by 
\begin{equation*}
\langle v_1, v_2 \rangle_n^\Delta 
=\frac{1}{n}\sum_{i=1}^n \underline{v}_{1i}^T
\Delta_{0i}\bm{V}_i^{-1} \Delta_{0i}\underline{v}_{2i}
\ {\rm and}\ 
\langle v_1, v_2 \rangle^\Delta =
\RE \{  \langle v_1, v_2 \rangle_n^\Delta  \},
\end{equation*}
where $\underline{v}_{1i}$ and  $\underline{v}_{2i}$ are defined 
in the same way as $\underline{T}_{i}$ and
\[
\Delta_{0i} = {\rm diag}\big(\mu'(\bm{X}_{i1}^T\bm{\beta}_0+
\bm{Z}_{i1}^T\bm{g}_0(T_{i1})), \ldots,
\mu'(\bm{X}_{im_i}^T\bm{\beta}_0+ \bm{Z}_{im_i}^T\bm{g}_0(T_{im_i}))\big).
\]
The associated norms are then defined by
\begin{equation*}
\| v\|_n^\Delta = ( \langle v, v \rangle_n^\Delta  )^{1/2}
\ {\rm and}\ 
\| v\|^\Delta = ( \langle v, v \rangle^\Delta  )^{1/2}.
\end{equation*}
We now define the projections, with respect to $\| \cdot
\|^{\Delta}$, of  the $k$th element of $\bm{X}$ onto $\bm{Z}^T\bm{G}$ and $\bm{Z}^T\bm{G}_B$ by
\begin{equation*}
\Pi_\Delta X_k =\argmin_{\bm{g}\in \bm{G}}
\| X_k -\bm{Z}^T\bm{g} \|^\Delta\ {\rm and}\ 
\Pi_{\Delta n}X_k =\argmin_{\bm{g}\in \bm{G}_B}
\| X_k -\bm{Z}^T\bm{g} \|^\Delta,
\end{equation*}
where 
\[
\| X_k -\bm{Z}^T\bm{g} \|^\Delta
= \frac{1}{n}\RE \Big\{
\sum_{i=1}^n( \underline{X}_{ik}-
\underline{(\bm{Z}^T\bm{g})}_i)^T\Delta_{0i}\bm{V}_i^{-1}
\Delta_{0i}( \underline{X}_{ik}-
\underline{(\bm{Z}^T\bm{g})}_i) \Big\},
\]
with $\underline{X}_{ik}=(X_{i1k},\ldots, X_{im_ik})^T$ and
$\underline{(\bm{Z}^T\bm{g})}_i=(
\bm{Z}_{i1}^T\bm{g}(T_{i1}), \ldots,
\bm{Z}_{im_i}^T\bm{g}(T_{im_i}))$.
We denote these projections by $\bm{\varphi}_{\Delta k}^*  = \Pi_{\Delta} X_k$ and $\overline{\bm{\varphi}}_{\Delta k}  =
\Pi_{\Delta n} X_k$, and define another one by
\begin{equation*}
\widehat{\bm{\varphi}}_{\Delta k} = \widehat{\Pi}_{\Delta n} X_k,
\end{equation*}
where
\[
\widehat{\Pi}_{\Delta n} X_k = \argmin_{\bm{g}\in \bm{G}_B}
\| X_k -\bm{Z}^T\bm{g} \|_n^\Delta.
\]
The arguments in Section \ref{sec:projection} also
apply to this $\bm{\varphi}_{\Delta k}^*$.

Some matrices are necessary to
present Proposition \ref{prop:prop5} and we define them
here. Let
\begin{align}
\widetilde{\bH}&= 
\begin{pmatrix}
\sum_{i=1}^n \underline{\bm{X}}_i^T\Delta_{0i}\bm{V}_i^{-1}
\Delta_{0i}\underline{\bm{X}}_i
& 
\sum_{i=1}^n \underline{\bm{X}}_i^T\Delta_{0i}\bm{V}_i^{-1}
\Delta_{0i}\underline{\bm{W}}_i
\\
\sum_{i=1}^n \underline{\bm{W}}_i^T\Delta_{0i}\bm{V}_i^{-1}
\Delta_{0i}\underline{\bm{X}}_i
&
\sum_{i=1}^n \underline{\bm{W}}_i^T\Delta_{0i}\bm{V}_i^{-1}
\Delta_{0i}\underline{\bm{W}}_i
\end{pmatrix}\nonumber \\
 &=
\begin{pmatrix}
\widetilde{\bH}_{11}& \widetilde{\bH}_{12}\\
\widetilde{\bH}_{21}& \widetilde{\bH}_{22}
\end{pmatrix} \quad {\rm (say)}, \nonumber\\ 
\widetilde{\bH}_{11 \cdot 2}& = \widetilde{\bH}_{11}- \widetilde{\bH}_{12}\widetilde{\bH}_{22}^{-1}\widetilde{\bH}_{21}\,, \quad
{\rm and}\quad \widetilde{\bH}^{11}=(\widetilde{\bH}_{11 \cdot 2})^{-1}\,.\nonumber
\end{align}
Let $\widetilde{\Omega}_{\bm{V}n}$ be a $p\times p$ matrix whose $(k,l)$th
element is 
\begin{equation*}
\frac{1}{n}
\sum_{i=1}^n
\RE \Big\{
( \underline{X}_{ik} - \underline{ (\bm{Z}^T
\bm{\varphi}_{\Delta k}^* )}_i )^T \Delta_{0i} \bm{V}_i^{-1}
\Delta_{0i}( \underline{X}_{il} - \underline{(\bm{Z}^T
\bm{\varphi}_{\Delta l}^* )}_i) 
\Big\}.
\end{equation*}
Note that $n^{-1}\widetilde{\bH}_{11 \cdot 2}$
is an estimate of $\widetilde{\Omega}_{\bm{V}n}$.
We assume that there exists a 
$p\times p$ positive definite matrix $\widetilde{\Omega}_{\bm{V}}$ such that 
\begin{equation}
\lim_{n\to \infty}\widetilde{\Omega}_{\bm{V}n}=
\widetilde{\Omega}_{\bm{V}}.
\label{eqn:e1236}
\end{equation}

We present Propositions \ref{prop:prop5}-\ref{prop:prop7}
before stating the assumptions for these propositions. 
By using Lemma \ref{lem:lem10} we can prove Proposition \ref{prop:prop5} based on the same arguments as those in \cite{CZH2014}.
\begin{prop}(Asymptotic normality of $\widehat{\bm{\beta}}_{\bm{V}}$)
\label{prop:prop5}
Under Assumption S in Section \ref{sec:efficiency} for the norm
here, (\ref{eqn:e1236}), and Assumptions A1$^\prime$, A2$^\prime$, A3, A4$^\prime$, A5$^\prime$, and A6$^\prime$, 
we have
\[
\widehat{\bm{\beta}}_{\bm{V}} = \bm{\beta}_0 + \widetilde{\bH}^{11}
\sum_{i=1}^n (\underline{\bm{X}}_i - \underline{\bm{W}}_i
\widetilde{\bH}_{22}^{-1}\widetilde{\bH}_{21})^T \Delta_{0i}\bm{V}_i^{-1}\underline{\epsilon}_i
+o_p\Big( \frac{1}{\sqrt{n}}\Big).
\]
We also have
\[
\widetilde{\Gamma}_{\bm{V}}^{-1/2}
(\widehat{\bm{\beta}}_{\bm{V}}- \bm{\beta}_0)\stackrel{d}{\to}
\RN(0, \bm{I}_p),
\]
where $\widetilde{\Gamma}_{\bm{V}}$
is 
\begin{equation*}
\widetilde{\bH}^{11}\sum_{i=1}^n
\Big\{ ( \underline{\bm{X}}_i - \underline{\bm{W}}_i
\widetilde{\bH}_{22}^{-1}\widetilde{\bH}_{21})^T \Delta_{0i}\bm{V}_i^{-1}\bm{\Sigma}_i
\bm{V}_i^{-1}\Delta_{0i}( \underline{\bm{X}}_i - 
\underline{\bm{W}}_i \widetilde{\bH}_{22}^{-1}\widetilde{\bH}_{21})\Big\}\widetilde{\bH}^{11}.
\end{equation*}
\end{prop}

We give in Proposition \ref{prop:prop6} the semiparametric efficiency bound for estimation 
of $\bm{\beta}_0$. It can be proved in the
same way as Lemma 1 of \cite{CZH2014} and the proof is omitted.
We denote the semiparametric efficient score function of $\bm{\beta}$ 
by
\[
\tilde{\bm{l}}_{\bm{\beta}}^*= (\tilde{l}_{\bm{\beta}1}^*,
\ldots, \tilde{l}_{\bm{\beta}p}^*)^T.
\]
Its expression is given in Proposition \ref{prop:prop6}.
When $\bm{V}_i=\bm{\Sigma}_i$, we denote $\bm{\varphi}_{\Delta k}^* (t)$ by
$\tilde{\bm{\varphi}}_{eff,k}^* (t) $.

\begin{prop}(Semiparametric efficiency bound)
\label{prop:prop6}
Under the same assumptions as in Proposition \ref{prop:prop5},
we have
\[
\tilde{l}_{\bm{\beta}k}^* = 
\sum_{i=1}^n ( \underline{X}_{ik} - \underline{ (\bm{Z}^T
\tilde{\bm{\varphi}}_{eff,k}^* )}_i )^T
\Delta_{0i} \bm{\Sigma}_i^{-1}\{ 
\underline{Y}_i - \underline{\mu}
( \underline{\bm{X}}_{i}\bm{\beta}_0+
\underline{(\bm{Z}^T\bm{g}_0)}_i )\},
\]
and the semiparametric efficient information matrix for $\bm{\beta}$
is given by
\[
\lim_{n \to \infty}\frac{1}{n}
\RE \{ \tilde{\bm{l}}_{\bm{\beta}}^* (\tilde{\bm{l}}_{\bm{\beta}}^*)^T \}
= \widetilde{\Omega}_{\bm{\Sigma}} \ with \ \bm{V}_i = \bm{\Sigma}_i\ in\ (\ref{eqn:e1236}). 
\]
\end{prop}

Proposition \ref{prop:prop7} is parallel to Proposition \ref{prop:prop3}. 
It can be proved in the same
way as Corollary 1 of \cite{CZH2014}, and it also
follows from Proposition \ref{prop:prop5} and Lemma \ref{lem:lem10} (vii). 
Thus the proof is omitted. 

\begin{prop}(Oracle efficient estimator)
\label{prop:prop7}
Under the same assumptions as in Proposition \ref{prop:prop5},
we have with $\bm{V}_i=\bm{\Sigma}_i$ in\ (\ref{eqn:e206})
\[
\sqrt{n}\,\widetilde{\Omega}_{\bm{\Sigma}}^{1/2}
(\widehat{\bm{\beta}}_{\bm{\Sigma}}- \bm{\beta}_0)\stackrel{d}{\to}
\RN(0, \bm{I}_p).
\]
\end{prop}

\setcounter{prop}{0}

Now we describe assumptions for the above propositions.
Here we need Assumption A6$^\prime$ since we need some results from
the empirical process theory in dealing with general link functions.\\

\noindent
{\bf Assumption A1$^\prime$}.
\begin{enumerate}
\item[(i)] $\mu (x)$ is twice continuously differentiable and
$\inf_{x\in R} \mu'(x) >0$.

\item[(ii)] For some positive constant $C_{B9}$, we have
$\displaystyle{\limsup_{|x|\to \infty} |\mu(x)|/|x|^{C_{B9}}
< \infty}$.
\end{enumerate}

\noindent
{\bf Assumption A2$^\prime$}. 
The joint density functions $f_{ij}(t)$ and $f_{ijj'}(s,t)$ are uniformly bounded and we have for some positive constants
$C_{B1}$ and $C_{B2}$,
\begin{align*}
C_{B1}& < \frac{1}{n} \sum_{i=1}^n \sum_{j=1}^{m_i} f_{ij}(t)<C_{B2}
\ {\rm on} \ [0,1]\\
\mbox{and} \quad C_{B1}& <\frac{1}{n}\sum_{i=1}^n\sum_{j\ne j'} f_{ijj'}(s,t)<C_{B2}
\  {\rm on} \ [0,1]^2.
\end{align*}


\noindent
{\bf Assumption A4$^\prime$}. For some positive constants
$C_{B5}$ and $C_{B6}$, we have uniformly in $i$,
\[
C_{B5}\le \lambda_{\rm min}(\bm{\Sigma}_i)
\le \lambda_{\rm max}(\bm{\Sigma}_i) \le C_{B6}.
\]

\noindent
{\bf Assumption A5$^\prime$}. For some positive constants
$C_{B7}$ and $C_{B8}$, we have uniformly in $i$,
\[
C_{B7} \le \lambda_{\rm min}(\bm{V}_i)
\le \lambda_{\rm max}(\bm{V}_i) \le C_{B8}.
\]

\noindent
{\bf Assumption A6$^\prime$}. For some positive constants
$C_{B10}$ and $C_{B11}$, we have uniformly in $i$,
\[
\max_{1\le i \le n} C_{B10}
\RE \{ \exp(|\underline{\epsilon}_i|^2/C_{B10}) -1 |
\underline{\bm{X}}_i, \underline{\bm{Z}}_i,
\underline{T}_i \}\le C_{B11}.
\]

To prove Proposition \ref{prop:prop5},
we have only to proceed as in \cite{CZH2014S} by
replacing their $\bm{Z}_{ij}$, $\bm{Z}_i$, and
$\bm{\varphi}_{k}^* (\bm{t})$
with $\bm{W}_{ij}$, $\underline{\bm{W}}_i$,
and $\bm{Z}^T\bm{\varphi}_{\Delta k}^* (t) $, respectively.
We just
state the relevant changes and remarks in the following:
\begin{description}
\item{(i)} Lemmas S.2-S.4 of \cite{CZH2014S}: We reorganize these lemmas in Lemma \ref{lem:lem10} given later.
Its (i)-(iii), (iv) and (vi) correspond to Lemma S.2, the latter half of Lemma S.3  
and Lemma S.4 of \cite{CZH2014S}, respectively. 
The former half of Lemma S.3 of \cite{CZH2014S} seems to be used in their Corollary 1. However,
it can be relaxed to (v) of Lemma \ref{lem:lem10} here.
\item{(ii)} Lemma S.8 of \cite{CZH2014S}: The regressors $\bm{X}_{ij}$ and $\bm{W}_{ij}$
still form a VC class and we can proceed completely in the same
way as in \cite{CZH2014S}.
\end{description}

We state Lemma \ref{lem:lem10} in the following. It can be proved
it in the same way as Lemma \ref{lem:lem0}. 

\begin{lem} Assume that Assumptions A1$^\prime$, A2$^\prime$, A3,  A4$^\prime$, A5$^\prime$ hold. 
Then we have the following results.
\label{lem:lem10}
\hspace{0.1cm}

\begin{enumerate}
\item[(i)] There are positive constants $C_1$ and $C_2$ such that
\[
C_1 \| \bm{g} \|_{G,2} \le
\| \bm{Z}^T\bm{g} \|^\Delta \le C_2 \| \bm{g} \|_{G,2}
\]
for any $\bm{g} \in \bm{G}$.

\item[(ii)] There are positive constants $C_3$ and $C_4$ such that
\[
\| \bm{g} \|_{G,\infty}^2 \le
C_3 K_n \| \bm{g} \|_{G,2}^2
\le C_4 K_n ( \| \bm{Z}^T\bm{g} \|^\Delta )^2
\]
for any $\bm{g} \in \bm{G}_B$.

\item[(iii)] There is a positive constant $C_5$ such that
for any $\bm{\beta} \in \RR^p$ and $\bm{g} \in \bm{G}_B$,
\[
\| \bm{X}^T\bm{\beta}+ \bm{Z}^T \bm{g} \|_\infty
\le C_5 K_n^{1/2} \| \bm{X}^T\bm{\beta}+ \bm{Z}^T
\bm{g} \|^\Delta,
\]
where $\|v \|_\infty = \max_{i,j}|v_{ij}|$. Besides we have for
some positive constant $C_6$, 
\[
\|v \|^\Delta \le C_6 \|v \|_\infty.
\]

\item[(iv)]
\[
\sup_{\bm{g}_1,\bm{g}_2\in \bm{G}_B}
\Big| 
\frac{ \langle \bm{Z}^T\bm{g}_1, \bm{Z}^T\bm{g}_2 
\rangle_n^\Delta - \langle \bm{Z}^T\bm{g}_1, 
\bm{Z}^T\bm{g}_2 \rangle^\Delta}{\| \bm{Z}^T\bm{g}_1
\|^\Delta \| \bm{Z}^T\bm{g}_2 \|^\Delta }
\Big|
=O_p(K_n\sqrt{\log n/ n}).
\]

\item[(v)] For any positive constant $M$, we have
\[
\langle X_j- \bm{Z}^T\bm{g}_j, X_k- \bm{Z}^T\bm{g}_k 
\rangle_n^\Delta - \langle X_j- \bm{Z}^T\bm{g}_j, 
X_k - \bm{Z}^T\bm{g}_k \rangle^\Delta = o_p(1)
\]
uniformly in $\bm{g}_j\in \bm{G}_B$ and
$\bm{g}_k\in \bm{G}_B$ satisfying
$\|\bm{g}_j\|_{G,2}\le M$ and $\|\bm{g}_k\|_{G,2}\le M$,
respectively.

\item[(vi)] For any stochastic process  $\delta_n$ taking values at $T_{ij}$ satisfying that $ \| \delta_n \|_\infty $ is uniformly bounded in $n$
and $\{ \delta_{n,ij} \}_{j=1}^{m_i}$ are mutually independent
in $i$, we have
\[
\sup_{\bm{g} \in \bm{G}_B}
\Big| 
\frac{  \langle \delta_n, \bm{Z}^T\bm{g} 
\rangle_n^\Delta - \langle \delta_n, 
\bm{Z}^T\bm{g} \rangle^\Delta }{
\| \bm{Z}^T\bm{g} \|^\Delta }
\Big|
=O_p(\sqrt{ K_n / n})  \| \delta_n \|_\infty .
\]

\item[(vii)] We also have Assumption S for the norm here. Then
we have for $k=1, \ldots, p$,\ \ 
$ \| \widehat{\bm{\varphi}}_{\Delta k}\|_\infty = O_p(1)$, 
\[
\| \bm{Z}^T( \bm{\varphi}_{\Delta k}^*  -
\widehat{\bm{\varphi}}_{\Delta k})  \|_n^\Delta = o_p(1), \quad
{\rm and}\quad 
\| \bm{Z}^T( \bm{\varphi}_{\Delta k}^*  -
\widehat{\bm{\varphi}}_{\Delta k}  ) \|^\Delta = o_p(1).
\]
\end{enumerate}
\end{lem}
\setcounter{lem}{0}

\newpage

\setcounter{page}{1}
\setcounter{section}{0}
\setcounter{equation}{0}
\setcounter{thm}{0}
\setcounter{prop}{0}
\setcounter{lem}{0}
\setcounter{table}{0}
\numberwithin{equation}{section}
\renewcommand{\thesection}{\arabic{section}}
\renewcommand{\thethm}{\arabic{thm}}
\renewcommand{\theprop}{\arabic{prop}}
\renewcommand{\thelem}{\arabic{lem}}
\renewcommand{\thetable}{\arabic{table}}

\begin{frontmatter}


\title{Efficient estimation in semivarying coefficient models for longitudinal/clustered data}
\runtitle{Efficient Estimation}


\begin{aug}

\author{\fnms{Ming-Yen} \snm{Cheng} \corref{}\thanksref{t1,m1}\ead[label=e1]{cheng@math.ntu.edu.tw}, \fnms{Toshio} \snm{Honda}\thanksref{t2,m2}\ead[label=e2]{t.honda@r.hit-u.ac.jp}},
\and
\author{\fnms{Jialiang} \snm{Li}\thanksref{t3,m3}\ead[label=e3]{stalj@nus.edu.sg}}

\affiliation{National Taiwan University\thanksmark{m1}, Hitotsubashi University\thanksmark{m2},   and}
\affiliation{National University of Singapore\thanksmark{m3}}

\address{M.-Y. Cheng\\Department of Mathematics\\ National Taiwan University\\ Taipei 106, Taiwan\\ \printead{e1}}

\address{T. Honda\\ Graduate School of Economics\\ Hitotsubashi University\\ Kunitachi, Tokyo 186-8601, Japan \\ \printead{e2}}

\address{J. Li\\ Department of Statistics \& Applied Probability\\ National University of Singapore\\ Singapore 117546\\ \printead{e3}}

\thankstext{}{This research was partially supported by the Hitotsubashi International Fellow Program and a Taiwan Ministry of Education grant.}

\thankstext{t1}{Corresponding author. Research was supported by the Ministry of Science and Technology grants 101-2118-M-002-001-MY3
and 104-2118-M-002-005-MY3.}

\thankstext{t2}{Research was supported by the JSPS Grant-in-Aids for Scientific Research (A) 24243031 and (C) 25400197.}

\thankstext{t3}{Research was supported by grants AcRF R-155-000-130-112 and NMRC/CBRG/0014/2012.}

\runauthor{Cheng et al.}
\end{aug}

\begin{abstract}
In semivarying coefficient modeling of longitudinal/clustered data,
of primary interest is usually the parametric component which involves unknown constant coefficients. 
First we study semiparametric efficiency bound for estimation of the constant coefficients in a
general setup. It can be achieved by spline regression using the true within-subject covariance
matrices, which are often unavailable in reality. Thus we propose an estimator when the covariance matrices are unknown and depend only on the index variable.
To achieve this goal, we estimate the covariance matrices using residuals obtained from a preliminary estimation based on working independence and both spline and local linear regression. Then, using the covariance matrix estimates, 
we employ spline regression again  to obtain our final estimator.
It achieves the semiparametric efficiency bound
under normality assumption and has the smallest asymptotic covariance matrix among a
class of estimators even when normality is violated. 
Our theoretical results hold either when the number of within-subject observations diverges or when it is uniformly bounded. In addition, the local linear estimator of the nonparametric component is superior to the spline estimator in terms of numerical performance.  The proposed method is compared with the working independence estimator and some existing method via simulations and application to a real data example. 
\end{abstract}

\begin{keyword}[class=MSC]
\kwd[Primary ]{62G08}
\end{keyword}

\begin{keyword}
Covariance matrix estimation; 
local linear regression; 
semiparametric efficiency bound;
spline functions.
\end{keyword}

\end{frontmatter}

\newpage

\section{Introduction}
\label{sec:intro}
Suppose we have a scalar response $Y$, and two $p$-dimensional and 
$q$-dimensional covariate vectors $\bm{X}$ and $\bm{Z}$.
Longitudinal data consist of $(Y_{ij},\bm{X}_{ij},\bm{Z}_{ij},T_{ij}), i=1,\ldots, n,
j=1,\ldots, m_i$, where $Y_{ij}$, $\bm{X}_{ij}=(X_{ij1}, \ldots, X_{ijp})^T$ and 
$\bm{Z}_{ij}=  (Z_{ij1}, \ldots, Z_{ijq})^T$
are respectively the values of  $Y$, $\bm{X}$ and $\bm{Z}$ 
of the $i$th subject at the $j$th observation 
time $T_{ij}\in [0,1]$. Such kind of data are commonly acquired for various purposes,
such as evidence based knowledge discovery and empirical study, in a wide range of subject areas.
When the subjects are changed to clusters and the $T_{ij}$'s are observations on some index variable other than time, they are usually called clustered data. 
We assume that all the covariates are uniformly bounded for technical
reasons. Besides, we let $Z_{ij1}\equiv 1$ and suppose $\bm{X}_{ij}$ has no constant
element for all $i$ and $j$. 

For $i=1,\ldots,n$, denote
\begin{align*}
\underline{\bm{X}}_i & = (\bm{X}_{i1}, \ldots,
\bm{X}_{im_i})^T,\,\,
\underline{\bm{Z}}_i =  (\bm{Z}_{i1}, \ldots,
\bm{Z}_{im_i})^T,
 \,\, {\rm and} \,\,
\underline{T}_i =  (T_{i1}, \ldots, T_{im_i})^T.
\end{align*}
A popular model for longitudinal data analysis is the semivarying coefficient model, which is specified by 
\begin{align}
\lefteqn{\RE( Y_{ij}| \bm{X}_{ij}, \bm{Z}_{ij}, T_{ij},
\underline{\bm{X}}_i, \underline{\bm{Z}}_i, 
\underline{T}_i ) }\label{eqn:e100}
\\
&= \RE( Y_{ij}| \bm{X}_{ij}, \bm{Z}_{ij}, T_{ij} )
\equiv \mu ( \bm{X}_{ij}^T\bm{\beta}+
\bm{Z}_{ij}^T\bm{g}(T_{ij}) )=\mu_{ij},
\nonumber
\end{align}
where $\bm{A}^T$ stands for the transpose of a matrix $\bm{A}$. In model (\ref{eqn:e100}), $\mu (x)$ is a known strictly increasing smooth
link function, $\bm{\beta}$
is an unknown regression coefficient vector, and $\bm{g}(t)=\big(g_1(t),\ldots,g_q(t)\big)^T$ is a vector of unknown smooth functions. Define 
\begin{equation}
\underline{\epsilon}_i
=(\epsilon_{i1}, \ldots, \epsilon_{im_i})^T
= \underline{Y}_i-\underline{\mu}_i\,, \,\,
{\rm and}\ 
\bm{\Sigma}_i = \var (\underline{\epsilon}_i| \underline{\bm{X}}_i,
\underline{\bm{Z}}_i, \underline{T}_i ),
\label{eqn:e102}
\end{equation}
where $\underline{Y}_i=(Y_{i1}, \ldots, Y_{im_i})^T$,
$\underline{\mu}_i=(\mu_{i1}, \ldots, \mu_{im_i})^T$,
and $ \bm{\Sigma}_i $ is an $m_i \times m_i$ positive definite matrix
depending on $ \underline{\bm{X}}_i$, $\underline{\bm{Z}}_i$,
and $\underline{T}_i$, $i=1,\ldots,n$. This is a standard marginal model 
in longitudinal data analysis \cite{WZ2006}. 

Model (\ref{eqn:e100}) consists of a parametric component, which provides information on the constant impacts 
of some important covariates, and a nonparametric component which captures the dynamic impacts  
of the other covariates. In this way the model is able to reflect unknown nonlinear structures in the data while retaining similar interpretability as the classical linear models at the same time. There is an extensive literature on the variable selection, structure identification, estimation, and inference issues \cite{CHLP2014, FL2004, HWZ2004,WQ2009,XZT2004}. 
In particular, often of primary interest is to have access to the parametric component while the nonparametric component is viewed as  
the nuisance part. In this regard, it is well known that assuming independence or some mis-specified working covariance structure yields less efficient estimation of the constant coefficients. 
Therefore, a substantial portion of the existing literature aimed at improving the efficiency via modeling and  estimating the within-subject  covariance structure \cite{CHLP2014, FHL2007, FW2008, QL2006, YL2013, ZFS2009,ZQ2012}, which is itself a challenging task.

In this article, we focus on the identity link function and make contributions to the efficient estimation problem for model (\ref{eqn:e100})
in three directions. First, we allow some of the $m_i$'s to tend to infinity. As far as we know, this setup has not been treated before and the problem is nontrivial. Our results also hold when the $m_i$'s are uniformly bounded  
and $\underline{\epsilon}_i$ satisfies the sub-Gaussian property. See the supplement \cite{HCL2015S} for the details.  When all of the $m_i$'s are diverging, that is, if we have densely observed data, 
it becomes a kind of functional data problem and is
out of the scope of this paper. 
Second, we study explicit expression of the semiparametric efficiency bound
for estimation of $\bm{\beta}$
and asymptotic normality of the generalized estimating equations (GEE) spline estimator
under general covariance structures and error distributions.
Using the true covariance matrices in the GEE estimation leads to optimality among all GEE estimators of the parametric component. Furthermore, it achieves the semiparametric efficiency bound when the errors are conditionally normal.
Our results are in parallel to that for partially linear and partially linear additive models given 
by \cite{HZZ2007} and \cite{CZH2014} respectively. 
Those models are among a rich variety of semiparametric ways of modeling longitudinal
data, and they differ from semivarying coefficient models in that their
nonparametric components admit more direct additive expressions.
Partially linear (additive) models were also considered by 
\cite{Li2011,LC2006, LWWC2004, Ma2012,WCL2005}, among which \cite{Li2011,LC2006,LWWC2004,WCL2005} used kernel method and 
\cite{Ma2012} used spline estimation.

Our third contribution is to deal with adaptive efficient estimation when the within-subject covariance matrices 
are estimated nonparametrically using the data at hand. Notice that \cite{CZH2014} ignored this practical issue and  did not consider estimation of the covariances, and \cite{HZZ2007} suggested using some 
parametric specification which can be estimated $\sqrt{n}$-consistently. 
We consider the case where the nonparametric within-subject covariance matrices depend only on the observation times but not on the other covariates.
Such assumptions are reasonable because we do not assume that the observation
times are regular across different subjects or they are dense.
Indeed, with irregular and/or sparse observation times, estimating the covariances in a completely
nonparametric way, by letting them to be dependent on all of the $T_{ij}$, $\bm{X}_{ij}$ and $\bm{Z}_{ij}$ nonparametrically, is particularly problematic and even unreliable as the curse-of-dimensionality problem arises.
Our covariance estimator is constructed based on residuals yielded by an initial estimation. 
The final estimator of the true value of $\bm{\beta}$ is then 
given by plugging-in the covariance estimates to the GEE spline estimation. 
We show the asymptotic equivalence of our final estimator to the oracle efficient estimator which uses the true covariance matrices in the GEE spline estimation.

The above result is partly motivated by the study of \cite{Li2011} on efficient estimation
in partially linear models under the same nonparametric covariance structure. However, the kernel profile method taken by \cite{Li2011} involves only local linear regression,  
thus, to achieve semiparametric efficiency it  requires some complicated iterative backfitting calculation except for 
the identity link function \cite{LC2006, LWWC2004}.
By comparison, our approach to estimating the parametric and nonparametric components in the mean function is different and much simpler. We ingeniously use both spline approximation and local linear estimation to avoid complicated calculation while allowing for the asymptotic equivalence property at the same time. To the best of our knowledge, there are no existing results for semivarying coefficient models,
especially when some of the $m_i$'s tend to infinity or when the $\bm{\Sigma}_i$Õs are estimated.

Our final estimator is some kind of feasible generalized least squares (FGLS) estimator since we
replace the within-subject covariance matrices with their nonparametric estimates. Even if our assumption on the covariance matrices fails to hold, it still possesses the asymptotic normality under mild conditions and
still makes use of some information of the covariance matrices. For example, if the
covariances depend on some time-dependent covariates, to some extent such effects
are still captured by our method. In this sense, compared with existing methods which use either parametrically
estimated or some ad-hoc covariance matrices  \cite{FHL2007,QL2006,TXL2014}, our approach is more adaptive
to the unknown covariance matrices. A promising cluster bootstrap inference method was proposed by \cite{CYH2013}; it assumes some  parametric within-cluster covariance structure, however. In the case where there 
is one observation for each subject/cluster, our assumption on the covariance matrices
reduces to that of \cite{SCMW2014}, which also suggested to improve
the efficiency in a similar manner.

Our simulation study shows that numerically the proposed method
outperforms the working independence approach and the
quadratic inference functions (QIF) method by \cite{QL2006}, and
it behaves close to the oracle estimator which uses the true covariance matrices.
Note that, while the QIF procedure is suitable when there is some kind of regularity and stationarity in the error process, our procedure adapts to both non-stationarity and irregularity.
We also applied our method to the CD4 count dataset and identified some
interesting new effects not detected by the working independence approach.

After the semiparametric efficient estimation, we can estimate and make inference on the nonparametric component in the same way as in dealing with  varying coefficient models, using the difference between the response and the estimated parametric part \cite{XZT2004}.
When $p$ and $q$ are both diverging and the model is sparse, \cite{CHLP2014} suggested a simultaneous variable selection and structure identification procedure and showed its consistency property. By combining the method with the proposed estimation procedure and by putting together the corresponding consistency and efficiency results, we have an efficient estimation procedure in this case. 

The organization of this paper is as follows. In Section \ref{sec:efficiency} 
we derive the semiparametric efficiency bound for the constant coefficient
vector $\bm{\beta}$ and asymptotic normality of GEE spline estimators. In 
Section \ref{sec:estimation}, we propose an efficient estimator of $\bm{\beta}$
when the errors have some general covariance structure and 
state its asymptotic equivalence to the oracle estimator which assumes the covariance matrices are known. Section \ref{sec:numerical} summarizes and discusses results
of our simulation and empirical studies used to assess numerical performance of the
proposed efficient estimator.
Section \ref{sec:details1} contains some technical assumptions and proof of the asymptotic equivalence. In
the supplementary material \cite{HCL2015S}  we give additional simulation results for estimation, proofs of the other theoretical results, some lemmas, and theoretical results when the $m_i$'s are uniformly bounded.

\section{Semiparametric efficiency bound for $\bm{\beta}$}
\label{sec:efficiency}

In this section, $\bm{V}_i$ is a given $m_i\times m_i$ inverse weight matrix
depending only on $\underline{\bm{X}}_i$, $\underline{\bm{Z}}_i$,
and $\underline{T}_i$, $i=1,\ldots,n$. We use a $K_n$-dimensional equispaced B-spline basis on $[0,1]$, denoted by $\bm{B}(t)$, to approximate
the function $\bm{g}(t)$. See \cite{Schumaker2007} for the definition and
properties of B-spline bases. 
We set
$
\bm{W}_{ij}=\bm{Z}_{ij} \otimes \bm{B}(T_{ij})\
{\rm and}\ \underline{\bm{W}}_i=
(\bm{W}_{i1}, \ldots, \bm{W}_{im_i})^T,
$ 
where $\otimes$ is the Kronecker product, and we denote the true values of $\bm{\beta}$ and $\bm{g}(t)$ by $\bm{\beta}_0$ and $\bm{g}_0(t)=(g_{01}(t), \ldots, g_{0q}(t))^T$ respectively. Then we estimate  
$\bm{\beta}_0$ and $\bm{g}_0(t)$ by minimizing with respect
to $\bm{\beta}$ and $\bm{\gamma}$ simultaneously the following objective function: 
\begin{equation}
\sum_{i=1}^n
(\underline{Y}_i- \underline{\mu}(
\underline{\bm{X}}_i\bm{\beta} + \underline{\bm{W}}_i\bm{\gamma}
))^T\bm{V}_i^{-1}
(\underline{Y}_i- \underline{\mu}(
\underline{\bm{X}}_i\bm{\beta} + \underline{\bm{W}}_i\bm{\gamma}
)),
\label{eqn:e200}
\end{equation}
where $\bm{\gamma} \in \RR^{qK_n}$ and the $j$ th element of
$\underline{\mu}(
\underline{\bm{X}}_i\bm{\beta} + \underline{\bm{W}}_i\bm{\gamma}
)$ is $\mu (\bm{X}_{ij}^T\bm{\beta} + \bm{W}_{ij}^T\bm{\gamma} )$.
 Thus the generalized estimating equations are 
\begin{align}
\sum_{i=1}^n \underline{\bm{X}}_i^T \Delta_i \bm{V}_i^{-1}
(\underline{Y}_i- \underline{\mu}(
\underline{\bm{X}}_i\bm{\beta} + \underline{\bm{W}}_i\bm{\gamma}
))& = 0, \nonumber 
\\
 \mbox{and} \quad \sum_{i=1}^n \underline{\bm{W}}_i^T \Delta_i \bm{V}_i^{-1}
(\underline{Y}_i- \underline{\mu}(
\underline{\bm{X}}_i\bm{\beta} + \underline{\bm{W}}_i\bm{\gamma}
))& = 0,\label{eqn:e206}
\end{align}
where $\Delta_i$ is an $m_i\times m_i$ diagonal matrix defined
by
$
\Delta_i = {\rm diag}(\mu'(\bm{X}_{i1}^T\bm{\beta}+
\bm{W}_{i1}^T\bm{\gamma}), \ldots,
\mu'(\bm{X}_{im_i}^T\bm{\beta}+ \bm{W}_{im_i}^T\bm{\gamma} )).
$
Denote the solution to 
(\ref{eqn:e206})
by $\widehat{\bm{\beta}}_{\bm{V}}$ and $\widehat{\bm{\gamma}}_{\bm{V}}\equiv\big(\widehat{\gamma}_{1V}^T,\ldots,\widehat{\gamma}_{qV}^T\big)^T$. Then the GEE spline estimator with weight matrices $\bm{V}_i^{-1}$, $i=1,\ldots,n$, for $\bm{\beta}_0$ is $\widehat{\bm{\beta}}_{\bm{V}}$ and that for $\bm{g}_0(t)$ is $\big(\widehat{\gamma}_{1V}^T\bm{B}(t),\ldots,\widehat{\gamma}_{qV}^T\bm{B}(t)\big)^T$.

Hereafter we focus on the identity link function and present the asymptotic normality of $\widehat{\bm{\beta}}_{\bm{V}}$ 
in Proposition \ref{prop:prop1} under general error distributions
as specified in Assumption A6 given in Section \ref{sec:details1}. 
We allow
some of the $m_i$'s to diverge in a way like 
$\sum_{i=1}^n m_i^5 = O(n) \quad {\rm and} \quad \max_{1\le i \le n}m_i = O(n^{1/8}).
$
See Assumptions A1 and A2 
for the specific conditions on the $m_i$'s.  We refer to the supplement 
\cite{HCL2015S} for the results for general link functions when the $m_i$'s
are uniformly bounded and the $\underline{\epsilon}_i$'s satisfy the sub-Gaussian
property.

First, we introduce some function spaces, inner products and projections. 
Let $L_2$ denote the space of square integrable functions on $[0,1]$ and recall $\bm{B}(t)$ is the equispaced B-spline basis on $[0,1]$. We define two function spaces: 
\begin{align*}
\bm{G} & = \{ (g_1, \ldots, g_q)^T\, | \, g_j \in L_2, \, j=1,\ldots,q\},\\
\mbox{and} \quad \bm{G}_B & = \{ (\bm{B}^T\bm{\gamma}_1, \ldots, \bm{B}^T\bm{\gamma}_q)^T
\, | \, \bm{\gamma}= (\bm{\gamma}_1^T, \ldots, \bm{\gamma}_q^T )^T
\in \RR^{qK_n} \}\,.
\end{align*}
Note that $\bm{G}_B \subset \bm{G}$.
Next, 
let $v_1$ and $v_2$ be two stochastic processes each taking scalar values at $T_{ij}$, $i=1,\ldots,n$, $j=1,\ldots,m_i$. 
Then we define two inner products of $v_1$ and $v_2$ by 
$
\langle v_1, v_2 \rangle_n^V
=\frac{1}{n}\sum_{i=1}^n \underline{v}_{1i}^T
\bm{V}_i^{-1} \underline{v}_{2i}
$
and 
$
\langle v_1, v_2 \rangle^V =
\RE \{  \langle v_1, v_2 \rangle_n^V  \},
$
where $\underline{v}_{1i}$ and  $\underline{v}_{2i}$ are defined 
in the same way as $\underline{T}_{i}$, and we define the associated norms by
$\| v\|_n^V = ( \langle v, v \rangle_n^V  )^{1/2}
$
and $\| v\|^V = ( \langle v, v \rangle^V  )^{1/2}.
$
The projections, with respect to $\| \cdot
\|^{V}$, of  the $k$th element of $\bm{X}$ onto $\bm{Z}^T\bm{G}$ and $\bm{Z}^T\bm{G}_B$ are given by
\begin{equation}
\Pi_{\bm{V}}X_k =\argmin_{\bm{g}\in \bm{G}}
\| X_k -\bm{Z}^T\bm{g} \|^V\,\, {\rm and}\,\, 
\Pi_{\bm{V}n}X_k =\argmin_{\bm{g}\in \bm{G}_B}
\| X_k -\bm{Z}^T\bm{g} \|^V,
\label{eqn:e218}
\end{equation}
where 
$
\| X_k -\bm{Z}^T\bm{g} \|^V
= \frac{1}{n}\RE \Big\{
\sum_{i=1}^n( \underline{X}_{ik}-
\underline{(\bm{Z}^T\bm{g})}_i)^T \bm{V}_i^{-1}
( \underline{X}_{ik}-
\underline{(\bm{Z}^T\bm{g})}_i) \Big\},
$
with $\underline{X}_{ik}=(X_{i1k},\ldots, X_{im_ik})^T$ and
$\underline{(\bm{Z}^T\bm{g})}_i=(
\bm{Z}_{i1}^T\bm{g}(T_{i1}), \ldots,
\bm{Z}_{im_i}^T\bm{g}(T_{im_i}))$.
Hereafter we write
$
\bm{\varphi}_{\bm{V}k}^*=\Pi_{\bm{V}} X_k\in \bm{G} \quad{\rm and}\quad
\overline{\bm{\varphi}}_{\bm{V}k} = \Pi_{\bm{V}n} X_k\in \bm{G}_B.
$

\newpage
\smallskip
\noindent
{\bf Assumption S}
\begin{enumerate}
\item[(i)] The projections $\bm{\varphi}_{\bm{V}k}^* (t)$, $k=1,\ldots,p$, and the varying coefficient function $\bm{g}_0$
are twice continuously differentiable on $[0,1]$, and they and their
second order derivatives are uniformly bounded in $n$. 

\item[(ii)] We take $K_n=\lfloor c_Kn^{1/5} \rfloor$ for some positive constant $c_K$, where
$\lfloor x\rfloor$ is the largest integer no greater than $x$.
\end{enumerate}

Assumption S(i) is a mild and standard assumption for semiparametric models.
We consider the existence and smoothness properties of $\bm{\varphi}_{\bm{V} k}^* (t)$
in Section \ref{sec:details1}. Recall that all the covariates are assumed
to be uniformly bounded.
Since the relevant functions are assumed to be at least twice continuously
differentiable, we recommend quadratic or cubic 
spline approximation. Then the order of $K_n$ specified in Assumption S(ii) is optimal.
If the smoothness of different functions varies, we refer to \cite{CW2011} for the convergence rate
interfere phenomenon.

The following matrices are necessary in order to
present asymptotic normality of $\widehat{\bm{\beta}}_{\bm{V}}$: 
\begin{align}
\bH&= 
\begin{pmatrix}
\sum_{i=1}^n \underline{\bm{X}}_i^T\bm{V}_i^{-1}
\underline{\bm{X}}_i
& 
\sum_{i=1}^n \underline{\bm{X}}_i^T\bm{V}_i^{-1}
\underline{\bm{W}}_i
\\
\sum_{i=1}^n \underline{\bm{W}}_i^T\bm{V}_i^{-1}
\underline{\bm{X}}_i
&
\sum_{i=1}^n \underline{\bm{W}}_i^T\bm{V}_i^{-1}
\underline{\bm{W}}_i
\end{pmatrix}
=
\begin{pmatrix}
\bH_{11}& \bH_{12}\\
\bH_{21}& \bH_{22}
\end{pmatrix} 
, \label{eqn:e230}\\
\bH_{11 \cdot 2}& = \bH_{11}- \bH_{12}\bH_{22}^{-1}\bH_{21}\,, \quad
{\rm and}\quad \bH^{11}=(\bH_{11 \cdot 2})^{-1}\,.\nonumber
\end{align}
Let $\Omega_{\bm{V}n}$ be a $p\times p$ matrix whose $(k,l)$th
element is 
\begin{eqnarray*}
\lefteqn{
\langle X_k- \bm{Z}^T \bm{\varphi}_{\bm{V}k}^*,
X_l - \bm{Z}^T \bm{\varphi}_{\bm{V}l}^* \rangle^V} 
\\
& = & \frac{1}{n}
\sum_{i=1}^n
\RE \Big\{
( \underline{X}_{ik} - \underline{ (\bm{Z}^T
\bm{\varphi}_{\bm{V}k}^* )}_i )^T \bm{V}_i^{-1}
( \underline{X}_{il} - \underline{(\bm{Z}^T
\bm{\varphi}_{\bm{V}l}^* )}_i) 
\Big\} . \nonumber
\end{eqnarray*}
Note that 
$n^{-1}\bH_{11 \cdot 2}$
is an estimate of $\Omega_{\bm{V}n}$.
We assume that there exists a 
$p\times p$ positive definite matrix $\Omega_{\bm{V}}$ such that 
\begin{equation}
\lim_{n\to \infty}\Omega_{\bm{V}n}=
\Omega_{\bm{V}}.
\label{eqn:e236}
\end{equation}

Now we are ready to state the asymptotic normality of $\widehat{\bm{\beta}}_{\bm{V}}$ under general error distributions
as specified in Assumption A6 given in Section \ref{sec:details1}. Its proof is given in the supplement \cite{HCL2015S}.
We denote the normal distribution with mean $\eta$ and covariance 
$\Omega$ by $\RN(\eta, \Omega)$, and by ``$\stackrel{d}{\to}$'' we mean 
convergence in distribution. Let $\bm{I}_l$ be the $l$-dimensional identity matrix. 

\begin{prop}(Asymptotic normality of $\widehat{\bm{\beta}}_{\bm{V}}$)
\label{prop:prop1}
Under Assumption S, (\ref{eqn:e236}), and Assumptions A1-6 given in
Section \ref{sec:details1}, we have
\[
\widehat{\bm{\beta}}_{\bm{V}} = \bm{\beta}_0 + \bH^{11}
\sum_{i=1}^n (\underline{\bm{X}}_i - \underline{\bm{W}}_i
\bH_{22}^{-1}\bH_{21})^T \bm{V}_i^{-1}\underline{\epsilon}_i
+o_p\Big( \frac{1}{\sqrt{n}}\Big).
\]
We also have
\[
\Gamma_{\bm{V}}^{-1/2}
(\widehat{\bm{\beta}}_{\bm{V}}- \bm{\beta}_0)\stackrel{d}{\to}
\RN(0, \bm{I}_p),
\]
where $\Gamma_{\bm{V}}$ is given by
\begin{equation}
\bH^{11}\sum_{i=1}^n
\Big\{ ( \underline{\bm{X}}_i - \underline{\bm{W}}_i
\bH_{22}^{-1}\bH_{21})^T \bm{V}_i^{-1}\bm{\Sigma}_i
\bm{V}_i^{-1} ( \underline{\bm{X}}_i - 
\underline{\bm{W}}_i \bH_{22}^{-1}\bH_{21})\Big\}\bH^{11}.
\label{eqn:e239}
\end{equation}
\end{prop}

Under (\ref{eqn:e236}), $\widehat{\bm{\beta}}_{\bm{V}}$ 
is $\sqrt{n}$-consistent for $\bm{\beta}_0$.
We can estimate its asymptotic covariance $\Gamma_{\bm{V}}$ given in (\ref{eqn:e239}) 
by replacing the $\bm{\Sigma}_i$'s
with some estimates based on $\widehat{\bm{\beta}}_{\bm{V}}$ and
$\widehat{\bm{\gamma}}_{\bm{V}}$. For example,  we can replace
$\bm{\Sigma}_i$ with $\widetilde{\underline{\epsilon}}_i
\widetilde{\underline{\epsilon}}_i^T$
where 
$
\widetilde{\underline{\epsilon}}_i = \underline{Y}_i
-\underline{\bm{X}}_i^T\widehat{\bm{\beta}}_{\bm{V}}
-\underline{\bm{W}}_i^T\widehat{\bm{\gamma}}_{\bm{V}}.
$
However, this approach may be too crude and it does not make use of the common information on the covariance structure contained in different subjects. Alternatively, we can estimate the $\bm{\Sigma}_i$'s by applying smoothing techniques to some residuals based on some assumption on the covariance structure. We investigate this problem in Section \ref{sec:estimation}.

Next, Proposition \ref{prop:prop2} gives the semiparametric efficiency bound for
estimation of $\bm{\beta}_0$. 
It can be proved in almost the same way as in Section 4.4 of \cite{HZZ2007} and Lemma 1 of \cite{CZH2014} and the proof is omitted. 
We denote the semiparametric efficient score function of
$\bm{\beta}$ by
$
\bm{l}_{\bm{\beta}}^*= (l_{\bm{\beta}1}^*,
\ldots, l_{\bm{\beta}p}^*)^T.
$ 
Its expression is given in Proposition \ref{prop:prop2}.
Then we denote $\bm{\varphi}_{\bm{\Sigma} k}^* (t)$ by
$\bm{\varphi}_{eff,k}^* (t) $ when $\bm{V}_i=\bm{\Sigma}_i$ in (\ref{eqn:e200}).

\begin{prop}(Semiparametric efficiency bound)
\label{prop:prop2}
Under the same assumptions as in Proposition \ref{prop:prop1},
we have
\[
l_{\bm{\beta}k}^* = 
\sum_{i=1}^n ( \underline{X}_{ik} - \underline{ (\bm{Z}^T
\bm{\varphi}_{eff,k}^* )}_i )^T
 \bm{\Sigma}_i^{-1}\{ 
\underline{Y}_i -
 \underline{\bm{X}}_{i}^T\bm{\beta}_0-
\underline{(\bm{Z}^T\bm{g}_0)}_i \},
\]
and the semiparametric efficient information matrix for $\bm{\beta}$
is given by
\[
\lim_{n \to \infty}\frac{1}{n}
\RE \{ \bm{l}_{\bm{\beta}}^* (\bm{l}_{\bm{\beta}}^*)^T \}
= \Omega_{\bm{\Sigma}} \ with \ \bm{V}_i = \bm{\Sigma}_i\ in\ (\ref{eqn:e236}). 
\]
\end{prop}

Proposition \ref{prop:prop3} gives the asymptotic normality of $\widehat{\bm{\beta}}_{\bm{\Sigma}}$, 
the so called oracle estimator, which uses the true covariance structure in the GEE spline regression.
It also asserts that $\widehat{\bm{\beta}}_{\bm{\Sigma}}$ achieves the semiparametric
efficiency bound derived from Proposition \ref{prop:prop2}. The proof is given in the supplement \cite{HCL2015S}.

\begin{prop}(Oracle efficient estimator)
\label{prop:prop3}
If we take $\bm{V}_i=\bm{\Sigma}_i$ in\ (\ref{eqn:e206}) then, under the same assumptions as in Proposition \ref{prop:prop1},
we have 
\[
\sqrt{n}\,\Omega_{\bm{\Sigma}}^{1/2}
(\widehat{\bm{\beta}}_{\bm{\Sigma}}- \bm{\beta}_0)\stackrel{d}{\to}
\RN(0, \bm{I}_p).
\]
\end{prop}

In practice, usually the $\bm{\Sigma}_i$'s are unknown and  we have no direct access
to the semiparametric efficient score function or the oracle estimator.
In the next section we study nonparametric estimation of the covariances so as to improve the efficiency.

\section{Efficient estimation}
\label{sec:estimation}

The semiparametric efficiency bound of
$\bm{\beta}$ given in Proposition \ref{prop:prop2} indicates that knowledge,
 or at least estimation, of the $\bm{\Sigma}_i$'s is necessary
in order to construct a semiparametric efficient estimator.
On the other hand, as discussed in the Introduction, when the $\bm{\Sigma}_i$'s are unknown it is almost impossible to estimate them in a fully nonparametric way. Fortunately, for longitudinal or clustered data sets,
it is reasonable to make some assumptions  
such as
\begin{equation}
\bm{\Sigma}_i =\bm{\Sigma} ( \underline{T}_i ), \,\, i=1,\ldots,n,
\label{eqn:e300}
\end{equation}
where the $(j,j)$th element of $\bm{\Sigma}_i$ is given by $\sigma^2 (T_{ij})$ and the $(j,j')$th
element is given by $\sigma (T_{ij}, T_{ij'})$ when $j\ne j'$, for some 
smooth functions $\sigma^2(t)$ and $\sigma (s,t)$. Based on (\ref{eqn:e300}), in Section \ref{method} we construct nonparametric estimates of the covariances and then use them to derive an FGLS procedure to improve the efficiency, and we show in Section \ref{estimation:asym} its asymptotic equivalence to the oracle estimator $\widehat{\bm{\beta}}_{\bm{\Sigma}}$. We also discuss estimation of the nonparametric component.

\subsection{Methodology}\label{method}

A preliminary estimation of $\bm{\beta}_0$ and $\bm{g}_0$ is necessary before we can estimate the covariances.
For simplicity and robustness, we utilize working independence in the GEE spline estimation.  
As noted following Proposition \ref{prop:prop1} we could then use the resultant residuals to estimate the covariance matrices directly. However  it is intuitively better to further make use of the covariance structure (\ref{eqn:e300}) by applying some nonparametric smoothing techniques to the residuals. In addition, alternative to the spline estimator, we could apply smoothing techniques to the pseudo responses $ \underline{Y}_{i}-\underline{\bm{X}}_{i}^T\widehat{\bm{\beta}}_{\bm{V}}$ to obtain another estimator of $\bm{g}_0$. We take this latter approach for technical and numerical reasons given in Remark \ref{rmk:rmk5}.
After the preliminary estimation, for each $i=1,\ldots,n$, we estimate $\bm{\Sigma}_i$ by applying local linear regression
and denote the resultant estimate by $\widehat{\bm{\Sigma}}_i$. Our final estimator of $\bm{\beta}_0$ 
is then obtained by taking $\bm{V}_i= \widehat{{\bm{\Sigma}}}_i$, $i=1,\ldots,n$, in the GEE spline estimation. Note that in the trivial case where $m_i$ is fixed for all $i$ and the $T_{ij}$'s are equi-spaced, we can estimate $\bm{\Sigma}_i$ without using any smoothing techniques. 

Let $K$ be a given kernel function. Our estimation procedure is formally specified as follows:
\begin{description}

\item[\bf Step 1.] 
Estimate $\bm{\beta}_0$ by the GEE spline method given in Section \ref{sec:efficiency}
with $\bm{V}_i=\bm{I}_{m_i}$, $i=1,\ldots,n$, and denote the resultant working independence estimate by $\widehat{\bm{\beta}}_I$.

\item[\bf Step 2.] 
Estimate $\bm{g}_0 (t)$ by applying local linear regression to $\big\{Y_{ij}-\bm{X}_{ij}^T\widehat{\bm{\beta}}_I,\,
i=1,\ldots,n, j=1,\ldots,m_i\big\}$, 
using bandwidth $h_1$. We denote the resultant estimate by
$\widehat{\bm{g}} (t)$, which is written as
\begin{equation}
\widehat{\bm{g}} (t) = D_q(A_{1n}(t))^{-1}
\frac{1}{N_1h_1}
\sum_{i=1}^n \sum_{j=1}^{m_i}
\bm{Z}_{ij}\otimes
\begin{pmatrix}
1\\
\frac{T_{ij}-t}{h_1}
\end{pmatrix}
K\Big( \frac{T_{ij}-t}{h_1} \Big)
( Y_{ij} -  \bm{X}_{ij}^T \widehat{\bm{\beta}}_I ),
\label{eqn:e303}
\end{equation}
where $N_1= \sum_{i=1}^n m_i$,
$D_q=\bm{I}_q \otimes (1\ \ 0)$, 
and
\begin{align*}
A_{1n}(t) & = \frac{1}{N_1h_1}
\sum_{i=1}^n \sum_{j=1}^{m_i}
( \bm{Z}_{ij} \bm{Z}_{ij}^T) \otimes
\begin{pmatrix}
1 & \frac{T_{ij}-t}{h_1}\\
\frac{T_{ij}-t}{h_1} & (\frac{T_{ij}-t}{h_1})^2
\end{pmatrix}
K\Big( \frac{T_{ij}-t}{h_1} \Big).
\end{align*}

\item[\bf Step 3.] 
Calculate the residuals, denoted as $\widehat{\epsilon}_{ij}$, given by
\[
\widehat{\epsilon}_{ij}= Y_{ij}- \bm{X}_{ij}^T\widehat{\bm{\beta}}_I
-\bm{Z}_{ij}^T\widehat{\bm{g}}(T_{ij}),\, i=1,\ldots,n, j=1,\ldots,m_i.
\]

\item[\bf Step 4.] 
Estimate the variance function $\sigma^2 (t)$ by applying to the squared residuals local linear regression 
with bandwidth $h_2$. Denote the resultant estimate by $\widehat{\sigma^2} (t)$; it 
can be expressed as
\begin{equation}
\widehat{\sigma^2} (t)
=(1\, 0)(A_{2n}(t))^{-1}
\frac{1}{N_1h_2}
\sum_{i=1}^n \sum_{ j=1}^{m_i}
\begin{pmatrix}
1\\
\frac{T_{ij}-t}{h_2}
\end{pmatrix}
K\big( \frac{T_{ij}-t}{h_2} \big)
(\widehat \epsilon_{ij})^2,
\label{eqn:e306}
\end{equation}
where
$
A_{2n} (t) =\frac{1}{N_1h_2}
\sum_{i=1}^n\sum_{j=1}^{m_i}
\begin{pmatrix}
1& \frac{T_{ij}-t}{h_2}\\
\frac{T_{ij}-t}{h_2} & (\frac{T_{ij}-t}{h_2})^2
\end{pmatrix}
K\big( \frac{T_{ij}-t}{h_2} \big).
$

\item[\bf Step 5.] 
Estimate the covariance function $\sigma (s,t)$ by applying to $\big\{\widehat \epsilon_{ij}\widehat \epsilon_{ij'}, j\neq j', i=1,\ldots,n\big\}$  local linear regression 
with bandwidth $h_3$. We denote the resultant estimate by $\widehat{\sigma}(s,t)$; it has the following expression:
\begin{align}
\widehat \sigma (s,t)
 & =(1\,0\,0)(A_{3n}(s,t))^{-1}
\label{eqn:e309}\\
 & \times \frac{1}{N_2h_3^2}
\sum_{i=1}^n \sum_{ j\ne j'}
\begin{pmatrix}
1\\
\frac{T_{ij}-s}{h_3}\\
\frac{T_{ij'}-t}{h_3}
\end{pmatrix}
K\big( \frac{T_{ij}-s}{h_3} \big)
K\big( \frac{T_{ij'}-t}{h_3} \big)
\widehat \epsilon_{ij}\widehat \epsilon_{ij'},
\nonumber
\end{align}
where $N_2= \sum_{i=1}^nm_i(m_i-1)$ and
\begin{eqnarray*}
\lefteqn{A_{3n}(s,t)}\\
& = & \frac{1}{N_2h_3^2}
\sum_i^n\sum_{ j\ne j'}
\begin{pmatrix}
1\\
\frac{T_{ij}-s}{h_3}\\
\frac{T_{ij'}-t}{h_3}
\end{pmatrix}
\begin{pmatrix}
1 & \frac{T_{ij}-s}{h_3} & \frac{T_{ij'}-t}{h_3}
\end{pmatrix}
K\big( \frac{T_{ij}-s}{h_3} \big)
K\big( \frac{T_{ij'}-t}{h_3} \big).
\end{eqnarray*}

\item[\bf Step 6.] 
Calculate $\widehat{\bm{\Sigma}}_i$ by combining the results from steps 4 and 5 by letting
\[
\widehat{\bm{\Sigma}}_i (j,j') =  \widehat\sigma(T_{ij},T_{ij'})I(j\neq j') + \widehat{\sigma^2}(T_{ij}) I(j=j') ,
\]
and then estimate 
$\bm{\beta}_0$ with $\bm{V}_i = \widehat{\bm{\Sigma}}_i$ in the GEE (\ref{eqn:e206}). Denote the resultant estimate of
$\bm{\beta}_0$ by $\widehat{\bm{\beta}}_{\widehat{\bm{\Sigma}}}$.   

\item[\bf Step 7.] 
Update the nonparametric estimator of $\bm{g}_0(t)$ given in Step 2 by replacing $Y_{ij}-\bm{X}_{ij}^T\widehat{\bm{\beta}}_I$ with $Y_{ij}-\bm{X}_{ij}^T\widehat{\bm{\beta}}_{\widehat{\bm{\Sigma}}}$, $i=1,\ldots,n, j=1,\ldots,m_i$. Denote the resultant estimator by $\widehat{\bm{g}}_{U}(t)$. Alternatively, we 
can estimate $\bm{g}_0(t)$ with splines, by replacing  $\bm{\beta}$  with $\widehat{\bm{\beta}}_{\widehat{\bm{\Sigma}}}$ and taking $\bm{V}_i = \widehat{\bm{\Sigma}}_i$ in the GEE (\ref{eqn:e206}). Denote the resultant estimator by $\widehat{\bm{g}}_{S}(t)$. 

\end{description}

In general the covariance function estimate $\widehat\sigma(s,t)$ given by step 5 may not be positive semidefinite.
We can modify it by truncating the
eigenfunctions in its spectral decomposition that have eigenvalues not exceeding some
nonnegative constant $\lambda_L$. Then we have positive definite covariance estimates if we replace $\widehat\sigma(s,t)$ with this modified version in step 6. 

\begin{rmk}
\label{rmk:rmk5}
When we calculate $\widehat{\bm{\beta}}_I$ in step 1,
we also have $\widehat{\bm{\gamma}}_I$ and get the set of residuals
$\{\widetilde{\epsilon}_{ij}=Y_{ij}-\bm{X}_{ij}^T \widehat{\bm{\beta}}_I-\bm{W}_{ij}^T
\widehat{\bm{\gamma}}_I\}$.
Then we could omit steps 2 and 3 of our 
procedure by exploiting this set of residuals when we estimate $\bm{\Sigma}_i$ in steps 4-6.
However, our simulation results summarized in Section \ref{sec:numerical}
indicate that this simplified approach is inferior to the proposed one. Intuitively speaking, to achieve the semiparametric efficiency in the GEE spline 
estimation of $\bm{\beta}_0$, to some extent the accompanying estimation of $\bm{g}_0(t)$ requires undersmoothing and thus it often 
exhibits spurious wiggling patterns.
Besides, it is difficult to  justify theoretically this simplified approach as the local property of spline estimators seems to be intractable.
\end{rmk}

\subsection{Asymptotic results}\label{estimation:asym}

First we establish the asymptotic equivalence between the data-driven estimator 
$\widehat{\bm{\beta}}_{\widehat{\bm{\Sigma}}}$ and the oracle estimator 
$\widehat{\bm{\beta}}_{\bm{\Sigma}}$ 
by exploiting some desirable properties of $\widehat{\bm{\Sigma}}_i$.
First, we specify our assumptions on the smoothness of $\bm{g}_0(t)$, 
$\sigma^2(t)$ and $\sigma (s,t)$. 
We need Assumption B given below, which is more restrictive than usual, in order to
evaluate the difference between $\widehat{\bm{\Sigma}}_i^{-1}$ and
$\bm{\Sigma}_i^{-1}$. 

\smallskip
\noindent
{\bf Assumption B}.
\begin{enumerate}
\item[(i)] Assumption (\ref{eqn:e300}) holds.

\item[(ii)] The true varying coefficient function $\bm{g}_0(t)$ is three times  continuously differentiable on $[0,1]$.

\item[(iii)] The variance function $\sigma^2(t)$ is three times continuously differentiable on $[0,1]$.

\item[(iv)] The covariance function $\sigma(s,t)$ is three times continuously differentiable on $[0,1]^2$.
\end{enumerate}

In the following  we collect our assumptions
on the kernel function $K$ and the three bandwidths used in the construction of the proposed estimator. Assumption H(i) on $K$ is a standard one. When Assumption B holds, our assumptions on
the bandwidths $h_1,\ h_2$ and $h_3$ are not restrictive. For example, the optimal order of $h_1$ and $h_2$ is $n^{-1/5}$ which falls into the specified range. A larger order 
is recommended only for $h_3$ due to the two-dimensional
smoothing in step 5. However, since the effective number of observations used  in step 5 
of the procedure is $N_2$ 
we anticipate that bandwidth choice will not seriously
affect the performance of our final estimator.

\smallskip
\noindent
{\bf Assumption H}.
\begin{enumerate}
\item[(i)] The kernel function $K$ is some continuously differentiable
symmetric density function with a compact support.

\item[(ii)] The bandwidths  $h_1$, $h_2$ and $h_3$ satisfy $h_1=c_1 n^{-a_h}$ for some $1/6< a_h\le 1/4$,
$h_2=c_2 n^{-b_h}$ for some $1/6< b_h\le 1/4$ and $h_3=c_3 n^{-c_h}$ for some $1/6 <c_h< 1/4$, where $c_1$, $c_2$ and
$c_3$ are some positive constants.
\end{enumerate}

The asymptotic expression of $\widehat{\bm{\Sigma}}_i$ is given in
Proposition \ref{prop:prop4}, which is verified 
in the supplementary material \cite{HCL2015S}. 
Note that we need more elaborate representations than those 
used by \cite{Li2011} since we deal with a $(p+qK_n)$-dimensional
linear regression model. Note also that the functions $B_j$, $j=1,\ldots,4$,
that appear in Proposition \ref{prop:prop4} are implicitly defined in the proof of the proposition 
and only their boundedness property is needed in the proof of
Theorem \ref{thm:thm1}.

\begin{prop}(Representations of the covariance estimators)
\label{prop:prop4}
Under the assumptions in Proposition \ref{prop:prop1}
with $\bm{V}_i=\bm{I}_{m_i}$, 
and Assumptions B and H,
we have the following representations of $\widehat{\sigma^2} (t)$
and $\widehat \sigma (s,t)$. Uniformly in $t$,
\begin{align*}
\widehat{\sigma^2} (t)- \sigma^2 (t)
&= B_1(t)h_2^2+ B_2(t)E_{1}(t)+
O_p(h_1^{3}+h_2^{3})  + O_p \Big( \frac{\log n}{nh_1}+\frac{\log n}{nh_2} \Big)
\end{align*}
where uniformly\ in\ t\, 
\begin{align*}
E_1(t) & = \frac{1}{N_1h_2}\sum_{i=1}^n\sum_{j=1}^{m_i}
\begin{pmatrix}
1\\
\frac{T_{ij}-t}{h_2}
\end{pmatrix}
K \Big( \frac{T_{ij}-t}{h_2} \Big) (\epsilon_{ij}^2 - 
\sigma^2( T_{ij} ) ) = O_p \Big(\sqrt{ \frac{\log n}{nh_2} } \Big), 
\end{align*}
and $B_1(t)$ and $B_2(t)$ are bounded functions. Uniformly in $s$ and $t$ $(s \ne t)$,
\begin{align*}
\widehat{\sigma} (s, t)- \sigma (s, t)&
= B_3(s,t)h_2^2+ B_4(s, t)
E_{2}(s, t)+O_p(h_1^{3}+h_3^{3})
+ O_p \Big( \frac{\log n}{nh_1} + \frac{\log n}{nh_3^2}\Big),
\end{align*}
where
\begin{align*}
E_2(s,t) & = \frac{1}{N_2h_3^2}\sum_{i=1}^n\sum_{j\ne j'}
\begin{pmatrix}
1\\
\frac{T_{ij}-s}{h_3}\\
\frac{T_{ij'}-t}{h_3}
\end{pmatrix}
K \Big( \frac{T_{ij}-s}{h_3} \Big) 
K \Big( \frac{T_{ij'}-t}{h_3} \Big) (\epsilon_{ij}
\epsilon_{ij'} - \sigma ( T_{ij}, T_{ij'} ) )\\
& = O_p \Big(\sqrt{ \frac{\log n}{nh_3^2} }
\Big) \, \quad uniformly\ in\ s\ and\ t,
\end{align*}
and $B_3(s,t)$ and $B_4(s,t)$ are bounded functions.
\end{prop}

We state in Theorem \ref{thm:thm1} the desirable equivalence property of
$\widehat{\bm{\beta}}_{\widehat{\bm{\Sigma}}}$ to the oracle estimator. The proof uses Proposition \ref{prop:prop4}; it is tedious and 
technical and thus is postponed to Section \ref{sec:details2}. We have not yet obtained a similar result 
for general link functions even when the $m_i$'s are uniformly bounded, and that  
is a future research topic.

\begin{thm}
\label{thm:thm1}
Under the assumptions in Proposition \ref{prop:prop4}, we have
\[
\widehat{\bm{\beta}}_{\widehat{\bm{\Sigma}}}
= \widehat{\bm{\beta}}_{\bm{\Sigma}} + o_p(n^{-1/2}).
\]
\end{thm}

Suppose (\ref{eqn:e300}) fails to hold, but $\var( \underline{\epsilon}_i \,|\, \underline{T}_i)$ still 
can be represented by some functions $\sigma^2(t)$ and $\sigma (s,t)$. Then 
Proposition \ref{prop:prop1} and Theorem \ref{thm:thm1} continue to hold 
$\bm{\Sigma_i}=\var( \underline{\epsilon}_i \,|\, \underline{\bm{X}}_i, \underline{\bm{Z}}_i, \underline{T}_i)$ is replaced by
$\var( \underline{\epsilon}_i \,|\, \underline{T}_i)$.
We are still exploiting the information on $\var( \underline{\epsilon}_i \,|\, \underline{T}_i)$. 

Besides, we can replace the three times continuously differentiability with
the twice continuously differentiability and the H\"older continuity
of the second derivatives of order $\alpha_1$, $\alpha_2$, and
$\alpha_3$ in assumptions B(ii), B(iii), and B(iv), respectively. In this case, the bandwidths
in steps 2, 4, and 5 of our method 
have to satisfy the condition
$
\sqrt{n}(h_1^{2+\alpha_1} + h_2^{2+\alpha_2} + h_3^{2+\alpha_3})
\to 0.
$
Note
that $\alpha_3$ must be positive because step 5
of our procedure requires two-dimensional smoothing.
Then we can prove similar results when $0\le \alpha_1<1$,
$0\le \alpha_2<1$, and $0< \alpha_3<1$. 
Specifically, the $O_p(h_j^3)$ terms in
Proposition \ref{prop:prop4}
will be replaced by $O_p(h_j^{2+\alpha_j})$, $j=1,2,3$.

\begin{rmk}
\label{rmk:rmk1}
In Proposition \ref{prop:prop2}, no assumptions on the structure of the 
$\bm{\Sigma}_i$'s or the conditional normality of the $\underline{\epsilon}_i$'s
is imposed. However, as mentioned before it is difficult to estimate the 
$\bm{\Sigma}_i$'s in a fully nonparametric way 
and thus we impose assumption (\ref{eqn:e300}). 
On the other hand, when (\ref{eqn:e300}) holds, we should use this information in calculating
the semiparametric efficient score function. Unfortunately, under general errors this task seems
intractable and we have no results in this regard. Nevertheless, when (\ref{eqn:e300}) and some regularity conditions hold, we 
come up with some remedies to improve the efficiency,  
as compared to using 
some working covariance structure. 
Indeed, $ \widehat{\bm{\beta}}_{\widehat{\bm{\Sigma}}} $ 
has the smallest asymptotic variance among all $ \widehat{\bm{\beta}}_{\bm{V}} $ in this case, 
based on Propositions \ref{prop:prop1}-\ref{prop:prop3}, Theorem \ref{thm:thm1}, 
and the fact that it is an FGLS estimator.
Furthermore, it is semiparametric efficient when $\underline{\epsilon}_i$ is
normally distributed conditionally on $\underline{\bm{X}}_i$,
$\underline{\bm{Z}}_i$ and $\underline{T}_i$, as discussed in A.1 of \cite{WCL2005}.
\end{rmk}

Suppose we use cubic splines in the final spline estimator given in Step 7. Then, under the assumptions in Proposition \ref{prop:prop4} and assume the minimum eigenvalue of
 $\bH_{22.1}=\bH_{22}-\bH_{21}\bH_{11}^{-1}\bH_{21}$ is bounded below by $Cn/K_n$ for some positive constant $C$, we can show the following asymptotic 
normality:
\[
\sqrt{n/K_n}\bm{\Psi}(t)^{-1/2} \big(\widehat{\bm{g}}_{S}(t)- \bm{g}_0(t)  \big) 
\stackrel{d}{\to}
\RN\big(0, \bm{I}_q \big),
\]
where $\bm{\Psi}(t)=\lim_{n\rightarrow\infty}nK_n^{-1}(\bm{I}_q\otimes\bm{B}(t)^T)\bH_{22.1}^{-1}(\bm{I}_q\otimes\bm{B}(t))$. As for the updated local linear estimator given in Step 7, let $\mu_2= \int u^2K(u)du$ and $\nu_0=\int K(u)^2du$, and suppose the assumptions in Proposition \ref{prop:prop4} hold and $h_1=Cn^{-1/5}$, then we have the following asymptotic 
normality:
\[
\sqrt{N_1h_1}\big(\widehat{\bm{g}}_{U}(t)- \bm{g}_0(t) -\frac{h_1^2}{2} \mu_2 \bm{g}_0''(t) \big) 
\stackrel{d}{\to}
\RN\big(0, \nu_0\bm{\Psi}_U(t)\big)
\]
where $\bm{\Psi}_U(t)=\bm{\Lambda}_1^{-1}\bm{\Lambda}_2\bm{\Lambda}_1^{-1}$, 
$\displaystyle{\bm{\Lambda}_1= \lim_{n\rightarrow\infty}\frac{1}{N_1}\sum_{i=1}^n\sum_{j=1}^{m_i}\RE(\bm{Z}_{ij}\bm{Z}_{ij}^T|T_{ij}=t)f_{ij}(t)}$, 
$\displaystyle{\bm{\Lambda}_2= \lim_{n\rightarrow\infty}\frac{1}{N_1}\sum_{i=1}^n\sum_{j=1}^{m_i}\RE(\bm{Z}_{ij}\bm{Z}_{ij}^T|T_{ij}=t)f_{ij}(t)}
\RE(\epsilon_{ij}^2|T_{ij}=t)$, and $f_{ij}(t)$ denotes the density of $T_{ij}$.

\section{Numerical studies}
\label{sec:numerical}

\subsection{Simulation study}
In our simulation study summarized in this section, the data were generated from the following model:
$$Y_{ij}={\bm X}_{ij}^T{\bm\beta}_0+{\bm Z}_{ij}^T{\bm g}_0(T_{ij})+\epsilon_{i}(T_{ij}),\,\,j=1,\ldots,m_i,\, i=1,\ldots,n,$$
with the first component of ${\bm Z}_{ij}$ being taken as 1.
The number of observation time points in the $i$th subject was set as $m_i=m_0+binomial(m_r, 0.65)$. Then the observation time points $T_{ij}$ were uniformly distributed over the interval $[(j-1)/(m_0+m_r),j/(m_0+m_r)]$, $j=1,\cdots,m_i$. We note that when $m_i=m_0+m_r$, the subject is observed at all follow-up time points; when $m_i<m_0+m_r$, the subject may be lost to follow up. This setup is intended to model real and more complicated scenarios that often happen in practice. We set $m_0=6$ and $m_r=6$.
We generated the other $(p+q-1)-$dimensional covariates from a multivariate Gaussian distribution, and we considered the following coefficients settings:
\begin{itemize}
\item[] 
$p=4$, $q=4$, ${\bm\beta}_0=(5,5,-5,-5)^T$ and \\ ${\bm g}_0(t)=\big(3.5\sin(2\pi t),5(1-t)^2, 3.5(\exp(-(3t-1)^2)+\exp(-(4t-3)^2))-1.5, 3.5t^{1/2}\big)^T$.

\end{itemize}
The random error process $\epsilon_{i}(t)$ was simulated from an ARMA($1,1$) Gaussian process with mean zero and covariance function $\mbox{cov}(\epsilon_{i}(s),\epsilon_{i}(t))= {\omega} {\sf \rho}^{|s-t|}$. We set ${\omega}=4.95$ and considered $\rho=0.4$ or $0.8$.

We considered two types of working covariance structure: working independence covariances and the proposed covariance estimates.  For the sake of comparison, we also considered using the true covariances and using the covariance estimator with the crude raw residuals obtained from Step 1.

Throughout the numerical studies, following \cite{FMD2014}, we used cubic splines and took the spline dimension $K_n$ as $K_n=\lfloor 2n^{1/5}\rfloor$. 
For the efficient estimator, $h_1$ and $h_2$ were selected via the commonly used leave-one-subject-out cross-validation, and the bandwidth $h_3$ was set as $h_3=2 h_1$. 
We report  in Table \ref{table-1} the average estimation bias and estimated standard error (SE) obtained from 200 repetitions. The empirical standard errors are very close to the estimated standard errors and thus are omitted. In general, the efficient estimator could yield smaller estimation bias and variance, compared to the naive estimator assuming working independence. In particular, the standard error for the efficient estimator is only $20\sim50\%$ of that of the working independence estimator, indicating a remarkable reduction. In addition, we note that the efficient estimator has very similar performance to that of the oracle estimator.
Regarding the crude estimator, as it is based on a simplified residual construction it produces relatively less accurate covariance estimation. Thus, its estimation bias and standard error are respectively larger than that for the efficient estimator. 

\begin{table}[htbp]
\begin{center}
\caption{Estimation results of 200 simulations. ``Independent'' corresponds to $\bm{V}_i=\bm{I}_{m_i}$; ``Efficient'' refers to using $\bm{V}_i=\widehat{\bm{\Sigma}}_i$; ``Oracle'' refers to using the true $\Sigma_i$ as $V_i$; ``Crude'' refers to using residuals directly from Step 1 to estimate the covariances. } \label{table-1}
{
\tiny
\begin{tabular}{rrrllllllllll}
\hline
&&&\multicolumn{2}{c}{Independent}&\multicolumn{2}{c}{Efficient}&\multicolumn{2}{c}{Oracle}&\multicolumn{2}{c}{Crude}&\multicolumn{2}{c}{Quadratic}\\
$n$& $\rho$ &  & bias & SE & bias & SE & bias & SE  & bias & SE  & bias & SE\\
\hline
100& 0.4& $\beta_{1}$ & .0214& .0726& .0128& .0366& .0133 & .0245&.0165&.0425  & .0154& .0421\\
     &&$\beta_{2}$ &-.0218 & .0727 & -.0186 & .0362 & -.0146& .0251&-.0165&.0442 & .0102& .0425 \\
     &&$\beta_{3}$ &-.0309& .0718& -.0126 & .0364& -.0147 & .0245&-.0127&.0435 & .0095& .0455\\
     &&$\beta_{4}$ &.0199& .0736& .0145& .0369 & .0132& .0246&.0210& .0438 & -.0113 & .0398\\
200& 0.4& $\beta_{1}$ &-.0072&.0525&-.0082&.0247& -.0028& .0176& -.0122&.0337 & .0049 & .0302\\
     &&$\beta_{2}$ &.0088& .0528& .0136& .0226& .0034& .0174&.0115& .0356 & .0089 & .0345\\
&&$\beta_{3}$ &-.0071& .0526& .0075& .0256& .0112& .0174&-.0146& .0354 & -.0076 & .0312 \\
&&$\beta_{4}$ &.0094& .0525& .0124& .0272& .0132& .0178&-.0204& .0355  & -.0075 & .0305 \\
100& 0.8& $\beta_{1}$ & .0257& .0723& .0245& .0334& -.0070 & .0109&.0347&.033 & .0112 & .0378\\
     &&$\beta_{2}$ &-.0179 & .0731 & -.0122 & .0328 & -.0112& .0106&.0436&.0332 & -.0109 & .0344 \\
     &&$\beta_{3}$ &.0388 & .0729& -.0257 & .0335& .0214 & .0107&.0279&.0332 & -.0179 & .0394 \\
     &&$\beta_{4}$ &-.0193& .0735& .0447& .0334 & -.0122& .0108&-.0345& .0326 & .0184& .0404\\
200& 0.8& $\beta_{1}$ &.0173&.0497&.0149&.0194& .0057& .0089& .0144&.0248 & .0089 & .0250\\
     &&$\beta_{2}$ &.0169& .0512& -.0146& .0196& -.0010& .0092&-.0167& .0242 & -.0064 & .0248\\
&&$\beta_{3}$ &-.0364& .0499& .0145& .0190& .0058& .0090&.0135& .0232 & -.0053 & .0212 \\
&&$\beta_{4}$ &.0289& .0496& -.0139& .0182& -.0035& .0089& -.0222&.0238 & .0083 & .0196 \\
\hline
\hline
\end{tabular}
}
\end{center}
\end{table}

There are also other existing methods based on estimating equations. We specifically considered the one based on quadratic inference function (QIF) \cite{QL2006} in which, to incorporate the longitudinal dependence, the correlation matrix is approximated using a matrix expansion. We used the same basis matrices as recommended by \cite{QL2006}, i.e., the first order basis matrix with 0 on the diagonal and 1 off-diagonal, which is suitable for unequal cluster sizes and irregular time points. Any negative eigenvalue was set to zero whenever it occurred.  From Table \ref{table-1}, we notice that this approach is more efficient than the estimator assuming working independence but is less efficient than our proposed method. The QIF approach indirectly models the correlations using some matrix approximation while our method directly models the covariances. The actual covariance dependence may differ from the pattern suggested by the basis matrices in the quadratic inference function. When that happens the estimation results using QIF method may be less satisfactory than our nonparametric approach. Therefore our method may incorporate a more accurate covariance structure in the estimation and thus achieve better efficiency. Besides, the covariance of the estimating function depends on the unknown parameters, and is estimated and integrated in the QIF. This may decrease the stability in solving the optimization problem.

We next considered the situation where $m_i$ might diverge for some subjects $i$. We randomly selected $n_0={\sf C}n^{3/8}$ subjects such that their observation points are ${\sf B}n^{1/8} m_i$ equally spaced on $[0,1]$  and we let  the ramaining $n-n_0$ subjects to have $m_i$ observations, where $m_i$ was generated in the same way as described above. All the other model settings are identical to that in the previous simulation studies. For different values of {\sf B} and {\sf C}, we obtained the results given in Table \ref{table-2}. We notice that all the considered estimators improve with relatively smaller biases and smaller standard errors as compared with the respective bounded $m_i$ case. The efficient estimator still performs much better than the independent estimator in all cases.
We do not report results for the QIF method by \cite{QL2006} here, as it is not tailored for the case of diverging $m_i$ and becomes relatively unstable in this case.

\begin{table}[htbp]
\begin{center}
\caption{Estimation results of 200 simulations. ``Independent'' corresponds to $\bm{V}_i=\bm{I}_{m_i}$; ``Efficient'' refers to using $\bm{V}_i=\widehat{\bm{\Sigma}}_i$; ``Oracle'' refers to using the true $\Sigma_i$ as $V_i$. {\sf B} adjusts the diverging $m_i$ and {\sf C} controls the proportion of cases with diverging $m_i$. } \label{table-2}
\scriptsize
\begin{tabular}{rrrllllll}
\hline
\multicolumn{2}{c}{${\sf B}=1.5, {\sf C}=4$}&&\multicolumn{2}{c}{Independent}&\multicolumn{2}{c}{Efficient}&\multicolumn{2}{c}{Oracle}\\
$n$& $\rho$ &  & bias & SE & bias & SE  & bias & SE \\
\hline
100& 0.4& $\beta_{1}$ & .0182&.0707&.0087&.0361&-.0017&.0204\\
&&$\beta_{2}$ & -.0186&.0717&-.0172&.0329&-.0055&.0205\\
     &&$\beta_{3}$ &-.0236&.0702& .0041& .0336& -.0056&.0205\\
     &&$\beta_{4}$ &.0100&.0702&-.0034&.0346& .0008&.0205\\
200& 0.4& $\beta_{1}$ &-.0130&.0517&-.0157&.0228&-.0037&.0153\\
     &&$\beta_{2}$ &.0146&.0516&.0177&.0227&.0028&.0151\\
&&$\beta_{3}$ &-.0151&.0512&.0041&.0224&.0011&.0152\\
&&$\beta_{4}$ &-.0076&.0517&.0065&.0229&.0038&.0153\\
100& 0.8& $\beta_{1}$ & .0181&.0683&-.0175& .0213&.0028&.0102\\
     &&$\beta_{2}$ &-.0111&.0682&-.0147&.0203&.0028&.0102\\
     &&$\beta_{3}$ &-.0030&.0674&-.0105&.0199&-.0015&.0100\\
     &&$\beta_{4}$ &.0260&.0675&.0125&.0208&.0028&.0101\\
200& 0.8& $\beta_{1}$ &-.0017&.0499&-.0024&.0132&.0014&.0076\\
     &&$\beta_{2}$ &-.0005&.0496&.0006&.0129&-.0001&.0076\\
&&$\beta_{3}$ &.0045&.0499&.0041&.0133&.0004&.0076\\
&&$\beta_{4}$ &-.0052&.0496&-.0059&.0130&-.0009&.0075\\
\hline
\multicolumn{2}{c}{${\sf B}=1.5, {\sf C}=4$}&&\multicolumn{2}{c}{Independent}&\multicolumn{2}{c}{Efficient}&\multicolumn{2}{c}{Oracle}\\
$n$& $\rho$ &  &  bias & SE& bias & SE& bias & SE\\
\hline
100& 0.4& $\beta_{1}$ & .0105&.0710&.0039&.0315&-.0026&.0174\\
&&$\beta_{2}$ & -.0180&.0715&-.0095&.0313&-.0046&.0174\\
     &&$\beta_{3}$ &-.0122&.0730&-.0104&.0323&.0010&.0176\\
     &&$\beta_{4}$ &.0141&.0707&.0105&.0317&.0034&.0174\\
200& 0.4& $\beta_{1}$ &-.0085&.0510&-.0060&.0223&-.0036&.0134\\
     &&$\beta_{2}$ &-.0066&.0513&-.0062&.0225&-.0018&.0135\\
&&$\beta_{3}$ &.0094&.0510&-.0015&.0225&-.0016&.0136\\
&&$\beta_{4}$ &.0062&.0514&.0001&.0224&.0006&.0137\\
100& 0.8& $\beta_{1}$ & -.0154&.0703&.0042&.0212&-.0040&.0087\\
     &&$\beta_{2}$ &-.0152&.0690&.0028&.0215&.0001&.0087\\
     &&$\beta_{3}$ &.0129&.0677&.0044&.0208&-.0002&.0092\\
     &&$\beta_{4}$ &-.0076&.0699&-.0032&.0215&.0008&.0088\\
200& 0.8& $\beta_{1}$ &-.0141&.0489&.0111&.0157&-.0001&.0067\\
     &&$\beta_{2}$ &-.0136&.0490&-.0145&.0147&-.0003&.0069\\
&&$\beta_{3}$ &.0058&.0491&.0016&.0142&-.0001&.0069\\
&&$\beta_{4}$ &.0071&.0483&.0041&.0150&-.0001&.0072\\
\hline
\end{tabular}
\end{center}
\end{table}
 
We also conducted additional simulations to examine performance of estimation of the nonparametric coefficients and estimation accuracy of parametric coefficients using modified approaches. For space consideration, we report the results in the supplement \cite{HCL2015S}.

\normalsize
\subsection{Real data example}

We now present an application of our method to the CD4 count data from the AIDS Clinical
Trial Group 193A Study \cite{H1998}. The data came from a randomized, double-blind
study of AIDS patients with CD4 counts of $\le 50$ cells/mm3. The patients were randomized to
one of four treatments with roughly equal group sizes; each consisted of a daily regimen of 600 mg of zidovudine. Treatment
1 is zidovudine alternating monthly with 400 mg didanosine; Treatment 2 is zidovudine plus
225 mg of zalcitabine; Treatment 3 is zidovudine plus 400 mg of didanosine; Treatment 4 is a
triple therapy consisting of zidovudine plus 400 mg of didanosine plus 400 mg of nevirapine.
Measurements of CD4 counts were scheduled to be collected at baseline and at eight week
intervals during the 40 weeks of follow-up. However, the real observation times were unbalanced
due to mistimed measurements, skipped visits and dropouts. The number of measurements of
CD4 counts during the 40 weeks of follow-up varied from 1 to 9, with a median of 4. The response
variable was taken as the log-transformed CD4 counts, $Y$ =log(CD4 counts + 1). There was also gender
and baseline age information about each patient.
A total of 1309 patients were enrolled in the study. We eliminated the 122 patients who dropped
out immediately after the baseline measurement.

\begin{figure}[h]
\begin{center}
\includegraphics[scale = 0.3, angle = 0]{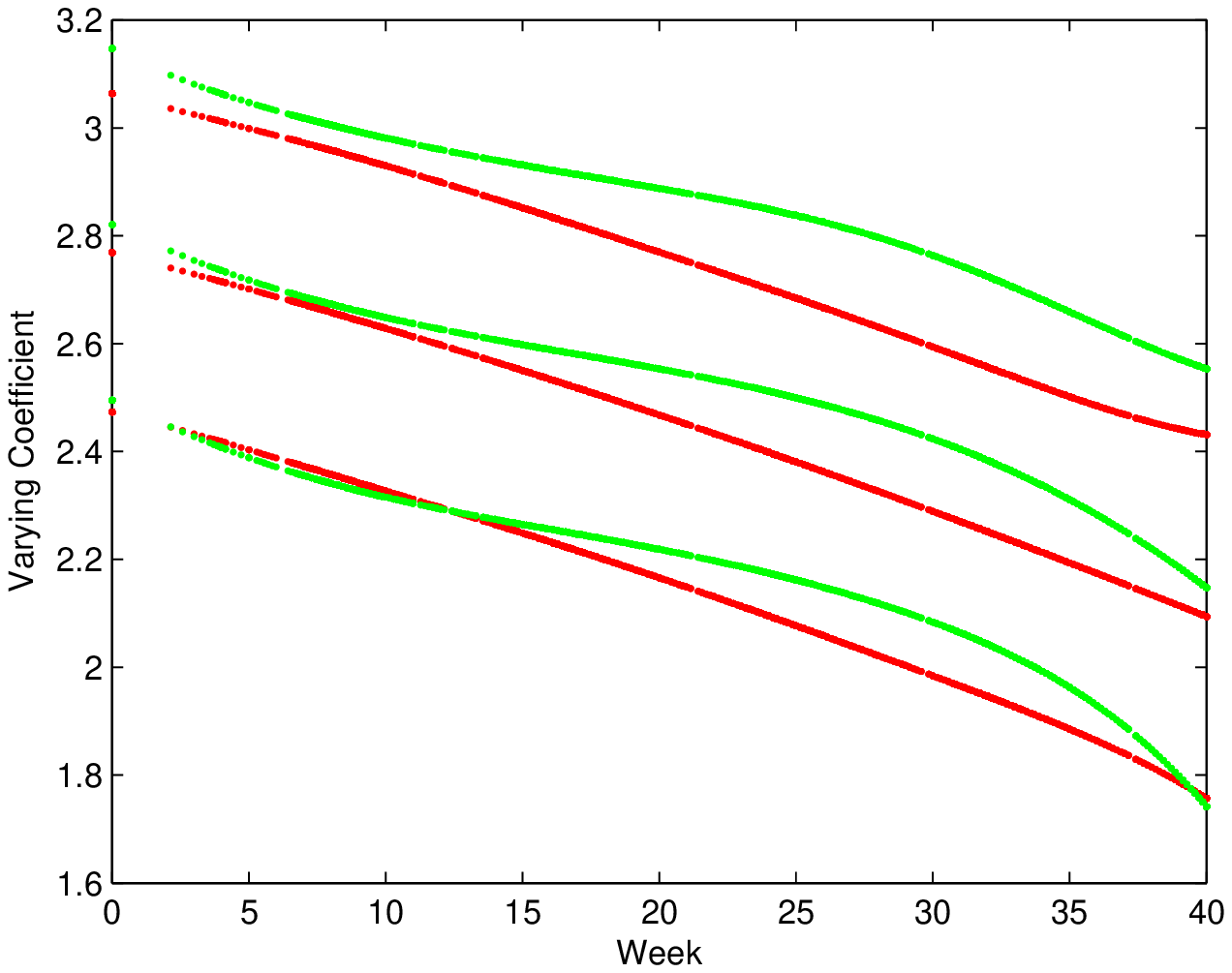}
\includegraphics[scale = 0.3, angle = 0]{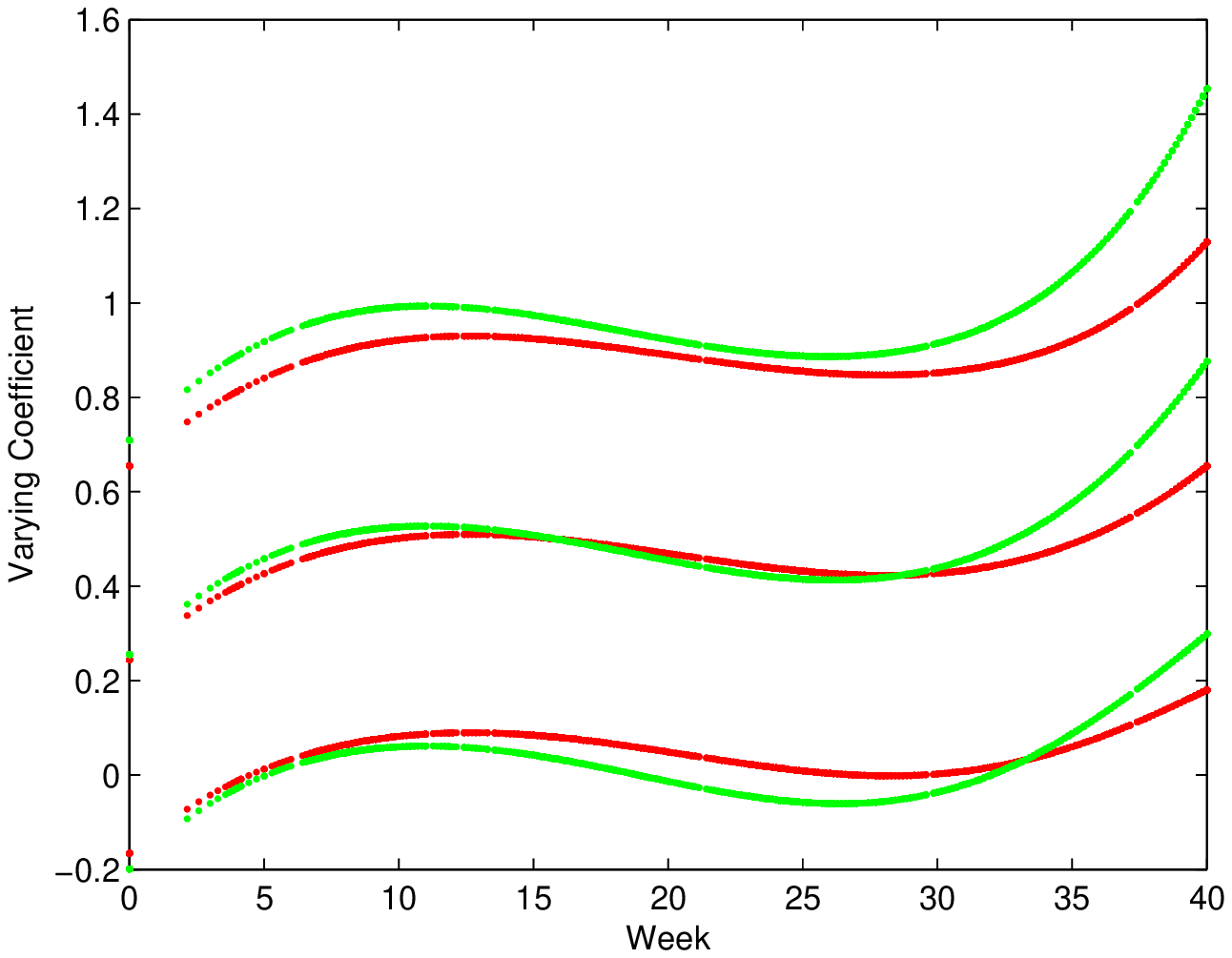}
\includegraphics[scale = 0.3, angle = 0]{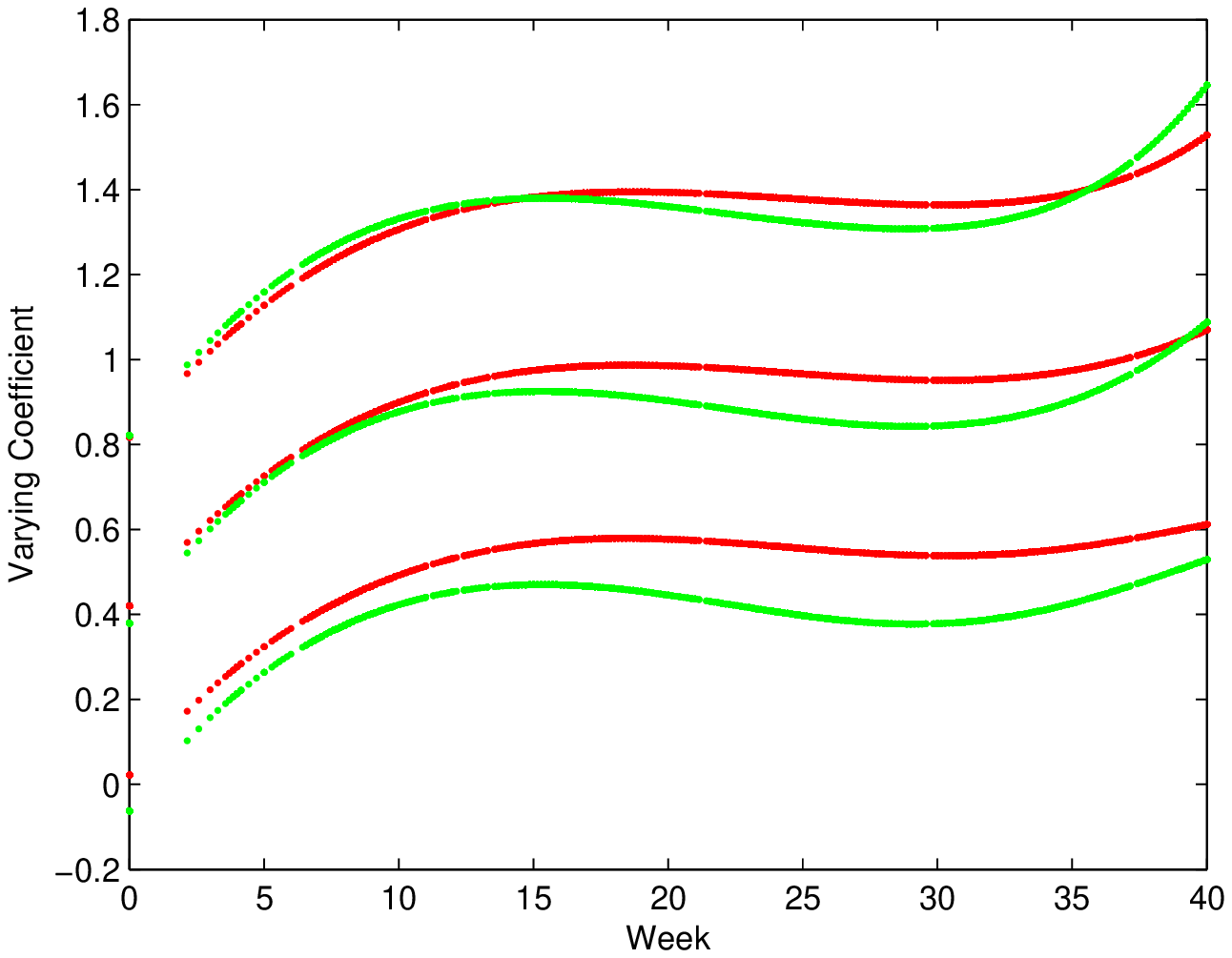}
\includegraphics[scale = 0.3, angle = 0]{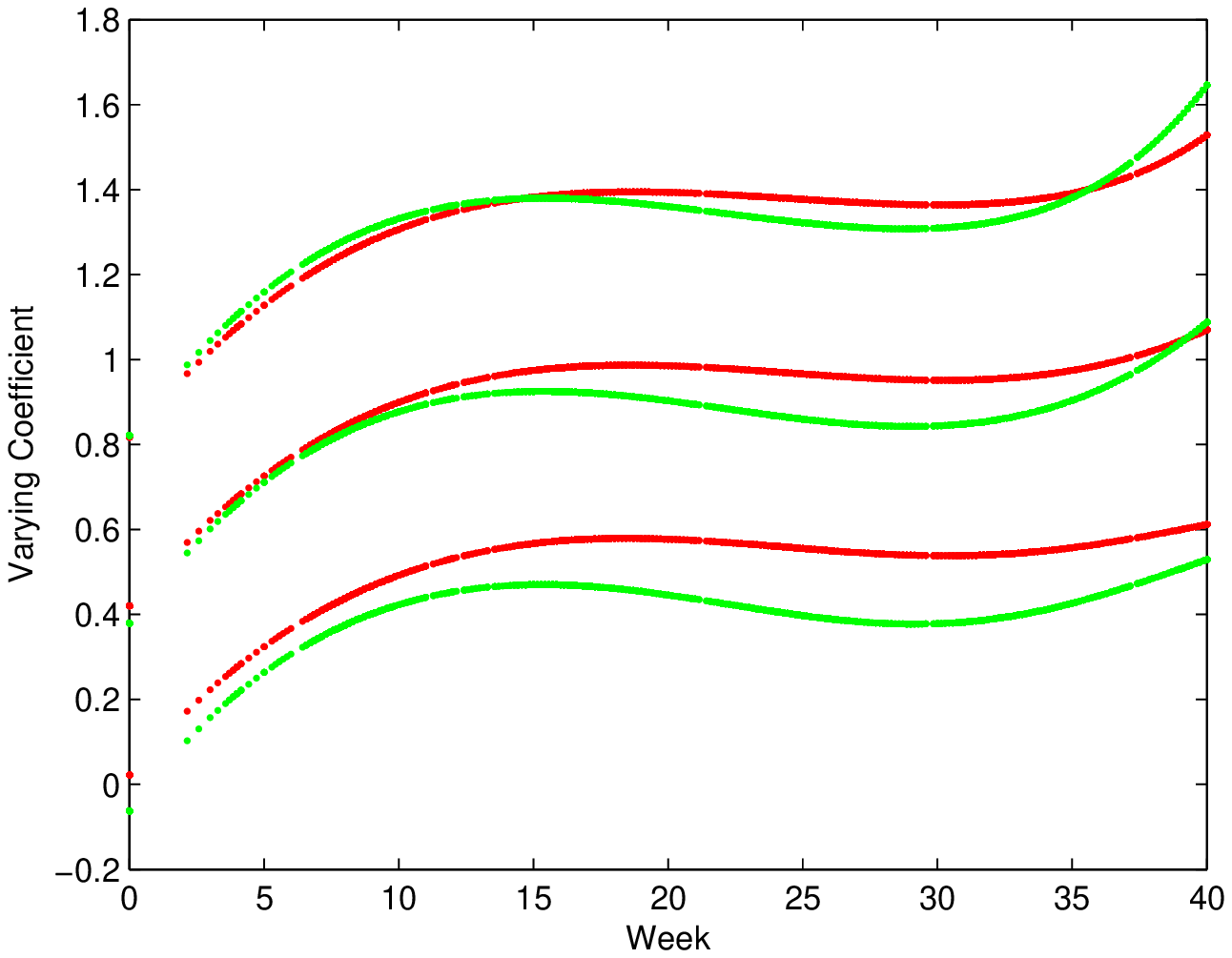}
\vspace{-6pt}
\caption{Estimated varying-coefficients along with 95\% confidence intervals for the intercept (upper left), treatment 3 (upper right), treatment 4 (lower left), and interaction between treatment 2 and sex (lower right). The red curves are efficient estimators while the green curves are estimators obtained under working independence.}\label{fig1}
\end{center}
\end{figure}

We considered the following available covariates: treatments 2, 3 and 4 (coded by three indicator variables for treatment groups 2, 3 and 4, respectively), age (years), sex (coded as 1 for male and 0 for female), and interaction effects between these covariates. Using the group SCAD structure identification procedure of Cheng et al. (2014), we found that the coefficients for treatment 3, treatment 4 and the interaction between treatment 2 and sex are varying, and the coefficients given in Table \ref{table-3} are constants. The group SCAD procedure also suggested that we remove all the other interaction effects. The estimated varying intercept (i.e. effect of treatment 1)  and the varying coefficients are displayed in Figure \ref{fig1} along with 95\% confidence intervals. The curves in the figures are updated local linear estimates without using the covariance function estimates. We used cross-validation to select the bandwidth. The constant coefficient estimates and their estimated standard errors are provided in Table \ref{table-3}. To facilitate a comparison, we reported the results using the estimators assuming working independence and the efficient estimator proposed in this paper.
Let ${\bm\theta}=(\bm\beta^T,\bm\gamma^T)^T$ and $\underline{{\bm U}}_i=(\underline{{\bm X}}_i, \underline{{\bm W}}_i)$. In practice, the variances for the efficient parameter estimates were obtained from the first $p$ diagonal elements of the following matrix:
$
\Big(\sum_{i=1}^n \underline{{\bm U}}_i^T\widehat{\bm{\Sigma}}_i^{-1}\underline{{\bm U}}_i\Big)^{-1},
$
and for the working independence parameter estimates the variances were obtained from the first $p$ diagonal elements of the following matrix:
$
\Big(\sum_{i=1}^n \underline{{\bm U}}_i^T\underline{{\bm U}}_i\Big)^{-1}\sum_{i=1}^n\underline{{\bm U}}_i^T\widehat{\bm{\Sigma}}_i\underline{{\bm U}}_i\Big(\sum_{i=1}^n \underline{{\bm U}}_i^T\underline{{\bm U}}_i\Big)^{-1}.
$

\begin{table}[htbp]
\begin{center}
\caption{Estimation results for CD4 count data. ``Independent'' corresponds to using $\bm{V}_i=\bm{I}_{m_i}$; ``Efficient'' refers to using $\bm{V}_i=\widehat{\bm{\Sigma}}_i$; ``Quadratic'' refers to the QIF based method. } \label{table-3}
{\scriptsize 
\begin{tabular}{rllllll}
\hline
&\multicolumn{2}{c}{Independent}&\multicolumn{2}{c}{Efficient}&\multicolumn{2}{c}{Quadratic}\\
 Covariates & Coefficients & SE & Coefficients & SE & Coefficients & SE  \\
\hline
treatment 2&.3614& .2257& .4038& .2027 & .3532&.1318\\
age&.0946& .0274 & .0818& .0245 & .0882&.0171\\
sex&.1704& .1768& .2246& .1587& .1187& .1034\\
treatment 3:sex&-.2922& .2472& -.2908& .2209& -.2625&.2485\\
treatment 4:sex&-.5321& .2416& -.5653& .2146& -.5580&.1574\\
\hline
\hline
\end{tabular}
}
\end{center}
\end{table}

\begin{figure}[h]
\begin{center}
\includegraphics[scale = 0.3, angle = 0]{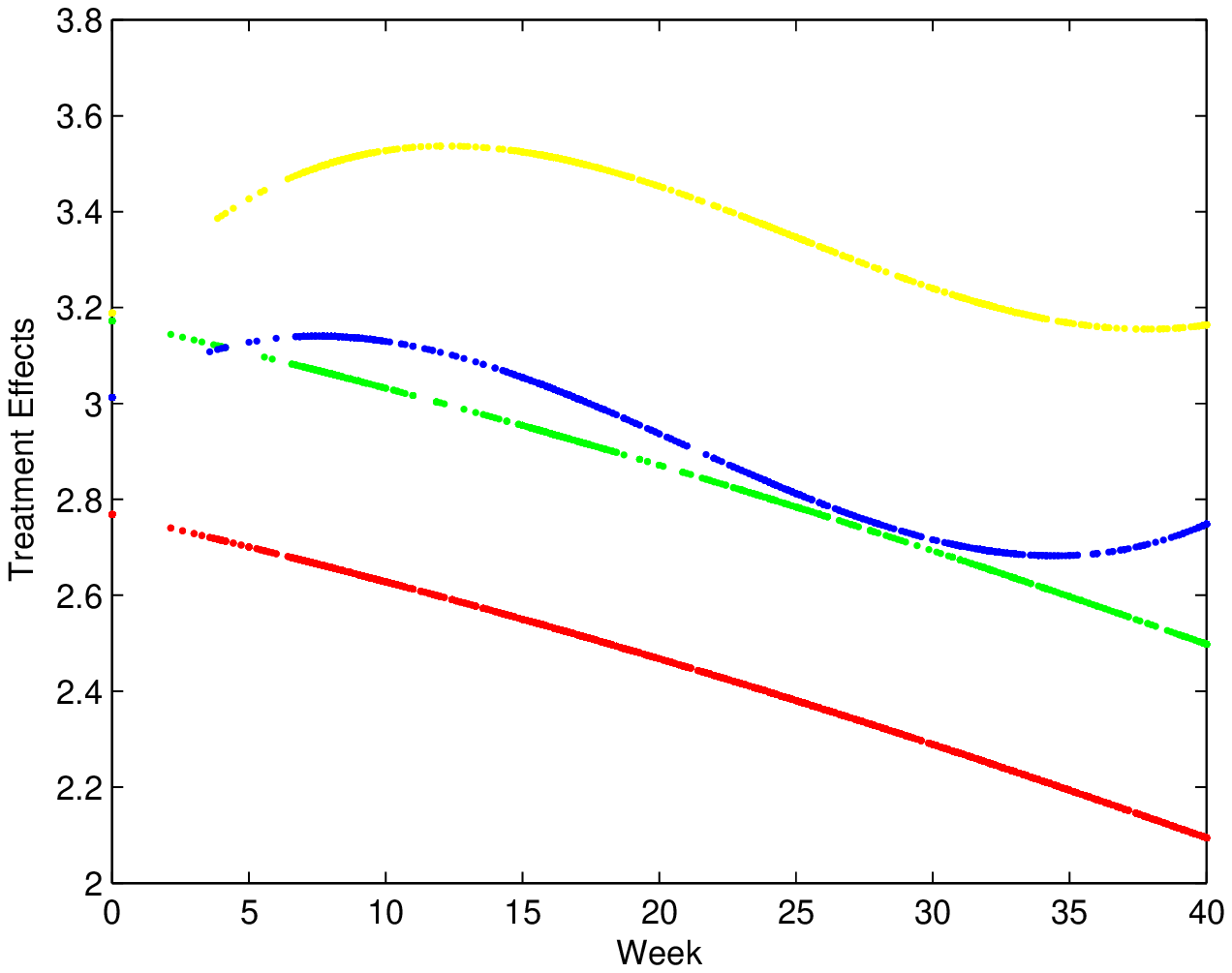}
\includegraphics[scale = 0.3, angle = 0]{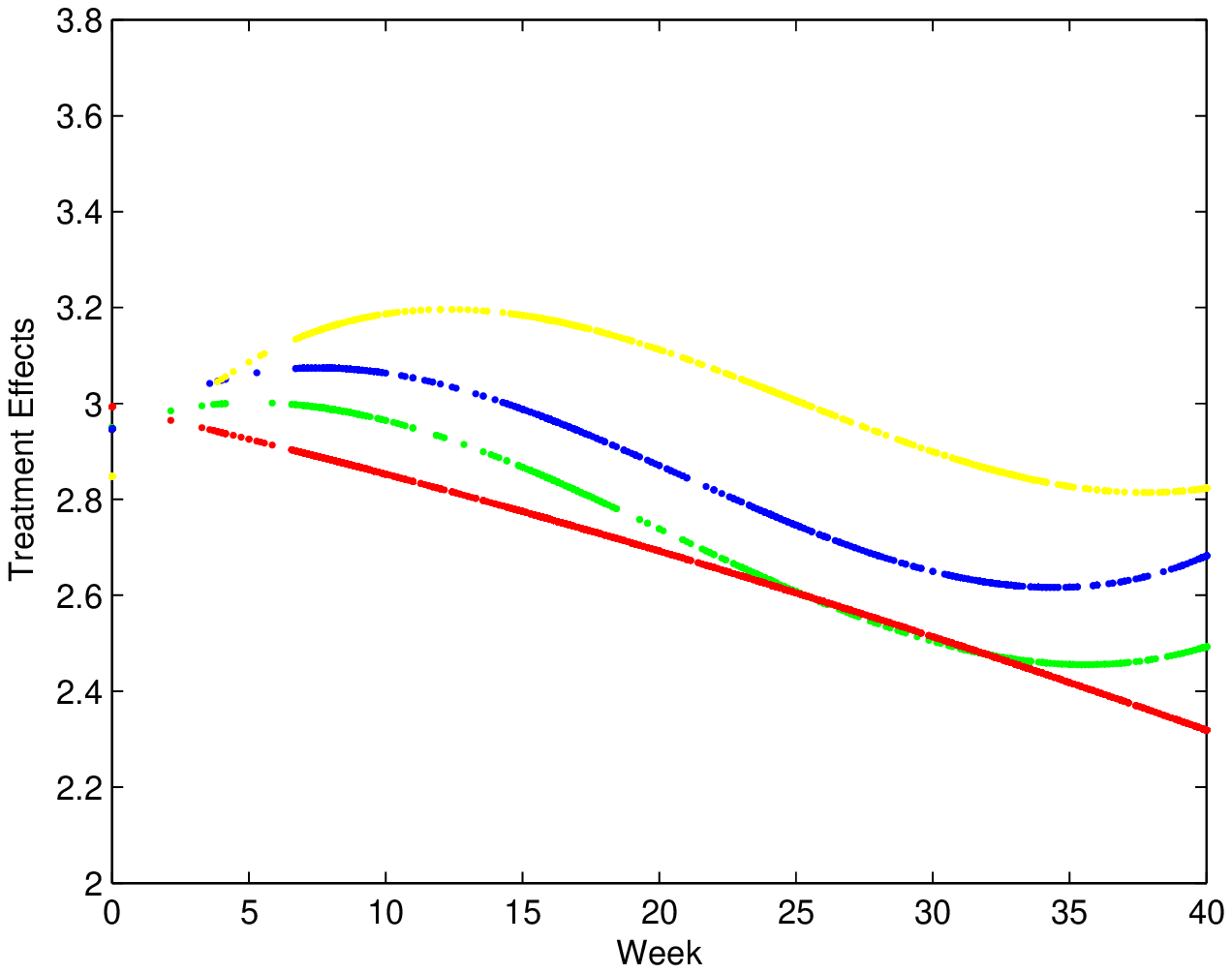}
\includegraphics[scale = 0.3, angle = 0]{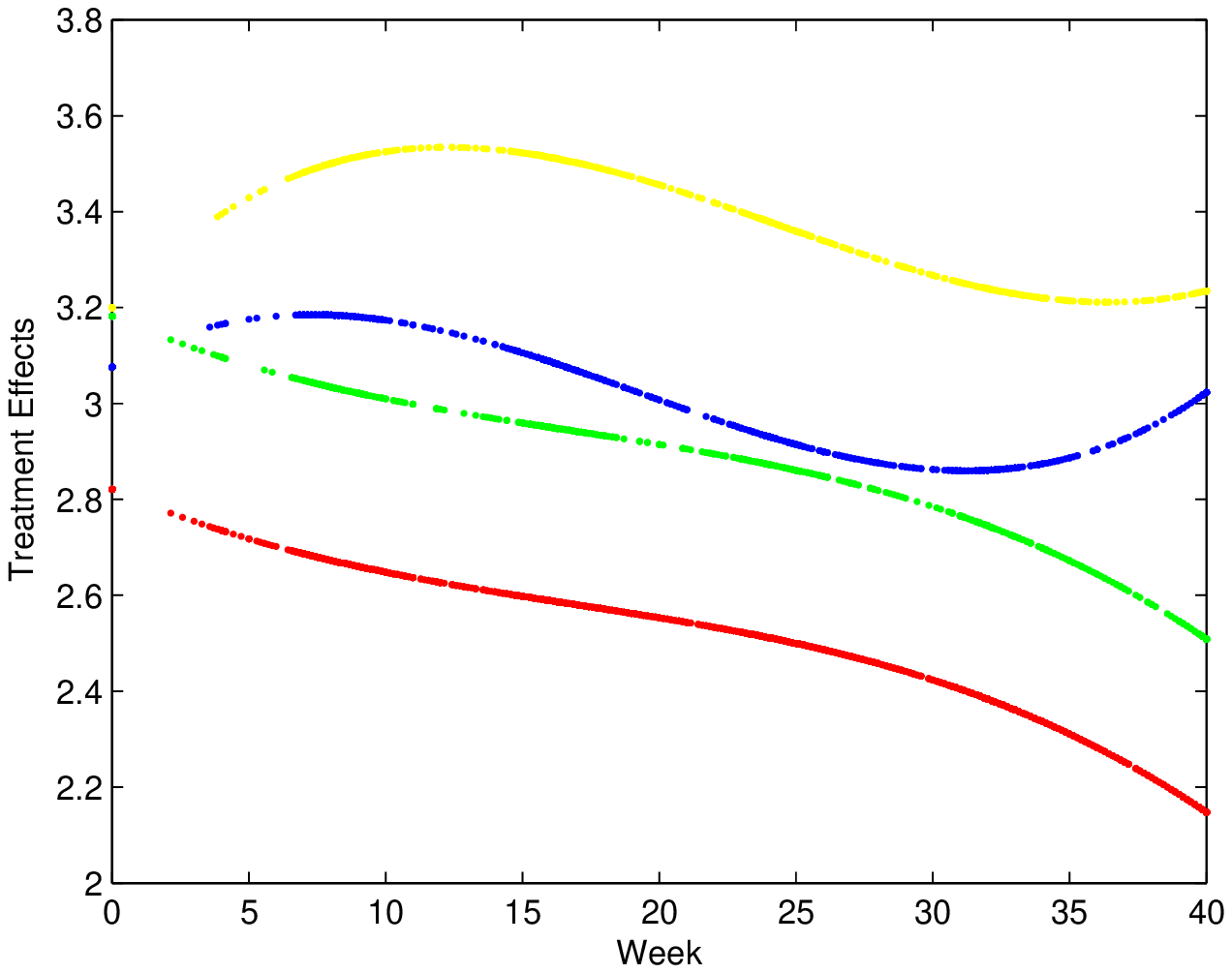}
\includegraphics[scale = 0.3, angle = 0]{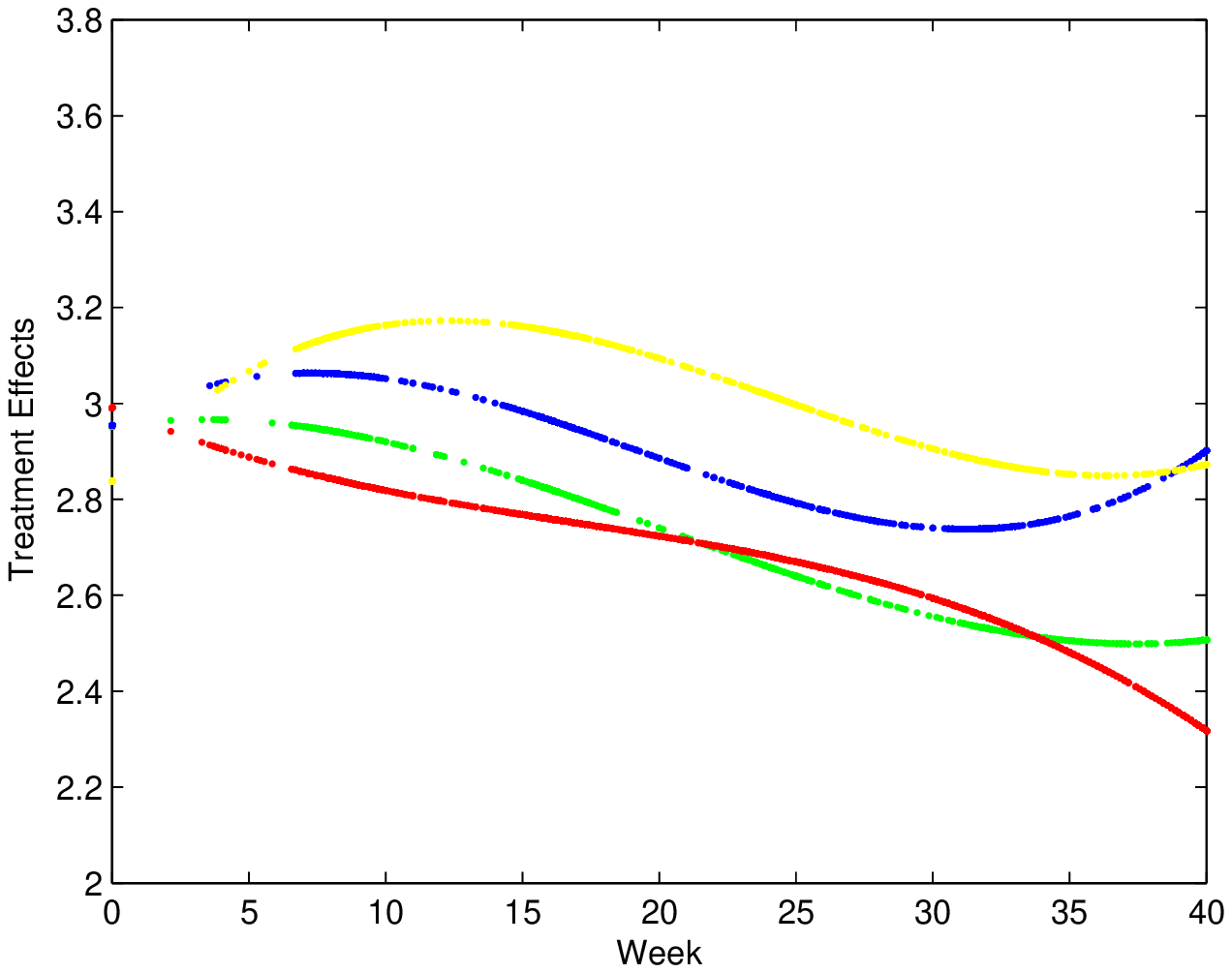}
\includegraphics[scale = 0.3, angle = 0]{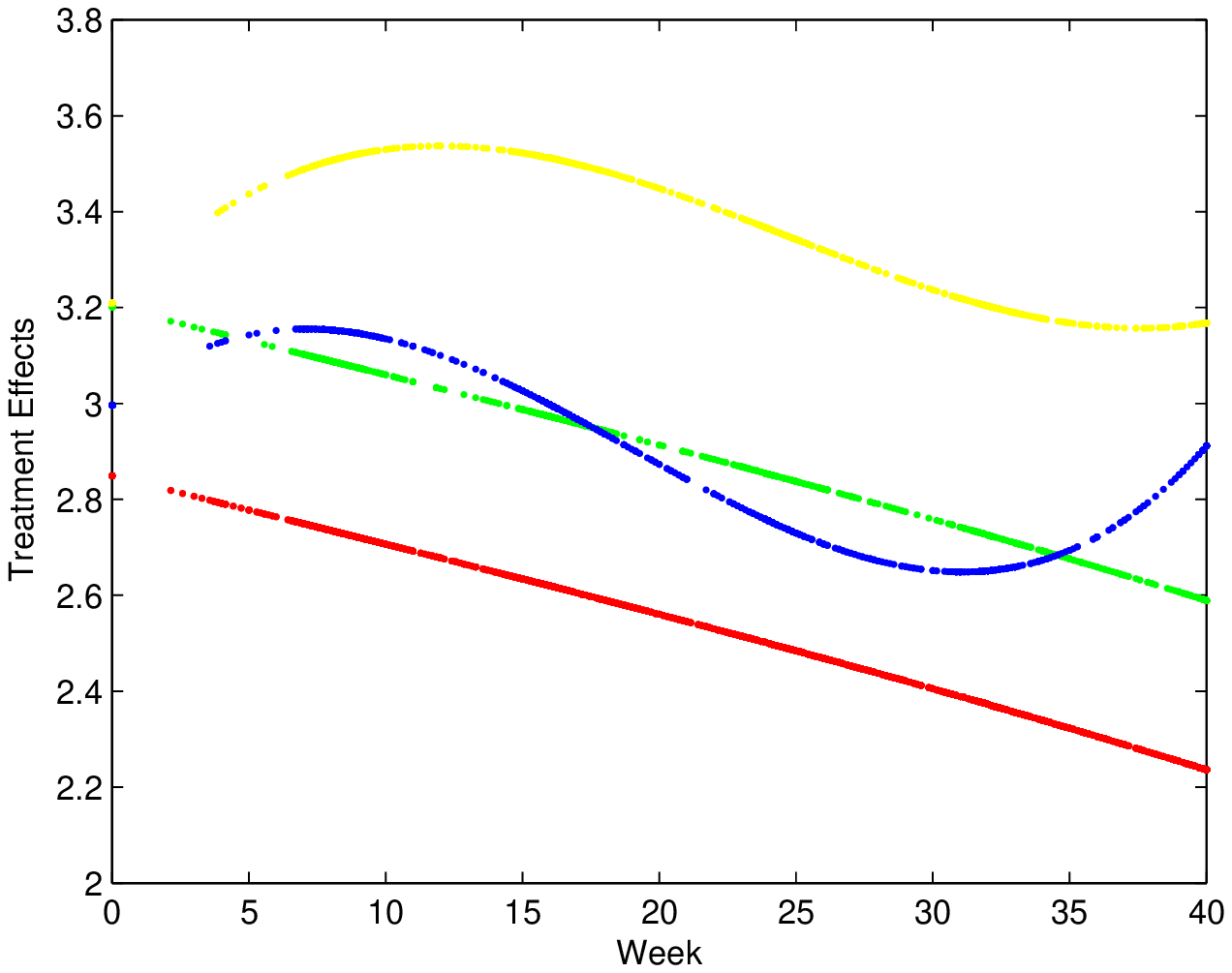}
\includegraphics[scale = 0.3, angle = 0]{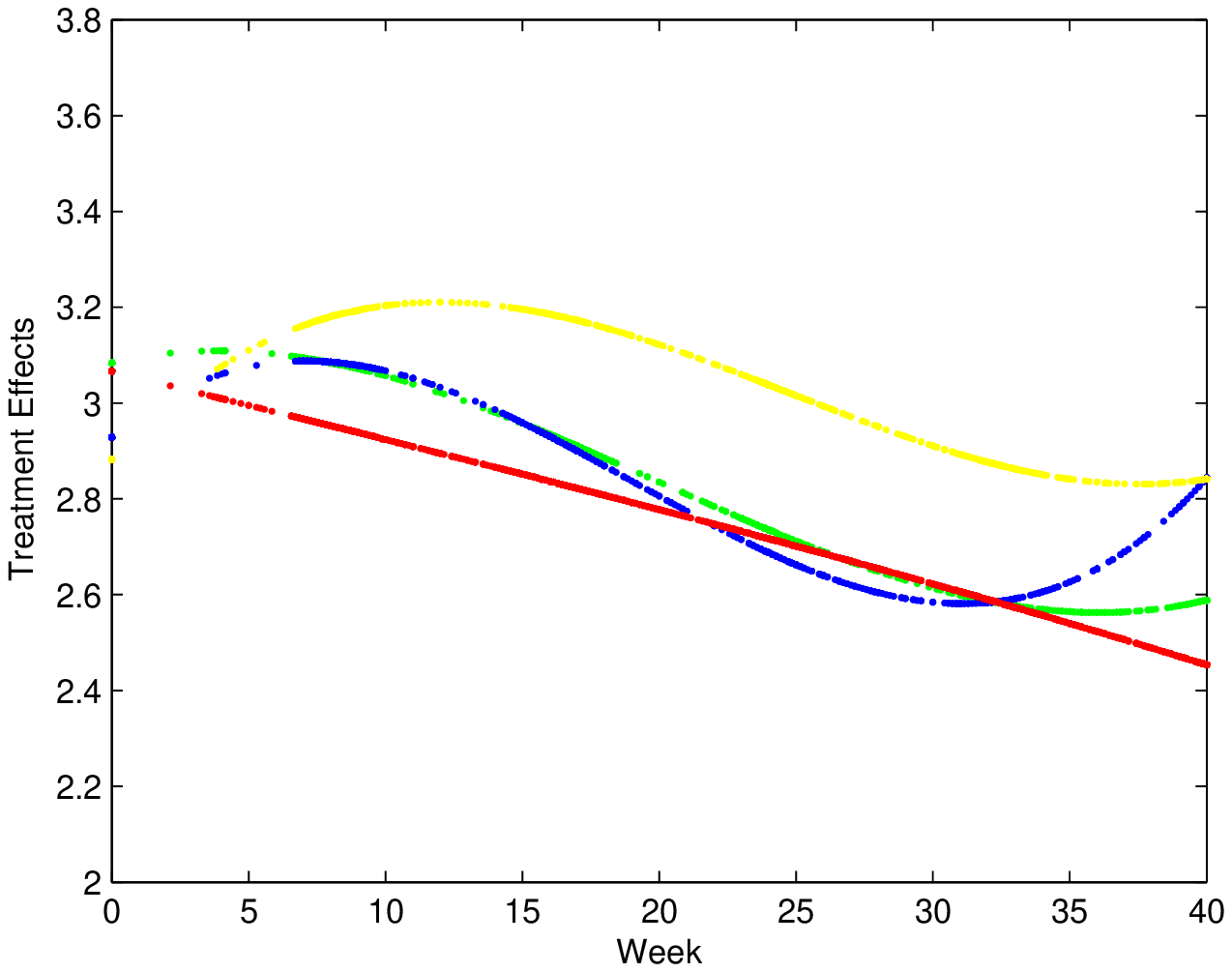}
\vspace{-6pt}
\caption{Estimated treatment effects for the four treatment groups. The panels in the top, middle and bottom rows are respectively the proposed efficient estimates, the estimates assuming independence and the estimates based on the QIF method. The panels in the left and right columns are respectively for the females and the males. Red, green, blue and yellow curves are for treatment groups 1, 2, 3 and 4, respectively.}\label{fig2}
\end{center}
\end{figure}

From Table \ref{table-3}, we note that the estimated constant coefficients for treatment 2, age, and the interaction between treatment 4 and sex are all quite significant. The constant coefficient estimates for sex are not significant but are still kept in the model since we include the interactions between treatments and sex. The efficient estimates for all the constant and varying coefficients have smaller standard errors than the respective estimates assuming working independence. In fact, the Wald test statistic for the coefficient of treatment 2 is $.3614/.2257=1.60<1.96$ under the working independence, failing to declare a significant difference. On the other hand, the Wald test statistic for the same coefficient is $.4038/.2027=1.99>1.96$ from the efficient estimation, leading to a significant treatment difference. Other than these, because the sample size in this study was rather large, the two types of estimates for all the constant and varying coefficients appear to be very similar. For the sake of comparison, we also present the estimation results for these regression coefficients from the  estimating equation methods based on the QIF method \cite{QL2006}. The conclusions on the estimation significance and effect direction remain the same as for the efficient estimation while the magnitude of the estimated coefficients slightly differs. For this particular dataset, sometimes the QIF estimator seems to have smaller standard error than the efficient estimator. An explanation is that it choses a covariance structure like compound symmetry in the matrix basis, thus it will be more efficient than our estimator when this structure is plausible (which is possibly the case here). Otherwise, it is generally not as good when the covariance structure is mis-specified.

In general, the CD4 count tends to increase with age in the fitted model. Our estimation results suggest that there exist interaction effects between treatment and sex. Specifically, for the females (sex=0), subjects receiving treatments 2, 3 and 4 tend to have increasingly higher CD4 counts than those under treatment 1. The effect for treatment 2 (as compared with treatment 1) is estimated as a constant and is significant, while those for the other two treatment groups are varying (the upper right and the lower left panels in Figure \ref{fig1}) with even greater positive differences from treatment 1. For the males (sex=1), subjects receiving treatments 2, 3 and 4 also tend to have higher mean CD4 counts than those receiving treatment 1. The interaction between treatment 2 and  sex is varying over time (the lower right panel in Figure \ref{fig1}) while those for treatments 3 and 4 are constant. The effects of treatments 3 and 4 are significantly different from that of treatment 1, judging from Table \ref{table-3}. Also, we notice that the differences between treatments seem to be greater between the females than between the males.

The estimated effects of the four treatment groups are plotted in Figure \ref{fig2} for the efficient estimator, the working independence estimator and the QIF estimator. Note that treatment effects given by the efficient estimator rarely cross each other, giving nice interpretation and ordering of the different treatments, whereas this is not the case for those given by the QIF or the working independence estimator. Previous authors identified a similar pattern on the order of magnitude of the time-varying treatment effects \cite{Li2011}. However, they ignored the interactions between the treatments and sex. Our findings suggest the treatment effect curves might be rather different between the males and the females.

\section{Proofs of the main results}
\label{sec:details1}

\subsection{Additional assumptions and  definitions}

We denote the Euclidean norm of a vector $a$
by $|a|$. Let $\lambda_{\rm min}(A)$ and $\lambda_{\rm max}(A)$
stand for the minimum and maximum eigenvalues
of a symmetric matrix $A$, respectively. 
Besides, $C$, $C_1$, $C_2$, $\ldots$
are generic positive constants whose values may vary from line to
line. 
Recall that the density function of $T_{ij}$ is denoted by by $f_{ij}(t)$, $i=1,\ldots,n$ and $j=1,\cdots, m_i$. 
Also, we denote the joint density function of $T_{ij}$ and 
$T_{ij'}$ $( j \ne j')$ by $f_{ijj'}(s,t)$. In Assumptions A1 and A2, we consider sparse and irregular observation times.
Note that 
we carry out two-dimensional smoothing in step 5 and there are three bandwidths involved in our method.  Therefore we impose these restrictive assumptions to avoid complicated assumptions involving $m_i$, $m_{\rm max}$, and the
bandwidths simultaneously. Roughly speaking, these assumptions imply we should have
$\sum_{i=1}^n m_i^5 = O(n)$.

\smallskip
\noindent
{\bf Assumption A1}. For some positive constant
$C_{A1}$, we have \\
$
\mx \equiv \max_{1\le i \le n} m_i = O(n^{1/8})
$ 
and $ \sum_{i=1}^n m_i < C_{A1}n$.

\noindent
{\bf Assumption A2}. The joint density functions $f_{ij}(t)$ and $f_{ijj'}(s,t)$ are uniformly bounded and we have for some positive constant
$C_{A2}$,
\begin{align*}
\frac{1}{C_{A2}}& < \frac{1}{n} \sum_{i=1}^n \frac{1}{m_i}
 \sum_{j=1}^{m_i} f_{ij}(t)\le 
\frac{1}{n} \sum_{i=1}^n m_i^4
 \sum_{j=1}^{m_i} f_{ij}(t)
<C_{A2}
\ {\rm on} \,\, [0,1],\quad \mbox{and}\\
\frac{1}{C_{A2}}& <\frac{1}{n}\sum_{i=1}^n\sum_{j\ne j'} f_{ijj'}(s,t)\le
\frac{1}{n}\sum_{i=1}^n m_i^3 \sum_{j\ne j'} f_{ijj'}(s,t)
<C_{A2}
\  {\rm on} \,\, [0,1]^2.
\end{align*}

\noindent
{\bf Assumption A3}. For some positive constants
$C_{A3}$ and $C_{A4}$, we have 
\[
C_{A3}\bm{I}_{p+q}\le \RE \left\{
\begin{pmatrix}
\bm{X}_{ij}\bm{X}_{ij}^T  & \bm{X}_{ij}\bm{Z}_{ij}^T \\
\bm{Z}_{ij}\bm{X}_{ij}^T  & \bm{Z}_{ij}\bm{Z}_{ij}^T 
\end{pmatrix}\, \middle| \,
\ T_{ij}
\right\} \le C_{A4}\bm{I}_{p+q},\,\, \mbox{uniformly in} \, i \,\mbox{and}\, j.
\]

\noindent
{\bf Assumption A4}. For some positive constants
$C_{A5}$ and $C_{A6}$, we have\\ 
$
C_{A5} \le \lambda_{\rm min}(\bm{\Sigma}_i)
\le \lambda_{\rm max}(\bm{\Sigma}_i) \le C_{A6}m_i \,,
$
uniformly in $i$.

\noindent
{\bf Assumption A5}. For some positive constants
$C_{A7}$ and $C_{A8}$, we have \\ 
$
C_{A7} \le \lambda_{\rm min}(\bm{V}_i)
\le \lambda_{\rm max}(\bm{V}_i) \le C_{A8}m_i\,, 
$
uniformly in $i$.

\noindent
{\bf Assumption A6}. For some positive constants
$C_{A9}$ and $C_{A10}$, we have\\ 
$
\RE \{ \exp( C_{A9} |\epsilon_{ij} | ) \, |\,
\underline{\bm{X}}_i, \underline{\bm{Z}}_i, 
\underline{T}_i \} < C_{A10}\,, 
$
uniformly in $i$ and $j$.

Assumption A3 is a standard one and is necessary for
identification of the constant coefficients and the varying coefficient functions.
When $\underline{\epsilon}_i$ consists of some stochastic process
and i.i.d. errors, we have
$
\bm{\Sigma_i} = \Xi ( \underline{T}_i ) + \eta^2 \bm{I}_{m_i},
$
where $\Xi(\underline{T}_i)$ is positive definite. Hence we impose
Assumptions A4 and A5 on $\bm{V}_i$ and $\bm{\Sigma}_i$, respectively.
In \cite{CZH2014}, it is assumed that $\underline{\epsilon}_i$
has the sub-Gaussian property in order to deal with general link functions.
The sub-Gaussian assumption prevents $m_i$ from tending to infinity.
Assumption A6, which is less restrictive, is enough for the identity link function
since we do not need to employ any results from the empirical process theory
in this case.

For $\bm{g}= (g_1, \ldots, g_q)^T \in \bm{G}$,
we define the sup and $L_2$ norms by
$
\| \bm{g} \|_{G,\infty}= \sum_{j=1}^q
\sup_{t\in [0,1]}|g_j(t)|
$
and
$
\| \bm{g} \|_{G,2}^2= \sum_{j=1}^q
\int_0^1 g_j^2 (t) dt. 
$
Assumptions A2 and A3 imply there are
positive constants $C_1$ and $C_2$ such that
\begin{equation}
C_1 \| \bm{g} \|_{G,2} \le
\| \bm{Z}^T\bm{g} \|^V
\le C_2 \| \bm{g} \|_{G,2}
\label{eqn:e503}
\end{equation}
for any $\bm{g} \in \bm{G}$. The details are given in
Lemma \ref{lem:lem0}.
In (\ref{eqn:e218}), we define two kinds of projections
of $X_k$. We define another one here: 
\begin{equation}
\widehat{\bm{\varphi}}_{\bm{V}k} = \widehat{\Pi}_{\bm{V}n} X_k = \argmin_{\bm{g}\in \bm{G}_B}
\| X_k -\bm{Z}^T\bm{g} \|_n^V.
\label{eqn:e506}
\end{equation}

\subsection{Spline approximation and projections}
\label{sec:projection}
Recall we assume all the relevant functions are at least twice
continuously differentiable and they and
their second order derivatives are uniformly bounded.
Hence the sup norm of approximation errors by spline functions
is bounded from above by $C_{approx} K_n^{-2}$, where
$C_{approx}$ depends on the relevant functions. See
Corollary 6.26 of \cite{Schumaker2007}.

\noindent

Note that $\langle \cdot, \cdot \rangle^V$
and $\| \cdot \|^V$ are defined on
$\{ v \, | \, \sum_{i,j}\RE (v_{ij}^2) < \infty \}$
and that $\{ \bm{Z}^T \bm{g} \}$ is a closed linear
subspace due to (\ref{eqn:e503}). Therefore the projections
$\bm{\varphi}_{\bm{V}k}^* = (\varphi_{\bm{V}k1}^*,
\ldots, \varphi_{\bm{V}kq}^*)^T$, $k=1,\ldots, p$, exist uniquely. Next, we 
set
$
\bm{V}_i^{-1}=( v_i^{j_1j_2} ).
$
Note that $\bm{\varphi}_{\bm{V}k}^* =\Pi_{\bm{V}}X_k$ defined in (\ref{eqn:e218}) satisfies that
$
\langle X_k - \bm{Z}^T\Pi_{\bm{V}} X_k, \bm{Z}^T\bm{g}
\rangle^V
 = 0 \ \  \forall \bm{g}\in \bm{G} \,.
$
By representing 
the above equality explicitly, we
can derive the following integral equations for
$\bm{\varphi}_{\bm{V}k}^* (t)$. For $d_1=1, \ldots, q$,
\begin{equation}
\sum_{d_2=1}^q
a_{d_2}^{(d_1)}(t)\varphi_{\bm{V}kd_2}^*(t)
= b^{(d_1)}(t) + \int_0^1 \sum_{d_2=1}^q
c_{d_2}^{(d_1)}(s,t)\varphi_{\bm{V}kd_2}^*(s)ds,
\label{eqn:e509}
\end{equation}
where
\begin{align*}
a_{d_2}^{(d_1)}(t) & = \frac{1}{n}\sum_{i=1}^n\sum_{j=1}^{m_i}
\RE \{  Z_{ijd_2}v_i^{jj}Z_{ijd_1}\, | \, T_{ij}=t
\}f_{ij}(t),\\
b^{(d_1)}(t) & = \frac{1}{n}\sum_{i=1}^n\sum_{1\le j_1,j_2 \le m_i}
\RE \{ X_{ij_1k}v_i^{j_1j_2}Z_{ij_2d_1}\, | \, 
T_{ij_2}= t \}f_{ij_2}(t),\\
c_{d_2}^{(d_1)}(s,t) & =
-\frac{1}{n}\sum_{i=1}^n\sum_{j_1 \ne j_2}
\RE \{  Z_{ij_1d_2}v_{i}^{j_1j_2}Z_{ij_2d_1}
\, | \, T_{ij_1}=s, T_{ij_2}=t \}f_{ij_1j_2}(s,t).
\end{align*}
Let $\bm{A}(t)$ be the $q \times q$ matrix whose $(d_1,d_2)$th
element is $a_{d_2}^{(d_1)}(t)$. Assumptions A2
and A3 imply that $|\bm{A}(t)|\ne 0$ on $[0,1]$ and we set
$
\psi_{\bm{V}kd_1}^*(t)=
\sum_{d_2=1}^q
a_{d_2}^{(d_1)}(t)\varphi_{\bm{V}kd_2}^*(t).
$
Then (\ref{eqn:e509}) reduces to (S.2) of \cite{CZH2014S}
and the same argument there applies. Therefore
$\bm{\varphi}_{\bm{V}k}^* (t)$ has the required smoothness
properties under similar regularity conditions.

\subsection{Remarks on the proofs of Propositions \ref{prop:prop1}--\ref{prop:prop3}}


We can proceed as in \cite{HZZ2007} (and \cite{CZH2014S})
by replacing
$Z_{ij}$, $\underline{Z}_i$, and $\varphi_{k}^* (t)$
in \cite{HZZ2007} (and
$\bm{Z}_{ij}$, $\bm{Z}_i$, and $\varphi_{k}^* (\bm{t})$
in \cite{CZH2014S})
with $\bm{W}_{ij}$, $\underline{\bm{W}}_i$,
and $\bm{Z}^T\bm{\varphi}_{\bm{V}k}^* (t) $, respectively.
They used several lemmas in their proofs. We reorganize
the corresponding lemmas in our setup into Lemma \ref{lem:lem0} given in the following.
Its proof and outlines of the proofs
of Propositions \ref{prop:prop1}-\ref{prop:prop3} are given in the supplement \cite{HCL2015S}.

\begin{lem} Assume that Assumptions A1-5 hold.
\label{lem:lem0}
\hspace{0.1cm}

\begin{enumerate}
\item[(i)] There are positive constants $C_1$ and $C_2$ such that for any $\bm{g} \in \bm{G}$,
$
C_1 \| \bm{g} \|_{G,2} \le
\| \bm{Z}^T\bm{g} \|^V \le C_2 \| \bm{g} \|_{G,2}\,.
$

\item[(ii)] There are positive constants $C_3$ and $C_4$ such that for any $\bm{g} \in \bm{G}_B$,
$
\| \bm{g} \|_{G,\infty}^2 \le
C_3 K_n \| \bm{g} \|_{G,2}^2
\le C_4 K_n ( \| \bm{Z}^T\bm{g} \|^V )^2\,.
$

\item[(iii)] There is a positive constant $C_5$ such that
for any $\bm{\beta} \in \RR^p$ and $\bm{g} \in \bm{G}_B$,
$
\| \bm{X}^T\bm{\beta}+ \bm{Z}^T \bm{g} \|_\infty
\le C_5 K_n^{1/2} \| \bm{X}^T\bm{\beta}+ \bm{Z}^T
\bm{g} \|^V,
$
where $\|v \|_\infty = \max_{i,j}|v_{ij}|$. Besides, for
some positive constant $C_6$, 
$
\|v \|^V \le C_6 \|v \|_\infty.
$

\item[(iv)]
\[
\sup_{\bm{g}_1,\bm{g}_2\in \bm{G}_B}
\Big| 
\frac{ \langle \bm{Z}^T\bm{g}_1, \bm{Z}^T\bm{g}_2 
\rangle_n^V - \langle \bm{Z}^T\bm{g}_1, 
\bm{Z}^T\bm{g}_2 \rangle^V}{\| \bm{Z}^T\bm{g}_1
\|^V \| \bm{Z}^T\bm{g}_2 \|^V }
\Big|
=O_p(K_n\sqrt{\log n/ n}).
\]

\item[(v)] For any positive constant $M$, we have
$
\langle X_j- \bm{Z}^T\bm{g}_j, X_k- \bm{Z}^T\bm{g}_k 
\rangle_n^V - \langle X_j- \bm{Z}^T\bm{g}_j, 
X_k - \bm{Z}^T\bm{g}_k \rangle^V = o_p(1)
$
uniformly in $\bm{g}_j\in \bm{G}_B$ and
$\bm{g}_k\in \bm{G}_B$ satisfying
$\|\bm{g}_j\|_{G,2}\le M$ and $\|\bm{g}_k\|_{G,2}\le M$.

\item[(vi)] For any process $\delta_n$ taking scalar values at $T_{ij}$ such that $ \| \delta_n \|_\infty $ is uniformly bounded in $n$
and $\{ \delta_{n,ij} \}_{j=1}^{m_i}$ are mutually independent
in $i$, 
\[
\sup_{\bm{g} \in \bm{G}_B}
\Big| 
\frac{  \langle \delta_n, \bm{Z}^T\bm{g} 
\rangle_n^V - \langle \delta_n, 
\bm{Z}^T\bm{g} \rangle^V }{
\| \bm{Z}^T\bm{g} \|^V }
\Big|
=O_p(\sqrt{ K_n / n})  \| \delta_n \|_\infty .
\]

\item[(vii)] We also suppose Assumption S holds. Then
for $k=1, \ldots, p$,\
$ \| \widehat{\bm{\varphi}}_{\bm{V}k}\|_\infty = O_p(1)$, 
$\| \bm{Z}^T( \bm{\varphi}_{\bm{V}k}^*  -
\widehat{\bm{\varphi}}_{\bm{V}k})  \|_n^V = o_p(1)$, and
$\| \bm{Z}^T( \bm{\varphi}_{\bm{V}k}^*  -
\widehat{\bm{\varphi}}_{\bm{V}k}  ) \|^V = o_p(1).
$
\end{enumerate}
\end{lem}

\subsection{Proof of Theorem \ref{thm:thm1}}
\label{sec:details2}

Since we consider the identity link function, we have
explicit 
expressions of $\widehat{\bm{\beta}}_{\bm{\Sigma}}-\bm{\beta}_0$
and $\widehat{\bm{\beta}}_{\widehat {\bm{\Sigma}}}-\bm{\beta}_0$:
\begin{align}
\widehat{\bm{\beta}}_{\bm{\Sigma}}-\bm{\beta}_0
= & \bH^{11}\sum_{i=1}^n (\underline{\bm{X}}_i
- \underline{\bm{W}}_i\bH_{22}^{-1}\bH_{21})^T
\bm{\Sigma}_i^{-1}\underline{\epsilon}_i\label{eqn:e600}\\
&  - \bH^{11}\sum_{i=1}^n (\underline{\bm{X}}_i
- \underline{\bm{W}}_i\bH_{22}^{-1}\bH_{21})^T
\bm{\Sigma}_i^{-1}( \underline{\bm{W}}_i \bm{\gamma}^*
-\underline{(\bm{Z}^T\bm{g}_0)}_i)\nonumber\\
= & I_1- I_2 \quad ({\rm say}),\nonumber
\end{align}
\begin{align}
\widehat{\bm{\beta}}_{\widehat{\bm{\Sigma}}}-\bm{\beta}_0
= & \widehat{\bH}^{11}\sum_{i=1}^n (\underline{\bm{X}}_i
- \underline{\bm{W}}_i\widehat{\bH}_{22}^{-1}\widehat{\bH}_{21})^T
\widehat{\bm{\Sigma}}_i^{-1}\underline{\epsilon}_i\label{eqn:e603}\\
& - \widehat{\bH}^{11}\sum_{i=1}^n (\underline{\bm{X}}_i
- \underline{\bm{W}}_i\widehat{\bH}_{22}^{-1}\widehat{\bH}_{21})^T
\widehat{\bm{\Sigma}}_i^{-1}( \underline{\bm{W}}_i \bm{\gamma}^*
-\underline{(\bm{Z}^T\bm{g}_0)}_i)\nonumber\\
= & \widehat{I}_1- \widehat{I}_2 \quad ({\rm say}),\nonumber
\end{align}
where $\widehat{\bH}^{11}$, $\widehat{\bH}_{22}$ and 
$\widehat{\bH}_{21}$ are defined as in (\ref{eqn:e230})
with $\bm{V}_i = \widehat{\bm{\Sigma}}_i$, $i=1,\ldots,n$, and $\bm{\gamma}^*
=(\bm{\gamma}_1^{*T}, \ldots, \bm{\gamma}_q^{*T})^T$
satisfies $| \bm{B}^T(t)\bm{\gamma}_j^{*}- g_{0j}(t)|
\le C_gK_n^{-2}$, $j=1,\ldots, q$, for some positive constant $C_g$  
depending on $\bm{g}_0(t)$. Proposition \ref{prop:prop4}
and Assumption A4 imply that with probability tending to 1, 
$
C_1 \bm{I}_{m_i} \le \widehat{\bm{\Sigma}}_i  \le C_2 m_i\bm{I}_{m_i}
$
uniformly in $i$ for some positive constants $C_1$ and $C_2$.
As for $\widehat{\bm{\Sigma}}_i^{-1}$, 
\begin{align}
\widehat{\bm{\Sigma}}_i^{-1}- \bm{\Sigma}_i^{-1}
& = \widehat{\bm{\Sigma}}_i^{-1}
(\bm{\Sigma}_i- \widehat{\bm{\Sigma}}_i)\bm{\Sigma}_i^{-1}
\nonumber\\
& =  \bm{\Sigma}_i^{-1}(\bm{\Sigma}_i- \widehat{\bm{\Sigma}}_i)\bm{\Sigma}_i^{-1}
+ \widehat{\bm{\Sigma}}_i^{-1}(\bm{\Sigma}_i- \widehat{\bm{\Sigma}}_i)
\bm{\Sigma}_i^{-1}(\bm{\Sigma}_i- \widehat{\bm{\Sigma}}_i)\bm{\Sigma}_i^{-1}.
\nonumber
\end{align}

It follows from Proposition \ref{prop:prop4}, Assumption A4, and
the above identity that
\begin{equation}
\widehat{\bm{\Sigma}}_i^{-1}- \bm{\Sigma}_i^{-1}
=\bm{\Sigma}_i^{-1}(\bm{\Sigma}_i- \widehat{\bm{\Sigma}}_i)\bm{\Sigma}_i^{-1}+m_i^2
O_p\Big(
h_2^4 + h_3^4+ \frac{\log n}{nh_2} + \frac{\log n}{nh_3^2} 
\Big).
\label{eqn:e609}
\end{equation}
The last term in the right-hand side of (\ref{eqn:e609})  is in the sense of eigenvalue evaluation.
By using Assumption A4 and Proposition \ref{prop:prop4},
we get an expression of each element
of $ \bm{\Sigma}_i^{-1}(\bm{\Sigma}_i- \widehat{\bm{\Sigma}}_i)\bm{\Sigma}_i^{-1} $.
This expression, along with the assumptions for Theorem \ref{thm:thm1} and
the local property of the B-spline basis, will be employed in the proofs of the following
lemmas.
These lemmas, assuming the same assumptions as in Theorem \ref{thm:thm1}, 
are needed in order to evaluate $\widehat{I}_1- I_1$ and their proofs are given
in the supplement \cite{HCL2015S}. 

\begin{lem}
\label{lem:lem1} 
Let $h_{12,kl}$ and $\widehat{h}_{12,kl}$ be the $(k,l)$ element of $\bH_{12}$ and $\widehat{\bH}_{12}$, respectively.
Then we have uniformly in $k$ and $l$,
\begin{align*}
&\frac{1}{n} h_{12,kl}  =  O_p ( K_n^{-1} ), \,\,   \frac{1}{n}( h_{12,kl} -
\widehat{h}_{12,kl})  = K_n^{-1}O_p\Big(
h_2^2 + h_3^2 
+ \sqrt{\frac{\log n}{nh_2}} +  \sqrt{\frac{\log n}{nh_3^2}}
\Big) ,\\
& \Big\{ \sum_{l=1}^{qK_n} ( n^{-1}h_{12,kl} )^2 \Big\}^{1/2}
 = O_p (K_n^{-1/2}),\\
&\Big[ \sum_{l=1}^{qK_n} \{n^{-1}(h_{12,kl}-
\widehat{h}_{12,kl} ) \}^2 \Big]^{1/2}
 = K_n^{-1/2}O_p \Big( h_2^2 + h_3^2
+ \sqrt{\frac{\log n}{nh_2}} +  \sqrt{\frac{\log n}{nh_3^2}}
\Big).
\end{align*}
\end{lem}

\begin{lem}
\label{lem:lem2}
With probability tending to 1, 
$
C_1 K_n^{-1} \le \lambda_{\rm min}(n^{-1}\bH_{22})
\le \lambda_{\rm max}(n^{-1}\bH_{22}) \le C_2 K_n^{-1}
$
for some positive constants $C_1$ and $C_2$. We also
have
\begin{eqnarray*}
\lefteqn{
\max \big\{ |\lambda_{\rm min}(n^{-1}(\widehat{\bH}_{22}
- \bH_{22}))|, |\lambda_{\rm max}(n^{-1}(\widehat{\bH}_{22}
- \bH_{22}))| \big\} }\\
& = &  K_n^{-1}
O_p \Big( h_2^2 + h_3^2 
+ \sqrt{\log n /(nh_2)} + \sqrt{\log n /(nh_3^2)}
\Big).
\end{eqnarray*}
Hence we have
$
\max \big\{ |\lambda_{\rm min}(n^{-1}
\widehat{\bH}_{22})|, |\lambda_{\rm max}(n^{-1}
\widehat{\bH}_{22})| \big\}= O_p(K_n^{-1})
$
and \\
$\max \big\{ |\lambda_{\rm min}( (n^{-1}\widehat{\bH}_{22})^{-1}
- (n^{-1}\bH_{22})^{-1} )|, |\lambda_{\rm max}(
(n^{-1}\widehat{\bH}_{22})^{-1} - (n^{-1}\bH_{22})^{-1})| \big\}$
is also bounded from above by
$
K_n O_p \Big( h_2^2 + h_3^2 
+ \sqrt{\log n /(nh_2)} + \sqrt{\log n /(nh_3^2)}\Big).
$
\end{lem}

\begin{lem}
\label{lem:lem3}
We have 
$\frac{1}{n}\widehat{\bH}_{11} = \frac{1}{n}\bH_{11}+ o_p(1)$ and 
$\frac{1}{n}\widehat{\bH}_{12}\big( \frac{1}{n}\widehat{\bH}_{22}
\big)^{-1}\frac{1}{n}\widehat{\bH}_{21}
 = \frac{1}{n}\bH_{12}\big( \frac{1}{n}\bH_{22}
\big)^{-1}\frac{1}{n}\bH_{21}+o_p(1),
$ 
where $o_p(1)$ means 
both componentwise and in the meaning of eigenvalue evaluation.
Hence we have
$
n\widehat{\bH}^{11} = n \bH^{11} + o_p(1).
$
\end{lem}

\begin{lem}
\label{lem:lem4}
We have, for some positive constants $C_1$ and $C_2$,
$
\frac{C_1}{K_n}\bm{I}_{qK_n} \le {\rm cov}\Big(
\frac{1}{\sqrt{n}}\sum_{i=1}^n \underline{\bm{W}}_i^T
\bm{\Sigma}_i^{-1}\underline{\epsilon}_i\Big)
\le \frac{C_2}{K_n}\bm{I}_{qK_n}.
$
In addition we have
\begin{eqnarray*}
\lefteqn{
\Big| \frac{1}{\sqrt{n}}\sum_{i=1}^n \underline{\bm{W}}_i^T
(\widehat{\bm{\Sigma}}_i^{-1}- \bm{\Sigma}_i^{-1})\underline{\epsilon}_i
\Big| }\\
& = &  \sqrt{\frac{n}{K_n}}
O_p \Big( \frac{\log n}{nh_1} +
\frac{ \log n }{nh_2} + \frac{\log n}{nh_3^2} \Big) +
\sqrt{\frac{n}{K_n}}
O_p(h_1^3+ h_2^3+ h_3^3)\\
&  & \quad 
+ O_p(h_2^2 + h_3^2 ) + 
O_p\Big(\frac{1}{\sqrt{nh_2}}+ \frac{1}{\sqrt{nh_3^2}} + 
\frac{1}{\sqrt{nK_n}h_2} + \frac{1}{\sqrt{nK_n} h_3^2} \Big).
\end{eqnarray*}
\end{lem}

\begin{lem}
\label{lem:lem5}
We have for some positive constants $C_1$ and $C_2$,
$
C_1 \bm{I}_{p} \le {\rm cov}\Big(
\frac{1}{\sqrt{n}}\sum_{i=1}^n \underline{\bm{X}}_i^T
\bm{\Sigma}_i^{-1}\underline{\epsilon}_i \Big)
\le C_2 \bm{I}_{p}.
$
In addition we have
\begin{eqnarray*}
\lefteqn{
\Big| \frac{1}{\sqrt{n}}\sum_{i=1}^n \underline{\bm{X}}_i^T
(\widehat{\bm{\Sigma}}_i^{-1}- \bm{\Sigma}_i^{-1})\underline{\epsilon}_i
\Big|}\\
& = &  
 \sqrt{n}O_p \Big( \frac{\log n}{nh_1} + \frac{\log n}{nh_2}
+ \frac{\log n}{nh_3^2} \Big) + \sqrt{n}O_p(h_1^3+ h_2^3+ h_3^3)\\
& & + O_p(h_2^2 + h_3^2 ) + 
O_p\Big( 1/(\sqrt{n}h_2) + 1/(\sqrt{n} h_3^2) \Big).
\end{eqnarray*}
\end{lem}

\smallskip
\noindent
Now we prove that $\widehat{I}_1- I_1= o_p(n^{-1/2})$.
Write
\[
I_1= \bH^{11}\sum_{i=1}^n \underline{\bm{X}}_i^T
\bm{\Sigma}_i^{-1}\underline{\epsilon}_i - \bH^{11}\bH_{12}
\bH_{22}^{-1}\sum_{i=1}^n \underline{\bm{W}}_i^T
\bm{\Sigma}_i^{-1}\underline{\epsilon}_i= \bH^{11}(I_{11}- I_{12})
\ ({\rm say}).
\]
We define $\widehat{I}_{11}$ and $\widehat{I}_{12}$
similarly.
From Proposition \ref{prop:prop1} and  
Lemma \ref{lem:lem3}, we have only to prove
\begin{equation}
\frac{1}{\sqrt{n}}(\widehat{I}_{11}- I_{11})=o_p(1)
\quad {\rm and}\quad
\frac{1}{\sqrt{n}}(\widehat{I}_{12}- I_{12})=o_p(1).
\label{eqn:e615}
\end{equation}
The former result in (\ref{eqn:e615}) can be handled in the same way as the latter
and we consider only the latter.
Write
\begin{align}
\frac{1}{\sqrt{n}}(\widehat{I}_{12}- I_{12})
 &= \frac{1}{n}\widehat{\bH}_{12}\big( \frac{1}{n}\widehat{\bH}_{22} \big)^{-1}
\frac{1}{\sqrt{n}}\sum_{i=1}^n \underline{\bm{W}}_i^T
(\widehat{\bm{\Sigma}}_i^{-1}- \bm{\Sigma}_i^{-1})\underline{\epsilon}_i
\nonumber\\
&  \qquad + \frac{1}{n}\widehat{\bH}_{12}
\big\{ \big( \frac{1}{n}\widehat{\bH}_{22} \big)^{-1}-
\big( \frac{1}{n}\bH_{22} \big)^{-1} \big\}
\frac{1}{\sqrt{n}}\sum_{i=1}^n \underline{\bm{W}}_i^T
\bm{\Sigma}_i^{-1}\underline{\epsilon}_i
\nonumber\\
&  \qquad + \big( \frac{1}{n}\widehat{\bH}_{12}- 
\frac{1}{n}\bH_{12} \big)
\big( \frac{1}{n}\bH_{22} \big)^{-1}
\frac{1}{\sqrt{n}}\sum_{i=1}^n \underline{\bm{W}}_i^T
\bm{\Sigma}_i^{-1}\underline{\epsilon}_i
\nonumber\\
& = DI_{12}^{(1)}+ DI_{12}^{(2)} + DI_{12}^{(3)}
\quad {\rm (say)}.\nonumber
\end{align}
Lemmas \ref{lem:lem1}, \ref{lem:lem2}, and \ref{lem:lem4} imply
\begin{align*}
DI_{12}^{(1)} = & \sqrt{n}O_p \Big( \frac{\log n}{nh_1} +
\frac{ \log n }{nh_2} + \frac{\log n}{nh_3^2} \Big) +
 \sqrt{n}O_p(h_1^3+ h_2^3+ h_3^3) \\
 &  + \sqrt{K_n}
O_p\Big(\frac{1}{\sqrt{nh_2}}+ \frac{1}{\sqrt{nh_3^2}} + 
\frac{1}{\sqrt{nK_n}h_2} + \frac{1}{\sqrt{nK_n} h_3^2} \Big)\\
 & + \sqrt{K_n}O_p(h_2^2 + h_3^2 ) = o_p(1),\\
DI_{12}^{(j)} = & \sqrt{K_n}
O_p \Big( h_2^2 + h_3^2 
+ \sqrt{\log n/(nh_2)} +  \sqrt{\log n/(nh_3^2)}\Big)= o_p(1), \, j=2,3.
\end{align*}
Hence we have established
\begin{equation}
\widehat{I}_1- I_1= o_p(n^{-1/2}).
\label{eqn:e625}
\end{equation}

\smallskip
Next we deal with $\widehat{I}_2-I_2$ and two more lemmas are necessary.

\begin{lem}
\label{lem:lem6}
\begin{eqnarray*}
\lefteqn{
\Big|
\frac{1}{\sqrt{n}}\sum_{i=1}^n
\underline{\bm{W}}_i^T
\bm{\Sigma}_i^{-1}( \underline{\bm{W}}_i \bm{\gamma}^*
-\underline{(\bm{Z}^T\bm{g}_0)}_i)\Big|
=O_p(\sqrt{n}K_n^{-5/2}),
\quad\mbox{and}}\\
\lefteqn{\Big|
\frac{1}{\sqrt{n}}\sum_{i=1}^n
\underline{\bm{W}}_i^T( \widehat{\bm{\Sigma}}_i^{-1}-
\bm{\Sigma}_i^{-1})( \underline{\bm{W}}_i \bm{\gamma}^*
-\underline{(\bm{Z}^T\bm{g}_0)}_i)\Big|}\\
& = &  \sqrt{n}K_n^{-5/2}
O_p 
\Big( h_2^2 + h_3^2 
 + \sqrt{\log n /(nh_2)} + \sqrt{\log n /(nh_3^2)}
\Big).
\end{eqnarray*}
\end{lem}

\begin{lem}
\label{lem:lem7}
\begin{eqnarray*}
\lefteqn{
\Big|
\frac{1}{\sqrt{n}}\sum_{i=1}^n
\underline{\bm{X}}_i^T
\bm{\Sigma}_i^{-1}( \underline{\bm{W}}_i \bm{\gamma}^*
-\underline{(\bm{Z}^T\bm{g}_0)}_i)\Big|
=O_p(\sqrt{n}K_n^{-2}) \quad\mbox{and}}\\
\lefteqn{ \Big|
\frac{1}{\sqrt{n}}\sum_{i=1}^n
\underline{\bm{X}}_i^T( \widehat{\bm{\Sigma}}_i^{-1}-
\bm{\Sigma}_i^{-1})( \underline{\bm{W}}_i \bm{\gamma}^*
-\underline{(\bm{Z}^T\bm{g}_0)}_i)\Big|}\\
& =   \sqrt{n}K_n^{-2}
O_p 
\Big( h_2^2 + h_3^2 
 + \sqrt{\log n /(nh_2)} + \sqrt{\log n /(nh_3^2)}
\Big).
\end{eqnarray*}
\end{lem}

\smallskip
\noindent
Now we can show that $\widehat{I}_2- I_2= o_p(n^{-1/2})$. 
Write
\begin{align*}
I_2
= & \bH^{11}\sum_{i=1}^n \underline{\bm{X}}_i^T
\bm{\Sigma}_i^{-1}
( \underline{\bm{W}}_i \bm{\gamma}^*
-\underline{(\bm{Z}^T\bm{g}_0)}_i) - \bH^{11}\bH_{12}
\bH_{22}^{-1}\sum_{i=1}^n \underline{\bm{W}}_i^T
\bm{\Sigma}_i^{-1}
( \underline{\bm{W}}_i \bm{\gamma}^*
-\underline{(\bm{Z}^T\bm{g}_0)}_i)\\
= & \bH^{11}(I_{21}- I_{22})\ ({\rm say}).
\end{align*}
We define $\widehat{I}_{21}$ and $\widehat{I}_{22}$
similarly and write $\widehat{I}_2 = \widehat{\bH}^{11}(\widehat{I}_{21}- \widehat{I}_{22})$.
From Proposition \ref{prop:prop1} and
Lemma \ref{lem:lem3}, we have only to prove
$
\frac{1}{\sqrt{n}}(\widehat{I}_{21}- I_{21})=o_p(1)
$
and
$\frac{1}{\sqrt{n}}(\widehat{I}_{22}- I_{22})=o_p(1)$.
The former result in the above can be handled in the same way as the latter
and we consider only the latter.
Write
\begin{eqnarray}
\frac{1}{\sqrt{n}}(\widehat{I}_{22}- I_{22})
&= &\frac{1}{n}\widehat{\bH}_{12}\Big( \frac{1}{n}\widehat{\bH}_{22} \Big)^{-1}
\frac{1}{\sqrt{n}}\sum_{i=1}^n \underline{\bm{W}}_i^T
(\widehat{\bm{\Sigma}}_i^{-1}- \bm{\Sigma}_i^{-1})
( \underline{\bm{W}}_i \bm{\gamma}^*
-\underline{(\bm{Z}^T\bm{g}_0)}_i)
\nonumber\\
&& \hspace{-30pt} + \frac{1}{n}\widehat{\bH}_{12}
\Big\{ \Big( \frac{1}{n}\widehat{\bH}_{22} \Big)^{-1}-
\Big( \frac{1}{n}\bH_{22} \Big)^{-1} \Big\}
\frac{1}{\sqrt{n}}\sum_{i=1}^n \underline{\bm{W}}_i^T
\bm{\Sigma}_i^{-1}
( \underline{\bm{W}}_i \bm{\gamma}^*
-\underline{(\bm{Z}^T\bm{g}_0)}_i)
\nonumber \\
&&  \hspace{-30pt} + \Big( \frac{1}{n}\widehat{\bH}_{12}- 
\frac{1}{n}\bH_{12} \Big)
\Big( \frac{1}{n}\bH_{22} \Big)^{-1}
\frac{1}{\sqrt{n}}\sum_{i=1}^n \underline{\bm{W}}_i^T
\bm{\Sigma}_i^{-1}
( \underline{\bm{W}}_i \bm{\gamma}^*
-\underline{(\bm{Z}^T\bm{g}_0)}_i).
\nonumber\\
& = & DI_{22}^{(1)} +  DI_{22}^{(2)} +
DI_{22}^{(3)} \quad {\rm (say)}\nonumber
\end{eqnarray}
Lemmas \ref{lem:lem1}, \ref{lem:lem2}, and \ref{lem:lem6} imply, for $j=1,2,3$,
\begin{align*}
DI_{22}^{(j)} = &  
\sqrt{n}K_n^{-2} O_p \Big( h_2^2 + h_3^2 
+ \sqrt{\log n/(nh_2)} +  \sqrt{\log n/(nh_3^2)}
\Big)= o_p(1).  \,\, 
\end{align*}
Hence we have established
$
\widehat{I}_2- I_2= o_p(n^{-1/2}).
$
The desired result follows from (\ref{eqn:e600}), (\ref{eqn:e603}),
(\ref{eqn:e625}) and the above result. 

\vspace{0.1in}

\noindent{\bf Acknowledgements.}
The authors thank the associate editor and three referees for their thoughtful and constructive comments on a previous
submission, which led to significant improvement of this paper.

\begin{supplement}
\sname{Supplement A}\label{suppA}
\stitle{Additional simulation results and technical material}
\slink[doi]{xx.xxxx/xx-AOSxxxxSUPP}
\sdescription{Additional simulation results, proofs of the propositions
and lemmas, and theory for the case of uniformly bounded cluster size and general link function.}
\end{supplement}

\bibliographystyle{imsart-number}

\bibliography{longitudinal2}


\end{document}